\documentclass[11pt]{article}
\usepackage[utf8]{inputenc}
\usepackage{bm}
\usepackage{hyperref}
\usepackage{xcolor}
\definecolor{winered}{rgb}{0.5,0,0}

\hypersetup
{
   colorlinks=true,
  linkcolor=winered,
  urlcolor={winered},
  filecolor={winered},
  citecolor={winered},
  allcolors={winered}
}

\title{Allocating Stimulus Checks in Times of Crisis}
\author{

Marios Papachristou \thanks{Supported by a Cornell University Graduate Fellowship.} \\ 
Cornell University \\ \small{\texttt{papachristoumarios@cs.cornell.edu}}

\and

Jon Kleinberg \thanks{Supported in part by a Simons Investigator Award,
a Vannevar Bush Faculty Fellowship,
MURI grant W911NF-19-0217, AFOSR grant FA9550-19-1-0183, 
ARO grant W911NF19-1-0057, a Simons Collaboration grant,
and a grant from the MacArthur Foundation.} \\ 
Cornell University \\ \small{\texttt{kleinberg@cornell.edu}} 
}

\date{}
\usepackage[margin=1in]{geometry}
\usepackage{xurl}
\usepackage{graphicx}
\usepackage{amsmath}
\usepackage{amsfonts}
\usepackage{amssymb}
\usepackage{amsthm}
\usepackage{appendix}
\usepackage{algorithmicx}
\usepackage{algorithm}
\usepackage{algpseudocode} 
\usepackage{subfigure}
\usepackage{color-edits}
\usepackage{caption}
\usepackage{paralist}
\usepackage{tikz-network}
\usepackage[noabbrev, capitalise]{cleveref}

\addauthor{mp}{red}
\addauthor{jk}{blue}

\newtheorem{theorem}[]{Theorem}
\newtheorem{corollary}[]{Corollary}
\newtheorem{lemma}[]{Lemma}

\newtheorem{assumption}[]{Assumption}
\newtheorem{condition}[]{Condition}

\theoremstyle{remark}
\newtheorem{example}{Example}

\newcommand{\ev}[2]{\mathbb E_{#1} \left [ #2 \right ]}

\newcommand{\opt}[0]{\mathrm {OPT}}
\newcommand{\sol}[0]{\mathrm {SOL}}
\newcommand{\optr}[0]{\mathrm {OPT_{Rel}}}

\newcommand{\one}[0]{\mathbf 1}
\newcommand{\zero}[0]{\mathbf 0}
\newcommand{\D}[0]{\mathcal D}

\renewcommand{\L}[0]{\mathcal L}
\newcommand{\B}[0]{\mathcal B}
\newcommand{\sop}[0]{\mathrm {SoP}}
\newcommand{\sot}[0]{\mathrm {SoT}}
\newcommand{\soip}[0]{\mathrm {SoIP}}

\newcommand{\fs}[0]{\mathrm {FS}}
\newcommand{\as}[0]{\mathrm {AS}}
\newcommand{\explain}[1]{\tag*{(#1)}}
\newcommand{\til}[0]{\widetilde}
\newcommand{\iid}[0]{\overset {\mathrm{i.i.d.}} {\sim}}

\newcommand{\evx}[1]{\ev {X \sim \D} {#1}}
\newcommand{\evz}[1]{\ev {Z \sim \mathrm{Be}(\til z^*)} {#1}}
\newcommand{\prz}[1]{\Pr_{Z \sim \mathrm{Be}(\til z^*)} \left [ #1 \right ]} 
\newcommand{\prx}[1]{\Pr_{X \sim \D} \left [ #1 \right ]}

\newcommand{\asref}[1]{\textcolor{winered}{A}\ref{#1}}
\newcommand{\condref}[1]{\textcolor{winered}{C}\ref{#1}}

\begin{document}

\maketitle

\begin{abstract}
We study the problem of allocating bailouts (stimulus, subsidy allocations) to people participating in a financial network subject to income shocks. We build on the financial clearing framework of Eisenberg and Noe~\cite{eisenberg2001systemic} that allows the incorporation of a bailout policy that is based on discrete bailouts motivated by the types of stimulus checks people receive around the world as part of COVID-19 economical relief plans. We show that optimally allocating such bailouts on a financial network in order to maximize a variety of social welfare objectives of this form is a computationally intractable problem. We develop approximation algorithms to optimize these objectives and establish guarantees for their approximation ratios. Then, we incorporate multiple fairness constraints in the optimization problems and establish relative bounds on the solutions with versus without these constraints. Finally, we apply our methodology to a variety of data, both in the context of a system of large financial institutions with real-world data, as well as in a realistic societal context with financial interactions between people and businesses for which we use semi-artificial data derived from mobility patterns. Our results suggest that the algorithms we develop and study have reasonable results in practice and outperform  other network-based heuristics. We argue that the presented problem through the societal-level lens could assist policymakers in making informed decisions on issuing subsidies.

\medskip


\end{abstract}

\section{Introduction}

The COVID-19 pandemic has spread uncertainty among financial entities. Governments around the world are faced with the problem of saving entities from economic ruin that are subject to financial shocks. A policy framework, applied in many places throughout the world, is \emph{stimulus checks}, i.e. cash injections so that consumption is stimulated~\cite{baker2020income, baker2018debt, carroll2020modeling, broda2014economic, johnson2006household, parker2013consumer, nygaard2020optimal, nygaard2021optimal, abebe2020subsidy}. A cardinal question faced by policymakers is \emph{who gets the subsidies}. In the US the CARES act \cite{act2020cares} gives a stimulus check within a certain range of income and increases the payments proportionally to the number of dependents. Other countries such as, for example, New Zealand\footnote{\url{https://www.theguardian.com/world/2020/mar/17/new-zealand-launches-massive-spending-package-to-combat-covid-19}} and Greece\footnote{\url{https://www.ekathimerini.com/economy/258852/measures-announced-for-relieving-lockdown-pressure-on-workers-households}} offer stimulus checks of fixed value to affected employees. A common pattern in these cases is that when somebody's income is below a certain threshold or satisfies some criterion then the household is entitled to a stimulus check of fixed value. 

An important limitation is that these rules ignore \emph{contagion} effects through the financial network. For example, if a business defaults that may translate to job loss for the employees who in their turn may not be able to pay their rent etc. The subject of who to bailout in a financial crisis is a subject of controversy~\cite{jihong2009save, congleton2009political, casey2015framework, brill2009sa, rosas2006bagehot, blau2013corporate, bechtel2017policy}, namely the policymakers are faced with the following quandary: do we need to bailout large businesses whose saving from default would help maintain job positions but can also make them richer and stronger with respect to the rest of the population, or do we need to bailout individuals (or groups thereof) and small businesses so as they are able to afford their rent and groceries with the fear of letting bigger companies collapse?

\noindent \textbf{Theoretical Contributions.} We develop a theoretical framework for optimally allocating discrete bailouts on the well-studied Eisenberg-Noe (EN) model \cite{eisenberg2001systemic} of contagion with the presence of income shocks, in order to maximize welfare objectives. We show that finding the optimal policy under the discrete bailouts scheme is an NP-Hard problem, develop approximation algorithms, and prove hardness-of-approximation results. We also find that thresholding entities with respect to their equity performs poorly. We then turn to a set of questions within the model that relate to fairness and equity considerations in the provision of bailouts.
In particular, we optimize the same objectives as above subject to an \emph{algorithmic fairness constraint} that regards stimulus distribution via the \emph{Gini Coefficient}~\cite{gini1921measurement}. We study three notions of fairness: (i) the classical Gini Coefficient that measures all-pairs inequality, (ii) a variant of the GC introduced in~\cite{mota2021fair} adjusted to our model, (iii) a novel Gini Coefficient index based on the attributes of each node in the financial network (e.g. minority status). We show that the Price of Fairness (PoF) can be unbounded in the discrete case and bounded in the fractional case, and devise bounds on the PoF for the fractional case. Our bounds relate to quantities from spectral graph theory, namely the \emph{conductance} of the underlying \emph{nominal liability network}. 

\noindent \textbf{Empirical Contributions.} We apply our algorithms on real-world banking datasets from~\cite{glasserman2015likely, chen2016financial} from bank stress-testing, and semi-artificial data such as social payments data from the Venmo platform, and  societal-granularity network topologies inferred by anonymized \emph{mobility data} from the SafeGraph platform, the US Census, and the US Economic Census. We compare the results of the proposed approximation algorithms with other network-based heuristics on these datasets. We empirically study the incorporation of the fairness constraint both between all pairs of nodes as well as in a minority/non-minority context on the semi-artificial data.

\noindent \textbf{Paper Organization.} The paper is organized as follows: \cref{sec:eisenberg_noe} gives a brief description of the Eisenberg-Noe framework for computing clearing payments and introduces the objectives that will be studied throughout the paper. \cref{sec:generalized_bailouts} describes the Generalized Discrete Bailouts Problem which is the general formulation of stochastic optimization problem we solve in the paper. \cref{sec:np_hardness} establishes the hardness results of the discrete bailouts problem. The remainder of \cref{sec:approximation} studies approximation algorithms with provable guarantees, and establishes inapproximability results for certain objectives. \cref{sec:fairness} relates metrics of algorithmic fairness to the problem and proposes ways to incorporate fairness when designing a stimulus allocation problem. \cref{sec:experiments} provides data-driven experiments with \emph{real-world}, and \emph{semi-artificial} datasets. \cref{sec:discussion} discusses the experiments as well as generalizations of the models studied in this paper, the societal impact of the work, as well as further related work. \cref{sec:conclusion} provides concluding remarks and future directions. Proofs of theoretical nature have been deferred to \cref{sec:ommited} and are presented in serial order. An extensive data analysis can be found in \cref{sec:data_analysis} and data ethics statement in \cref{sec:data_ethics}.

\noindent \textbf{Basic Notation.} $[n]$ denotes the set $\{ 1, \dots, n \}$. $\| x \|_p$ denotes the $p$-norm of the vector $x$. When $p = 2$ the subscript may be omitted. $\zero$ (resp. $\one$) denote the all zeros (resp. all ones) column vector. $\one_S$ represents the indicator column vector of the set $S$. For a matrix $A$, $\| A \|_p$ denotes the induced $p$-norm of $A$. $x \wedge y$ (resp. $x \vee y$) denotes the coordinate-wise minimum (resp. maximum) of $x, y$. $(x)^+ = \zero \vee x$ denotes the non-negative part of $x$. $A \sqcup B$ denotes the disjoint union of $A$ and $B$ ($A \cap B = \emptyset$). $\ge, \le, >, <$ denote coordinate-wise ordering. $\til O ( \cdot )$ ignores poly-logarithmic factors. $A \odot B$ is the Hadamard (element-wise) product of $A, B$. $A^{(k)}$ is defined as $A^{(0)} = I$ and $A^{(k + 1)} = A^{(k)} \odot A$ for $k \ge 1$. For a vector $v > \zero$ we define $\zeta(v) = \tfrac {v_{\max}} {v_{\min}} \ge 1$ to be a measure of \emph{conditioning} of the vector $v$ in the sense that $\zeta(v)$ is the condition number of the $n \times n$ matrix $\mathrm{diag}(v_1, \dots, v_n)$. When clear from the context, we use the shorthand notation $\zeta = \zeta(v)$.

\subsection{The Eisenberg-Noe Model} \label{sec:eisenberg_noe}

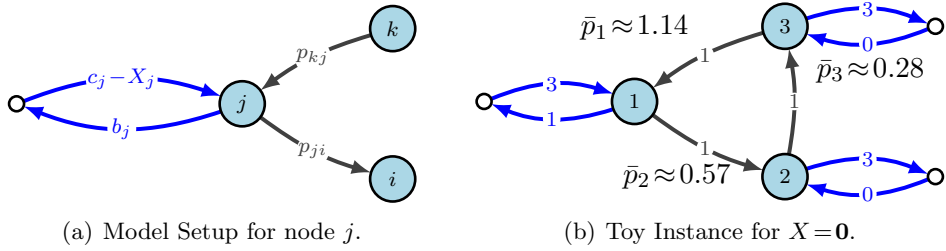
\begin{figure}
    \centering
    \subfigure[Model Setup for node $j$.]{
    \begin{tikzpicture}
        \Vertex[x=0, y=0, label=$j$]{i}
        \Vertex[x=2, y=-1, label=$i$]{j}
        \Vertex[x=2, y=1, label=$k$]{k}
        \Edge[Direct, bend=-10, label=$p_{ji}$](i)(j)
        \Edge[Direct, bend=-10, label=$p_{kj}$](k)(i)
        \Vertex[x=-3, y=0, size=0, opacity=0]{bi}
        \Edge[Direct, bend=20, label=$b_j$, color=blue](i)(bi)
        \Edge[Direct, bend=20, label=${c_j \! - \! X_j}$, color=blue](bi)(i)
    \end{tikzpicture}
    } \quad
    \subfigure[Toy Instance for $X \! = \! \zero$.]{
        \begin{tikzpicture}
        \Vertex[x=0, y=0, label=$1$]{i}
        \Text[x=0, y=0, distance=0.75, position=above]{$\bar p_1 \! \approx \! 1.14$}
        \Vertex[x=2, y=-1, label=$2$]{j}
        \Text[x=2, y=-1, position=left, distance=0.65]{$\bar p_2 \! \approx \! 0.57$}
        \Vertex[x=2, y=1, label=$3$]{k}
        \Text[x=2, y=1, position=below right, distance=0.5]{$\bar p_3 \! \approx \! 0.28$}
        \Edge[Direct, bend=-10, label=$1$](i)(j)
        \Edge[Direct, bend=-10, label=$1$](j)(k)
        \Edge[Direct, bend=-10, label=$1$](k)(i)
        \Vertex[x=-2, y=0, size=0, opacity=0]{bi}
        \Edge[Direct, bend=20, label=$1$, color=blue](i)(bi)
        \Edge[Direct, bend=20, label=$3$, color=blue](bi)(i)
        \Vertex[x=4, y=-1, size=0, opacity=0]{bj}
        \Edge[Direct, bend=20, label=$3$, color=blue](j)(bj)
        \Edge[Direct, bend=20, label=$0$, color=blue](bj)(j)
        \Vertex[x=4, y=1, size=0, opacity=0]{bk}
        \Edge[Direct, bend=20, label=$3$, color=blue](k)(bk)
        \Edge[Direct, bend=20, label=$0$, color=blue](bk)(k)
    \end{tikzpicture}
    
    }
    
    \caption{Eisenberg-Noe Model.}
    \label{fig:en_model}
\end{figure}

\noindent \textbf{Setup.} The EN model considers a directed network of payments $G([n], E)$ with the following structure. The network has $n$ nodes and internal liabilities which are represented on the directed edges $E$. Each node $j \in [n]$ of the payment network has an external influx of assets $c_j \ge 0$ and external liabilities $b_j \ge 0$ which correspond to the node's external exchanges. A direct way to think about external liabilities are \emph{taxes} that natural and legal people owe to their governments. Between two nodes $(j, i) \in E$ there is a liability $p_{ji} \ge 0$ from $j$ to $i$, which denotes how much $j$ owes to $i$ (in terms of monetary units, e.g. USD). The initial \emph{wealth}, or \emph{equity}, of node $j$ is given by 

\begin{equation*}
    w_j = c_j + \sum_{i : i \sim j} p_{ij} - p_j
\end{equation*}

where $p_j = b_j + \sum_{i: j \sim i} p_{ji}$ are the cumulative liabilities of node $j$. The network contains no isolated nodes, i.e. every node $j$ has $p_j > 0$ (either external liabilities, or internal liabilities or both).  With $\beta_j = (p_j - b_j) / p_j$ we refer to the \emph{financial connectivity} of $j$, i.e. the fraction of total liability that $j$ owes \emph{within} the network (see also \cite{glasserman2015likely}). Subsequently, $1 - \beta_j = b_j / p_j$ denotes the fraction of total liabilities $j$ owes \emph{outside} the network. If $w_j \ge 0$ then $j$ is said to be \emph{solvent}, i.e. it can pay all of its obligations, and has a profit of $w_j - p_j$. Otherwise the node is \emph{default} (or insolvent) in which case the node cannot satisfy all of its payments. In this case, the node defaults to its creditors and, according to the EN model, rescales its liabilities in order to pay its creditors to a reduced rate, according to the amount of money it owes to them. The relative liabilities matrix $A$ is defined to have entries $a_{ji} = p_{ji} / p_j$ whenever $p_j > 0$ and $a_{ji} = 0$ otherwise. We define $\beta_{\min} = \min_{j \in [n]} \beta_j$ and $\beta_{\max} = \max_{j \in [n]} \beta_j$. Later in the paper, the quantity $\beta_{\max}$ will play an important role regarding the algorithms we develop. Moreover, $\beta_{\max}$ determines the contraction properties of the clearing mapping. The clearing payment vector $\bar p$ is given by the EN equilibrium equations 

\begin{equation} \label{eq:en}
    \bar p = p \wedge \left ( A^T \bar p + c \right )
\end{equation}

The set of default nodes is defined to be the set $D = \{ j : \bar p_j < p_j \}$ whereas the set of solvent nodes is the set $R = \{ j : \bar p_j = p_j \}$, such that $V = D \sqcup R$\footnote{Throughout the rest of the paper we will use superscripts to denote the context under which we refer to the sets of default and solvent nodes and the clearing payment vector.}. When $\beta_{\max} = 1$ it can be shown that the fixed point operator $\bar p \overset {\Phi} {\mapsto} p \wedge \left ( A^T \bar p + c \right ) $ is a non-expansion and thus the EN model has \emph{multiple equilibria} which are bounded by the best-case equilibrium $\bar p^+$ and the worst-case equilibrium $\bar p^-$ where $\bar p^+ \ge \bar p^-$ due to the Knaster-Tarski Theorem~\cite{hayashi1985self}. For asserting the uniqueness of the equilibrium vector, namely imposing that $\bar p^+ = \bar p^-$,~\cite{glasserman2015likely} make the following assumption and prove uniqueness under it: 

\begin{assumption}[Uniqueness of Clearing Vector~\cite{glasserman2015likely}.] \label{assumption:glasserman_contraction}
    If every node $i$ can access a node $j$ such that $j$ has $b_j > 0$, then the EN model has a unique equilibrium $\bar p$, i.e. $\bar p^- = \bar p^+ = \bar p$. 
\end{assumption}

 Under this assumption one can apply the mapping $\Phi(\bar p) = p \wedge \left ( A^T \bar p + c \right )$ to iteratively compute the equilibrium vector. Under \cref{assumption:glasserman_contraction}, $\Phi$ is a contraction since the spectral radius of $A$ is less than 1, and the uniqueness of the equilibrium is asserted from to Banach's Fixed Point Theorem Moreover, $\Phi$ is increasing, positive, and concave~\cite{eisenberg2001systemic}.  In our paper, we rely on a very realistic assumption, that every node is directly obliged to the external sector, i.e.

\begin{assumption} \label{assumption:beta_max_lt_one}
    $\| A^T \|_1 = \beta_{\max} < 1$. The following are equivalent: (i) $A$ has strictly substochastic rows, and (ii) for all $j \in [n]$ there are obligations to the external sector. Subsequently the spectral radius of $A$ satisfies $\rho(A) \le \beta_{\max} < 1$. 
\end{assumption}

In the context of studying topologies involving people, \cref{assumption:beta_max_lt_one} is a reasonable assumption as we are able to model the external liabilities with a tax (e.g. income tax) or other fixed costs (e.g. rent/utilities payment) that is owed to an external entity (e.g. government, physical or legal person). 

At equilibrium, the set of solvent nodes is defined to be $R = \{ j : \bar p_j = p_j \}$ and the set of default nodes is defined to be $D = \{  j : \bar p_j < p_j \} = [n] \setminus R$. The vector $\bar p$ can also be found via solving an optimization problem of the following form~\cite[p. 10]{eisenberg2001systemic}: 

\begin{equation} \label{eq:gen_opt_en}
    \begin{split}
        \max \quad & f(\bar p) \\
        \text{s.t.} \quad & \bar p \le A^T \bar p + c \\
         & \zero \le \bar p \le p,
    \end{split}
\end{equation}

where $f$ is a \emph{strictly increasing} function of $\bar p$. We will also refer to the convex polytope $\mathcal P(P, b, c) = \{ \; \bar p : \zero \le \bar p \le p, (I - A^T) \bar p \le c \; \}$ as the \emph{domain} of the network parametrized by $P, b, c$ with $A$ being the relative liability matrix computed as a function of $P$ and $b$. 

\noindent \textbf{Objectives.} The objectives that we study in this paper (in terms of maximization) are of two kinds: The former type of objective is a linear objective that is a sum (with positive coefficients) of the payments in equilibrium, and the latter one is the total number of solvent nodes. Maximizing such objectives is reasonable both from a societal and a technical viewpoint (see also  \cite{ahn2019optimal, abebe2020subsidy, NisaRougTardVazi07, liu2010sensitivity}). From a societal viewpoint, the maximization of a sum (with positive coefficients) of the total payments in the network results in the (assumingly honest) nodes being incentivized to clear their liabilities for the benefit of society, and, thus, each node's \emph{utility} corresponds to the liabilities it is able to pay off. From a technical viewpoint, the positively-weighted linear objectives are known to yield clearing vectors, and the number of solvent nodes objective can also be slightly modified to yield a clearing vector.

\begin{compactenum} \label{enum:objectives} 

\item A linear objective parametrized by a vector $v > \zero$, i.e.

    \begin{equation} \label{eq:linobj} \tag{L-OBJ}
        f(\bar p) = v^T \bar p.
    \end{equation}

    This form of objective will be largely used throughout the paper since most welfare measures have a natural formulations in this context. Examples of such measures constitute the following:

\begin{compactenum} 
 
    \item \textbf{Sum of Payments (SoP).} The SoP objective is just the sum of the clearing payments, namely

    \begin{equation} \label{eq:sop} \tag{SoP}
        f_{\sop} (\bar p) = f(\bar p; \one) = \one^T \bar p. 
    \end{equation}
    
    \item \textbf{Sum of Internal Payments (SoIP).} The SoIP objective equals the sum of clearing payments with respect to within-network liabilities, that is
    
    \begin{equation} \label{eq:soip} \tag{SoIP}
        f_{\soip} (\bar p) = f(\bar p; \beta) = \beta^T \bar p. 
    \end{equation}
    
    \item \textbf{Sum of Taxes (SoT).} The SoT objective equals the sum of clearing payments with respect to outside liabilities (i.e. taxes), that is
    
    \begin{equation} \label{eq:sot} \tag{SoT}
        f_{\sot} (\bar p) = f(\bar p; \one - \beta) = (\one - \beta)^T \bar p.
    \end{equation}

    Note that $f_\sop = f_\soip + f_\sot$. 

    \item \textbf{Fractional Solvency (FS).} The objective is defined as
    
    \begin{equation} \label{eq:fs} \tag{FS}
        f_{\fs} (\bar p) = \sum_{j \in [n]} \frac {\bar p_j} {p_j}.
    \end{equation}
 
\end{compactenum}
 
    \item \textbf{Absolute Solvency (AS).} The objective is defined as 
    
    \begin{equation} \label{eq:as} \tag{AS}
        f_{\as} (\bar p) = \sum_{j \in [n]} \one \{ \bar p_j = p_j \} = | R|.
    \end{equation}

\end{compactenum}

In the above definitions there are some slight technical details that we now clarify. First, the~\eqref{eq:as} objective is \emph{not} strictly increasing. To obtain an $\epsilon$-approximation we augment the objective to include a small amount of the total payments to be added to the objective controlled by an approximation parameter $\epsilon > 0$. We defer the statement to \cref{sec:transform}. To make \eqref{eq:soip} strictly monotone we need $\beta > \zero$. Secondly, in the~\eqref{eq:soip} objective there may exist $\beta_j $ such that $\beta_j = 0$. In this case, we follow the same procedure to calculate the clearing vector if we choose to do so via optimization. Lastly, note that since the equilibrium is unique, the optimal solution to all the equations is the \emph{same} clearing vector, since it can be computed independently by  iteratively applying $\Phi$ (see Lemma 4 of~\cite{eisenberg2001systemic}) and then evaluating $f$ on the fixed point of $\Phi$.

\noindent \textbf{Shock Model.} In the presence of \emph{shocks}, Glasserman and Young \cite{glasserman2015likely} proposed the following alteration to the EN model: Suppose that a shock $x = (x_1, \dots, x_n)^T$ hits the network where each node $j$ receives a shock $x_j \in [0, c_j]$. Then, the adjusted EN equilibrium equations become

\begin{equation}
    \bar p = p \wedge \left ( A^T \bar p + c - x \right )
\end{equation}

where the notation holds as in the case of \eqref{eq:en}. The set of \emph{default nodes} is again defined to be $D = \{ j : \bar p_j < p_j \}$ and the set of \emph{solvent} nodes is defined to be $R = \{ j : \bar p_j = p_j \}$. Following the paradigm of~\cite{glasserman2015likely}, we impose randomness on the shocks. More concretely, the shocks are drawn from a distribution $\D$ with support the set $[\zero, c]$. In the context of randomization, we are interested in the influence of the shocks on the objectives we study \emph{on expectation under the distribution $\D$}. In the majority of our proofs, we will first fix a shock, i.e. condition on the event $\{ X = x \}$, and then extend the result by computing $\evx {\cdot}$ from the law of total expectation. 

Finally, we will use the following Lemma as a tool in many of our proofs. Briefly, the Lemma states that if the external assets vector is increased (point-wise) then the corresponding clearing vector is (point-wise) greater than the previous one. Formally, we prove the following.

\begin{figure}
    \centering
    \subfigure[Before shock]{
       \begin{tikzpicture}[transform shape,scale=0.75]
           \Vertex[size=0,opacity=0,y=2]{0}
           \Vertex[label=1,size=1,opacity=0.7]{1}
           \Vertex[label=2,x=3,size=1,opacity=0.7]{2}
           \Vertex[y=-2,opacity=0,size=0]{3}
           \Vertex[y=-2,x=3,opacity=0,size=0]{4}
           \Edge[Direct,label=3/2,fontcolor=blue](0)(1)
           \Edge[Direct,label=1,fontcolor=red](1)(2)
           \Edge[Direct,label=1/2,fontcolor=orange](1)(3)
           \Edge[Direct,label=1,fontcolor=orange](2)(4)
           \Text[x=1.2,y=1]{$\bar p_1 = \tfrac 3 2$}
           \Text[x=3,y=1]{$\bar p_2 = 1$}
       \end{tikzpicture} 
    } \quad
    \subfigure[After shock]{
       \begin{tikzpicture}[transform shape,scale=0.75]
           \Vertex[size=0,opacity=0,y=2]{0}
           \Vertex[label=1,size=1,opacity=0.7]{1}
           \Vertex[label=2,x=3,size=1,opacity=0.7]{2}
           \Vertex[y=-2,opacity=0,size=0]{3}
           \Vertex[y=-2,x=3,opacity=0,size=0]{4}
           \Edge[Direct,label=1/2,fontcolor=blue](0)(1)
           \Edge[Direct,label=1/3,fontcolor=red](1)(2)
           \Edge[Direct,label=1/6,fontcolor=orange](1)(3)
           \Edge[Direct,label=1/3,fontcolor=orange](2)(4)
           \Text[x=1.2,y=1]{$\bar p_1 = \tfrac 1 2$}
           \Text[x=3,y=1]{$\bar p_2 = \tfrac 1 3$}
       \end{tikzpicture} 
    } \quad
    \subfigure[After Bailouts]{
       \begin{tikzpicture}[transform shape,scale=0.75]
           \Vertex[size=0,opacity=0,y=2]{0}
           \Vertex[label=1,size=1,opacity=0.7]{1}
           \Vertex[label=2,x=3,size=1,opacity=0.7]{2}
           \Vertex[y=-2,opacity=0,size=0]{3}
           \Vertex[y=-2,x=3,opacity=0,size=0]{4}
           \Edge[Direct,label=3/2,fontcolor=blue](0)(1)
           \Edge[Direct,label=1,fontcolor=red](1)(2)
           \Edge[Direct,label=1/2,fontcolor=orange](1)(3)
           \Edge[Direct,label=1,fontcolor=orange](2)(4)
           \Text[x=1.2,y=1]{$\bar p_1 = \tfrac 3 2$}
           \Text[x=3,y=1]{$\bar p_2 = 1$}
       \end{tikzpicture} 
    } 
    \caption{\cref{example:simple} instance.}
    \label{fig:example}
\end{figure}

\begin{lemma}[Comparison Lemma] \label{lemma:comparison}
    Let $\mathcal N_1 = (P, c_1, b)$ and $\mathcal N_2 =(P, c_2, b)$ be two EN instances with $c_1 \ge c_2$ (w.l.o.g. with $X = 0$). Then for the clearing vectors $\bar p_1$ and $\bar p_2$ it holds that $\bar p_1 \ge \bar p_2$ and $v^T \bar p_1 \ge v^T \bar p_2$ for every vector $v > \zero$. Moreover, if $D_1$ is the set of the default nodes of $\mathcal N_1$, then the vector $\tilde \delta = \one_{D_1} \odot (c_1 - c_2) \ge \zero$ satisfies $\bar p_1 \ge \bar p_2 + \tilde \delta$, and subsequently $v^T (\bar p_1 - \bar p_2) \ge v^T \tilde \delta$. 
\end{lemma}

\noindent \textbf{Remark.} In contrast to earlier works, we do a slight change in notation. Earlier works, denote the clearing vector with $p$ instead of $\bar p$ and the total liabilities with $\bar p$ instead of $p$. We make this change in notation to improve readability, and ease the reader in distinguishing the input variables, from the clearing vectors in the discrete bailouts case, and the fractional case. 

\subsection{Analytical Overview of Contributions} \label{sec:contributions}

Having set up the model we are ready to state our contributions in detail: 

\begin{compactenum}
    \item We set up the optimization problem of discrete bailouts regarding various objectives (\cref{sec:generalized_bailouts}), and prove that the corresponding optimization problems are NP-Hard (\cref{sec:np_hardness} and \cref{theorem:hardness}). The problems correspond to general Mixed-Integer-Linear-Stochastic-Program\-ming Problems (MILSP). 
    
    \item We propose two families of approximation algorithms:  
    
    \begin{compactenum}
    
    \item Based on the \emph{continuous relaxations} of the aforementioned problems we design algorithms with the following approximation guarantees based on simple \emph{randomized rounding schemes} (\cref{sec:approx_lp}): More specifically, for every linear objective we introduce a family of algorithms with approximation ratios equal to $\tfrac {1 - \beta_{\max}} \zeta - o(1)$. 
    
    \item For a linear objective $v^T \bar p$ with $v > \zero$, a greedy \emph{hill-climbing} algorithm that bails out the node with the largest marginal gain achieves an approximation ratio of $1 - e^{-(1 - \beta_{\max}) / \zeta} - o(1)$ under the reasonable condition that a \emph{Small-Bailout Regime} holds. This subsequently provides approximation guarantees for all the linear  objective functions presented in \cref{sec:eisenberg_noe} (\cref{sec:approx_greedy}, \cref{lemma:approximation_greedy_induction}). 
    
    \item For the~\eqref{eq:as} objective we prove that it is inapproximable within a factor $\tfrac {k + n \cdot a(| \mathcal I|)} {k + n} = \Omega \left ( a (| \mathcal I |) \right )$, where $a(| \mathcal I |) \in \mathbb N^*$ is a poly-time computable function of the input instance size $| \mathcal I |$, unless P = NP. (\cref{sec:inapproximability} and \cref{theorem:as_inapproximability}). 
    
    \end{compactenum}
    
    \item We impose fairness constraints to our model in terms of the Gini Coefficient. We study three alternative indices on bailouts related to the overall inequality of the network, inequality over groups with a certain property (e.g. minority status), as well spatial inequality (i.e. each node with respect to its neighbors). We define a Price of Fairness with respect tho these measures, and study the behaviour of the PoF in these cases, connecting it with ideas from spectral graph theory. We show that in the discrete case there exist instances where the PoF is unbounded. On the contrary, in the fractional case, the PoF is always finite for increasing objectives $f(\bar p)$ with $f(\zero) = 0$ (such as in the case of linear objectives with $v > \zero$). 
    
    \item We run our algorithms on real-world and semi-artificial data and get insights about the corresponding networks under the presence of random shocks and quantify their behaviour. Moreover, we track the evolution of the variables of the continuous relaxations with respect to the changes in number of to-be-bailed-out nodes and devise \emph{fairness measures} regarding the fair allocation of stimulus checks. (\cref{sec:experiments}) 
\end{compactenum}

\subsection{The Generalized Discrete Bailouts Problem} \label{sec:generalized_bailouts}

We define Generalized Discrete Bailouts Problem (GDBP) as the solution to the following Mixed-Integer-Linear-Stochastic-Programming Problem, parametrized by the strictly increasing function $f$ (see \cref{sec:eisenberg_noe}), the budget $\Lambda$, the relative liabilities $A$, the total liabilities $p$, a stimulus vector $L > \zero$, and the shock distribution $\D$, with decision variables the continuous clearing payments $\bar p_i$, and the bailout indicator variables $\bar z_i \in \{ 0, 1 \}$. 

\begin{equation} \label{eq:generalized_bailouts} \tag{GDBP}
    \begin{split}
        \max \quad & \evx {f(\bar p)} \\
        \text{s.t.} \quad & \bar p \le A^T \bar p + c - X + L \odot \bar z \\
        & \zero \le \bar p \le p \\
        & L^T \bar z \le \Lambda \\ 
        & \bar z \in \{ 0, 1 \}^n.
    \end{split}
\end{equation}

We denote the overall vector as $\bar \xi = (\bar p, \bar z)^T \in [\zero, p] \times \{ 0, 1 \}^n$. Note that bailing out a solvent node does not increase the objective, so the support of $\bar z$ is a subset of the default nodes. The continuous relaxation \eqref{eq:generalized_bailouts_relaxation} of \eqref{eq:generalized_bailouts} allows $\bar z_i$ to range in the continuous $[0, 1]$ interval, where the variables that refer to the relaxation are denoted by $\til \xi = (\til p, \til z)$;

\begin{equation} \label{eq:generalized_bailouts_relaxation} \tag{RGDBP}
    \begin{split}
        \max \quad & \evx {f(\til p)} \\
        \text{s.t.} \quad & \til p \le A^T \til p + c - X + L \odot\til z \\
        & \zero \le \til p \le p \\
        & L^T \til z \le \Lambda \\ 
        & \zero \le \til z \le \one . 
    \end{split}
\end{equation}

Given an optimal solution $\opt_f$ to \eqref{eq:generalized_bailouts} and an optimal solution $\optr_f$ to \eqref{eq:generalized_bailouts_relaxation} the optimality conditions impose that $\optr_f \ge \opt_f$ since the relaxed problem has a larger feasible region. A feasible solution of \eqref{eq:generalized_bailouts} is denoted by $\sol_f$ and satisfies $\sol_f \le \opt_f$. Finally, we denote a feasible solution by $\bar \xi$ (resp. $\til \xi$), an optimal solution by $\bar \xi^*$ (resp. $\til \xi^*$). For warm-up we give the following example (see ~\cref{fig:example}). Note that if $L = \ell \cdot \one$ then the problem reduces to finding at most $k = \Lambda / \ell$ nodes such that the corresponding objective is maximized.

\begin{example}[see~\cref{fig:example}] \label{example:simple}
Let $\Lambda = 1$, $L = \one$ and the network that has nodes $\{ 1, 2 \}$, external assets $c = (1.5, 0)^T$, external liabilities  $b = (0.5, 1)^T$ and internal liability matrix $P$ with $p_{ij} = \one \{ i = j = 1 \}$. A point mass shock $x = (1, 0)^T$ hits the network and causes node 1 to default, which in turn causes node 2 to default. Thus node 1 can only pay $1 / 3$ to node 2 and $1/6$ to the external creditors and, in turn, node 2 can only pay $1/3$ to its external creditors.  The optimal solution is to bailout node 1 with 1 unit of cash at which case everyone is solvent, since the shock is fully averted. We have that $\opt_{\sop} = 5/2$, $\opt_{\as} = 2$, $\opt_{\fs} = 2$, $\opt_{\sot} = 3/2$, and $\opt_{\soip} = 1$. 
\end{example}

\noindent \textbf{Decision version of~\eqref{eq:generalized_bailouts}.} The decision version of the problem takes as input a network with liability matrix $P$, external assets $c$, internal assets $b$, a stimulus vector $L > \zero$, a stimulus value a budget $\Lambda \ge 0$ to be used for the bailouts, a shock distribution $\mathcal D$, and a lower bound $f^*$ and answers YES if and only if there exists a set of nodes which are bailed out with $L$ such that in equilibrium we have that $\evx {f(\bar p)}  \ge f^*$. 

\noindent \textbf{Construction of $L$.} The construction of the bailout vector $L$ differs according to cases. For instance, the stimulus $L_j$ of an node $j \in [n]$ under the CARES Act  was  determined as follows: First a \$1,200 check was allocated to individuals without children, with income at most \$75K whose amount faded out until the value of \$100K where for incomes greater than \$100K the value of the stimulus was zero. Married couples filing tax returns jointly received \$2.4K if their income was less than \$150K. For each dependent (child) an extra of \$500 was allocated. In other cases, such as in the case of Greece, an employee of a business that shut down due to COVID-19 received a bailout of a fixed amount. The above policies suggest computing the stimuli as a product of a non-negative value that depends on the node's features and a binary decision variable that captures some other property. In a  system with ``societal granularity'' the bailouts tend not to be fractionally dependent on the structure of the underlying financial network between people, mainly for equity issues, since, for instance, two households with identical features (e.g. number of dependents) shall get \emph{equal} bailouts. In large-scale financial systems (e.g. banks, economy sectors) fractional bailouts are considered and fractional allocation of bailouts subject to the EN model has been attempted in the past (see~\cref{sec:related_work}). This choice of bailouts creates a very interesting computational \emph{dichotomy}, namely the fractional bailout allocation problem is efficiently solvable, whereas the discrete bailout allocation problem is not, as we prove in~\cref{sec:np_hardness}.

\section{Approximation Algorithms} \label{sec:approximation}

We prove that the problem of determining the optimal bailouts when these are dictated by discrete decisions is NP-Hard. To mitigate this problem, we develop approximation algorithms. 
Our first algorithm is based on \emph{randomized rounding}. More specifically, we solve the relaxation problem given in~\eqref{eq:generalized_bailouts_relaxation} and then we apply a randomized rounding scheme by rounding  the decision variables randomly with coin flips of bias equal to their optimal values. While the rounding scheme itself is straightforward, the analysis is more subtle, and shows that this algorithm achieves an approximation guarantee of $\tfrac {1 - \beta_{\max}} {\zeta} - o(1)$ for every linear objective given by a vector $v > \zero$ of coefficients with $\zeta = \zeta(v)$.  Our second result is a greedy algorithm that always chooses the node with the maximum gain in welfare subject to respecting the budget constraints. This type of algorithm has been extensively used for maximizing monotone submodular functions subject to cardinality constraints~\cite{nemhauser1978best}, such as influence maximization problems~\cite{kempe2003maximizing}, outbreak detection~\cite{leskovec2007cost}, facility location problems~\cite{cornuejols1983uncapicitated}, and many more, and has an approximation ratio equal to $1 - 1/e$. Although the family of linear objectives we analyze in this paper is not submodular in general, we are able to adapt the style of analysis using the properties of $\Phi$ and~\cref{lemma:comparison}, as we describe in~\cref{sec:approx_greedy}. The algorithm achieves an approximation guarantee of $1 - e^{-\frac {1 - \beta_{\max}} \zeta} - o(1)$ under a reasonable condition. In addition, the hill-climbing algorithm is efficient and performs very well, outperforming all the network-based heuristics. The rounding algorithm has a better worst-case approximation ratio in our theoretical bounds than the greedy algorithm, however it performs slightly worse than it empirically. Finally, we conclude our results by an inapproximability result for the (non-linear)~\eqref{eq:as} objective which resembles inapproximability results appearing in generalized influence maximization problems where the objective in question is not submodular.

\subsection{NP-Hardness Results} \label{sec:np_hardness}

In this Section, we prove that maximizing the objectives defined in \cref{sec:contributions} is computationally intractable. To give intuition on the reduction and its construction, we proceed by giving a polynomial-time reduction from a variant of the Set-Cover problem to the decision version for the maximization of the~\eqref{eq:linobj} objective, and similarly for the \eqref{eq:as} objective. Our reduction resembles the reduction presented in~\cite{kempe2003maximizing} for the Influence Maximization problem, and can be generalized to the other objectives described in \cref{sec:contributions}, whose proof sketches we defer to \cref{sec:ommited}. The reduction is outlined below:

\noindent \textbf{The 3-Set-Cover Problem.} The (NP-Hard) 3-Set-Cover problem is a variation of the classical Set-Cover problem~\cite{karp1975computational}, and is described as follows: 

\begin{quote}
    Given a collection of $n$ items $\mathcal U = \{ u_1, \dots, u_n \}$ a collection of $m \le n$ sets $S_1, \dots, S_m \subseteq \mathcal U$ such that $|S_i| = 3$ for every $i \in [m]$, does there exist a collection $\mathcal I$ of $| \mathcal I | = k$ sets such that $\bigcup_{i \in \mathcal I} S_i = \mathcal U$? 
\end{quote}

\begin{figure}[t]
    \centering
    \begin{tikzpicture}[transform shape,scale=0.75]
        \Vertex[size=0,opacity=0,x=-3]{c_2}
        \Vertex[size=0,opacity=0,x=-3, y=1]{c_1}
        
        \Vertex[RGB,color={127,201,127}, label=$S_2$]{S_2}
        \Vertex[RGB,color={127,201,127}, y=1, label=$S_1$]{S_1}
        \Edge[Direct,label=${c_1 \! = \! x_1 \! = \! 3}$, fontcolor=blue, bend=15](c_1)(S_1)
        \Edge[Direct,label=${c_2 \! = \! x_2 \! = \! 3}$, fontcolor=blue, bend=-15](c_2)(S_2)
        \Edge[Direct,label=${b_1 \! = \! \alpha}$,fontcolor=orange, bend=15](S_1)(c_1)
        \Edge[Direct,label=${b_2 \! = \! \alpha}$,fontcolor=orange,bend=-15](S_2)(c_2)
        \Vertex[RGB,color={190,174,212}, label=$u_1$, x=4, y=2]{u_1}
        \Vertex[RGB,color={190,174,212}, label=$u_2$, x=4, y=1]{u_2}
        \Vertex[RGB,color={190,174,212}, label=$u_3$, x=4, y=0]{u_3}
        \Vertex[RGB,color={190,174,212}, label=$u_4$, x=4, y=-1]{u_4}
        \Edge[Direct,color=red](S_1)(u_1)
        \Edge[Direct,color=red](S_1)(u_2)            \Edge[Direct,color=red](S_1)(u_3)
        \Edge[Direct,color=red](S_1)(u_4)
        \Edge[Direct,color=red](S_2)(u_2)
        \Edge[Direct,color=red](S_2)(u_3)
        \Edge[Direct,color=red](S_2)(u_4)
        \Vertex[opacity=0,size=0, x=6, y=2]{bu_1}
        \Edge[Direct,color=red](u_1)(bu_1)
        \Vertex[opacity=0,size=0, x=6, y=1]{bu_2}
        \Edge[Direct,color=red](u_2)(bu_2)
        \Vertex[opacity=0,size=0, x=6, y=0]{bu_3}
        \Edge[Direct,color=red](u_3)(bu_3)
        \Vertex[opacity=0,size=0, x=6, y=-1]{bu_4}
        \Edge[Direct,color=red](u_4)(bu_4)

    \end{tikzpicture}
    \caption{Reduction Construction of~\cref{theorem:hardness} for two sets $S_1 = \{ u_1, u_2, u_3 \}, \; S_2 = \{ u_2, u_3, u_4 \}$, and four items $\{ u_1, u_2, u_3, u_4 \}$. Here $\alpha \in (0, 3)$. Red edges represent a liability of $1 - \alpha / 3$. White nodes represent the external sector (either as assets or liabilities). The financial connectivities are $\beta_{S_j} = 1 - \alpha / 3 \in (0, 1)$ for $j \in [2]$, and $\beta_{u_j} = 0$, for $j \in [4]$, and $\beta_{\max} = 1 - \alpha / 3 < 1$.}
    \label{fig:reduction}
\end{figure}
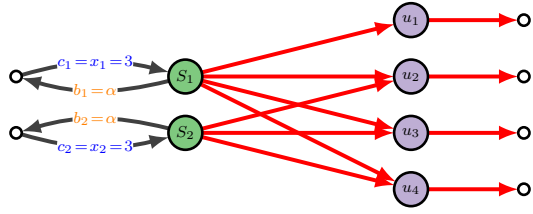

We prove that:

\begin{theorem}[see \cref{fig:reduction}] \label{theorem:hardness}
    Under \asref{assumption:beta_max_lt_one}, maximizing the \eqref{eq:linobj} and \eqref{eq:as} objectives are NP-Hard.
\end{theorem}

\subsection{Threshold Rules Can Have Bad Performance} \label{sec:thresholds}

A natural policy to allocate subsidies is through a \emph{threshold policy}, regarding one's existing equity, that is, if the initial equity $w_j$ of a node is below a threshold $\theta$ then this node is bailed out with a stimulus check of value $L_j$. Thus, the designer's choice should be to order nodes by their equities in ascending (or descending) order and bailout as many nodes as possible, in this order, subject to the budget, breaking ties consistently. Such a policy can perform arbitrarily bad. To observe this, consider a network $G$ on $n$ nodes which is built as follows: It consists of a node $v_1$ with $c_1 = b_1 = 1$ (note that this node is not isolated since it has $b_1 > 0$) and a directed path $v_2 \to v_n$ where $b_j = \varepsilon / 2$ for $2 \le j \le n - 1$, $b_n = 1 - (n - 1) \varepsilon + \varepsilon / 2$ and $c_2 = 1$. We let $X = c, L = \one, v = \one$ and $\Lambda = 1$. The optimal policy bails out $v_2$ and achieves a value of $O(n)$ whereas the naive policy bails out node $v_1$ and achieves an objective of $1$. The gap grows unbounded as $n \to \infty$. The same (bad) example can be altered to fool the policy which bails out nodes in descending order of their equity. 

\subsection{Approximation Algorithms Based on Randomized Rounding} \label{sec:approx_lp}

In this Section, we prove approximation guarantees based on randomized rounding. Briefly, we solve the relaxation problem where the indicator variables $\bar z_i$ are allowed to range in $[0, 1]$, as presented in \eqref{eq:generalized_bailouts_relaxation}. If $\til \xi^* = (\til p^*, \til z^*)^T$ is the optimal solution to the relaxation problem, we construct the corresponding rounded solution based on setting each rounded variable $Z_i$ equal to 1 with probability $\til z_i^*$ and 0 with probability $1 - \til z_i^*$\footnote{In fact, any rounding scheme such that $\ev {} {Z_i} \ge \til z_i^*$ would yield the same approximation ratio guarantee on expectation.}. To make our algorithm efficient we will allow $\evz { \sum_{i \in [n]} Z_i}$ to deviate from $\Lambda$ by some factor. For the approximation ratio analysis of the most algorithms, the following technical Lemmas are useful; their proofs, along with the proofs of the other results in this section, are deferred to \cref{sec:ommited}. 

We start by giving an approximation guarantee for the expected costs of the rounded solutions. When we condition upon a shock $X = x$ we overload the notation $\sol$, $\opt$, $\optr $ with $\sol(x)$, $\opt(x)$, $\optr (x)$, so that $\sol = \evx {\sol(X)}$ (resp. $\opt = \evx {\opt (X)}$, and $\optr = \evx {\optr (X)}$). 
Again, we refer to \cref{sec:ommited} for the proofs of the results in this section. Our result follows:

\begin{theorem} \label{theorem:apx_guarantees}
    The following results hold for the expected costs under the rounded variables $Z$ given a fixed shock $X = x$, for a linear objective $f(\bar p) = v^T \bar p$ with $v > \zero$:
    
    $$\evz {\sol_f(x)} \ge \left ( \frac {1 - \beta_{\max}} \zeta \right ) \cdot \opt_f (x). $$
    
\end{theorem}

We remind that $\zeta = \zeta(v) = \tfrac {v_{\max}} {v_{\min}} \ge 1$ is related to how ``well-conditioned'' the linear objective is. Taking expectations over $X \sim \D$, we get the approximation guarantees for the objectives. 



    
    


\noindent \textbf{Runtime Analysis.} To simulate the algorithms based on randomized rounding with the approximation guarantees devised in ~\cref{theorem:apx_guarantees}, we draw $m$ samples from $\D$ and with realizing each sample we do $T$ runs, such that at the end of the $T$ runs the computed solution is correct with high probability, i.e. probability that goes to 0 as $n \to \infty$. Given an oracle that solves the corresponding relaxed problems in time $\mathcal T(n)$ is $O(m (\mathcal T + kT))$ where $k$ is the maximum number of items that the algorithm selects and the $O(k)$ cost is paid to check the feasibility of the budget constraint. If the bailouts are equal, i.e. $L = \ell \cdot \one$ for some $\ell > 0$ we know that $k = \Lambda / \ell$. 

\cref{theorem:runtime} analyzes the runtime of the randomized rounding algorithm. We break the runtime analysis into two pieces to improve readability: First, we analyze the rounding algorithm for equal bailouts using independent rounding in which case we can directly make use of Chernoff bounds for the concentration of $L^T Z$. Second, we analyze the same algorithm for different values of the bailouts. The difference with the case of equal bailouts is that if $Z$ has independent components, then $L^T Z$ may not be concentrated and the runtime of the independent rounding scheme may explode. To remedy this problem, we make use of the oracle presented in~\cite{srinivasan2001distributions} to do dependent rounding on the variables. We adjust the oracle in a way that (i) the result of \cref{theorem:apx_guarantees} continues to hold after the dependent rounding procedure, (ii) $L^T Z$ obeys Chernoff-like concentration. In both of our rounding schemes, the planner is allowed to use an extra budget of $\til O(\sqrt \Lambda)$ added to the available budget of $\Lambda$. Moreover, due to the simulating the expected value of the rounded and the optimal solution, the approximation factor drops by an additive factor of $O(\varepsilon)$ for some sufficiently small $\varepsilon$. We present the Theorem below:

\begin{theorem} \label{theorem:runtime}
    Let $x^{(1)}, \dots, x^{(m)}$ be $m$ samples drawn from $\D$, let $\mathcal A$ be a randomized rounding algorithm as in ~\cref{theorem:apx_guarantees} to maximize an objective $f$ with $\| f \|_\infty < \infty$,  for which $\evz {\sol_f (x^{(i)})} \ge (1 - \gamma_f) \cdot \opt_f(x^{(i)}) $ for some $\gamma_f \in (0, 1)$,  and for all $i \in [m]$. Then the estimator $\overline {\sol_f} = \tfrac 1 m \sum_{i \in [m]} \sol_f (x^{(i)})$ satisfies $\overline {\sol_f} \ge (1 - \gamma_f - 2 \varepsilon ) \cdot \opt_f$ for $m = \tfrac {\log n} {\varepsilon^2}$ with probability $1 - O(\log n / (\varepsilon^2 n))$, runs in $\til O \left ( \tfrac {\mathcal T} {\varepsilon^2} + \tfrac k {\varepsilon^4} \right )$ time, and uses at most $\Lambda + \til O (\sqrt \Lambda)$ budget.
\end{theorem}

\noindent \textbf{Integrality Gap.} The integrality gap of the problem can be used to quantify the worst case ratio between a fractional optimal solution and an integral optimal solution. 
The integrality gap is given by $\sigma_f = \max_{\text{instances } \mathcal I} \frac {\optr_f (\mathcal I)} {\opt_f (\mathcal I)}$ and is well studied in the theory of approximation algorithms (see~\cite{williamson2011design} and the references therein). From a first glance, we expect that fractionally allocating stimulus is in general much more efficient than giving the stimulus to certain individuals; even in the optimal case. Indeed, this intuition is true theoretically, as we show that for every linear objective the integrality gap can become unbounded: 

\begin{theorem}[Integrality Gap] \label{thm:integrality_gap}
    For every $\varepsilon \in (0, 1), n \in \mathbb N^*$ and $k = o(n)$ there exists an instance for which the integrality gap for the linear objective $f(\bar p) = v^T \bar p$, with $v > \zero$, is unbounded as $n \to \infty$ and $\varepsilon \to 1$. Subsequently $\sigma_f \to \infty$.
\end{theorem}

        


\subsection{Greedy Approximation Algorithms} \label{sec:approx_greedy}

We consider a family of \emph{greedy hill-climbing  algorithms} in order to find the optimal bailout set for a linear objective $f(\bar p) = v^T \bar p $, where $v > \zero$. These algorithms run in $k \le \Lambda / L_{\min}$ steps, and at each step they pick the (feasible) element with the largest marginal gain until the budget constraint is violated, namely given a fixed shock $X = x$ where the current set of bailouts is $S_t$, with $S_0 = \emptyset$, we have $u_{t + 1} \in \mathrm{argmax}_{u \in [n] \setminus S_t \text { feasible}} \left \{ v^T \bar p_{S_t \cup \{ u \}} - v^T \bar p_{S_t} \right \}  $. This algorithm resembles the hill-climbing nature of~\cite{nemhauser1978best} for constrained monotone submodular maximization which guarantees an $(1 - 1 / e)$-approximation ratio. Below we prove that 
under a \emph{Small-Bailout Regime} our algorithms achieve an approximation ratio of $1 - e^{-(1 - \beta_{\max}) / \zeta} - o(1)$. We first state the Small-Bailout Regime condition, that suggests that whenever a node is bailed out, it remains default after the bailout (however with a greater value). It is important to note here that the hill climbing family of algorithms we consider would still work if run on an instance that violates the Small-Bailout Regime, and, in practice, the greedy hill climbing algorithm is the best-performing one, however the theoretical guarantee would hold only when this condition holds. For the rest of the analysis we fix a point-mass shock $X = x$. Given an approximation guarantee for the fixed shock $x$ we can later argue that this guarantee is achieved in expectation. We state our assumption

\begin{condition}[Small-Bailout Regime] \label{assumption:small_bailout}
    Under the randomness of $\D$ with probability 1, for every step $t$ the node $u_t$ selected by the algorithm yields a clearing vector $\bar p_{S_t}$ such that $\bar p_{S_t, u_t} < p_{u_t}$.
\end{condition}

This condition says that every node $u_t$ we include at iteration $t$ does not get ``saturated'' in the context of the constraint $\bar p_{S_t, u_t} = p_{u_t}$ holding and the default constraint being inactive. Under this condition, the approximation ratio of the greedy algorithm follows:

\begin{theorem} \label{lemma:approximation_greedy_induction}
    Under \condref{assumption:small_bailout}, for every linear objective the greedy hill-climbing algorithm achieves an approximation guarantee (under the randomness of $\D$, i.e. $X \sim \D$) of
    
    \begin{equation*}
        \sol_f \ge \left ( 1 - e^{-\tfrac {1 - \beta_{\max}} {\zeta}} \right ) \cdot \opt_f.
    \end{equation*}

\end{theorem}




    
    

\noindent \textbf{Runtime Analysis.} To run the algorithm in a simulation environment we calculate the expectation at each round using the sample average over $m$ i.i.d. samples from $\D$. Since by the problem constraints imply that $\zero \le \bar p \le p$ then approximating the expectation within an accuracy $\varepsilon > 0$ with probability of failure at most $\delta > 0$ requires $O(\| f \|_{\infty}^2 \log (1 / \delta) / \varepsilon^2)$ samples to be averaged, where $\| f \|_{\infty} = f(p)$.  The approximation ratio is off by an additive factor of $o(1)$ due to the approximation of the expectation via samples. We can apply a slight variation of Lemma 3.6 of~\cite{borgs2014maximizing} to get a similar guarantee.

\begin{theorem} \label{theorem:greedy_runtime}
    Let $x^{(1)}, \dots, x^{(m)} \iid \D$, let $\mathcal A$ be a greedy algorithm of the form of \cref{lemma:approximation_greedy_induction} to maximize a linear objective $f$ in $\bar p$ with $\| f \|_\infty < \infty$ and for which $\sol_f \left (x^{(i)} \right ) \ge (1 - \gamma_f) \cdot \opt_f \left (x^{(i)} \right )$ for some $\gamma_f \in (0, 1)$, and for all $i \in [m]$. Then the estimator $\overline {\sol_f} = \tfrac 1  m \sum_{i \in [m]} \sol_f(x^{(i)})$ is an $(1 - \gamma_f - 2 \varepsilon)$-approximation for $m = \til O( \Lambda / L_{\min} )$ samples, succeeds with probability $1 - O(1 / n)$, and runs in time $\til O \left ( \frac {(n + |E|) \Lambda^2} {L_{\min} \varepsilon^2 }\right )$ via invoking the fixed point operator $\Phi$.
    
\end{theorem} 

\noindent \textbf{Comparison with Randomized Rounding.} We have proved that for a linear objective $v^T \bar p$ the greedy algorithm yields an approximation ratio (on expectation) of $1 - e^{-(1 - \beta_{\max}) / \zeta}$ (under \condref{assumption:small_bailout}) whereas the randomized rounding algorithm achieves an approximation ratio (on expectation) of $(1 - \beta_{\max}) / \zeta$. Using the fact that $e^{-x} \ge 1 - x$ for all $x \in \mathbb R$ we have that always $(1 - \beta_{\max}) / \zeta \ge 1 - e^{-(1 - \beta_{\max}) / \zeta}$, so the approximation ratio of the randomized rounding algorithm is always better than the approximation ratio of the greedy algorithm through the present analysis. On the contrary, on the experimental instances of~\cref{sec:experiments}, the greedy algorithm nearly always outperforms randomized rounding.
     

\subsection{An Inapproximability Result for~\eqref{eq:as}} \label{sec:inapproximability}

In the previous Section we devised an $ \left ( \tfrac {1-\beta_{\max}} {\zeta} - o(1) \right )$-approximation rounding algorithm for the \eqref{eq:linobj} objective. However, if we ask a similar question for the \eqref{eq:as} objective --- namely, whether there is an approximation algorithm for optimizing the \eqref{eq:as} objective --- we obtain a negative answer. We prove that it is NP-hard to approximate a solution to the~\eqref{eq:as} objective within any poly-time computable function of the input. 

\begin{figure}[t]
    \centering
    \begin{tikzpicture}[transform shape,scale=0.75]
        \Vertex[size=0,opacity=0,x=-3, y=-1.5]{c_2}
        \Vertex[size=0,opacity=0,x=-3, y=1.5]{c_1}
        \Text[y=0]{$\vdots$}
        \Vertex[RGB,color={127,201,127}, label=$v_m$, y=-1.5]{S_2}
        \Vertex[RGB,color={127,201,127}, y=1.5, label=$v_1$]{S_1}
        \Edge[Direct,label=${c_1 \! = \! x_1 \! = \! 3}$, fontcolor=blue, bend=15](c_1)(S_1)
        \Edge[Direct,label=${c_m \! = \! x_m \! = \! 3}$, fontcolor=blue, bend=-15](c_2)(S_2)
        \Edge[Direct,label=${b_1 \! = \! \alpha}$,fontcolor=orange, bend=15](S_1)(c_1)
        \Edge[Direct,label=${b_m \! = \! \alpha}$,fontcolor=orange,bend=-15](S_2)(c_2)
        \Vertex[RGB,color={190,174,212}, label=$u_{1,1}$, x=2.5, y=2,size=0.85]{u_1}
        \Vertex[RGB,color={190,174,212}, label=$u_{1,2}$, x=2.5, y=1,size=0.85]{u_2}
        \Vertex[RGB,color={190,174,212}, label=$u_{1,n \! - \! 1}$, x=2.5, y=-1,size=0.85]{u_3}
        \Vertex[RGB,color={190,174,212}, label=$u_{1,n}$, x=2.5, y=-2, size=0.85]{u_4}
        \Text[x=2.5, y=0]{$\vdots$}
        \Edge[Direct,color=red](S_1)(u_1)
        \Edge[Direct,color=red](S_1)(u_2)            \Edge[Direct,color=red](S_1)(u_3)
        \Edge[Direct,color=red](S_1)(u_4)
        \Edge[Direct,color=red](S_2)(u_2)
        \Edge[Direct,color=red](S_2)(u_3)
        \Edge[Direct,color=red](S_2)(u_4)
        \Vertex[RGB,color={190,174,212}, label=$u_{2,1}$, x=5, y=2,size=0.85]{u_21}
        \Vertex[RGB,color={190,174,212}, label=$u_{2,2}$, x=5, y=1,size=0.85]{u_22}
        \Vertex[RGB,color={190,174,212}, label=$u_{2,n \! - \! 1}$, x=5, y=-1,size=0.85]{u_23}
        \Vertex[RGB,color={190,174,212}, label=$u_{2,n}$, x=5, y=-2, size=0.85]{u_24}
        \Text[x=5, y=0]{$\vdots$}
        \Edge[Direct,color=gray](u_1)(u_21)
        \Edge[Direct,color=gray](u_1)(u_22)           \Edge[Direct,color=gray](u_1)(u_23)
        \Edge[Direct,color=gray](u_1)(u_24)
        \Edge[Direct,color=gray](u_4)(u_21)
        \Edge[Direct,color=gray](u_4)(u_22)           \Edge[Direct,color=gray](u_4)(u_23)
        \Edge[Direct,color=gray,](u_4)(u_24)
        
        \Vertex[RGB,color={190,174,212}, label=$u_{a,1}$, x=7.5, y=2,size=0.85]{u_a1}
        \Vertex[RGB,color={190,174,212}, label=$u_{a,2}$, x=7.5, y=1,size=0.85]{u_a2}
        \Vertex[RGB,color={190,174,212}, label=$u_{a,n \! - \! 1}$, x=7.5, y=-1,size=0.85]{u_a3}
        \Vertex[RGB,color={190,174,212}, label=$u_{a,n}$, x=7.5, y=-2, size=0.85]{u_a4}
        \Text[x=7.5, y=0]{$\vdots$}
       
        \Text[x=6, y=0]{$\dots$}
    
        \Vertex[opacity=0,size=0, x=10, y=2]{bu_1}
        \Edge[Direct,color=red](u_a1)(bu_1)
        \Vertex[opacity=0,size=0, x=10, y=1]{bu_2}
        \Edge[Direct,color=red](u_a2)(bu_2)
        \Vertex[opacity=0,size=0, x=10, y=-1]{bu_3}
        \Edge[Direct,color=red](u_a3)(bu_3)
        \Vertex[opacity=0,size=0, x=10, y=-2]{bu_4}
        \Edge[Direct,color=red](u_a4)(bu_4)
        
        \Text[y=-3, x=5, color=gray]{$a(| \mathcal I|)$ Fully-connected Layers}
        
    \end{tikzpicture}
    \caption{Proof of~\cref{theorem:as_inapproximability} (see~\cref{sec:ommited} for the full proof). Here $\alpha \in (0, 3)$ and $a(| \mathcal I|) \in \mathbb N^*$ is a poly-time computable function of the input instance size $| \mathcal I|$. The red edges represent liabilities of value $1 - \alpha / 3$. Gray edges represent liabilities of value $\tfrac {1 - \alpha / 3} {n}$. The maximum financial connectivity is $\beta_{\max} = 1 - \alpha / 3 < 1$. A YES answer to the 3-Set-Cover problem implies at least $k + a(| \mathcal I|) \cdot n$ solvent nodes, whereas a NO answer implies at most $k + n$ solvent nodes.}
    \label{fig:inapproximatbility}
\end{figure}
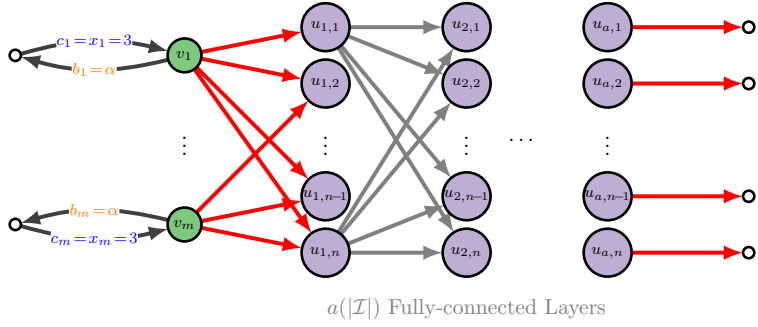

\begin{theorem}[\eqref{eq:as} Inapproximability (\cref{fig:inapproximatbility})] \label{theorem:as_inapproximability}
    The \eqref{eq:as} problem cannot be approximated within a factor of $\tfrac {k + a(| \mathcal I|) n} {k + n} = \Omega (a(|\mathcal I|))$ for every poly-time computable function $a(| \mathcal I|) \in \mathbb N^*$ of the input instance size $|\mathcal I|$, unless P = NP. 
\end{theorem}

\section{Fairness} \label{sec:fairness}

\subsection{Fairness Metrics} 

In this paper, we say that an allocation is \emph{fair} across the nodes if the allocation (discrete or fractional) obeys the following property: the amount of bailouts that a certain node gets ``does not differ a lot from its neighbors'' where the notion of \emph{neighborhood} here is general and does not necessarily refers to the network ``neighborhood`` as devised by the EN model. For instance, the ``neighborhood'' may refer to measuring inequality among all pairs of nodes, nodes who have a specific property, and nodes in the actual network. All metrics have to be homogeneous, i.e. multiplying all the bailouts by a positive number should not affect their value. We consider the following constraints to incorporate fairness in our model:  

    \noindent \textbf{Gini Coefficient.} The Gini Coefficient~\cite{gini1921measurement} measures the fairness of the stimulus allocations between all pairs of nodes and is defined to be 

    \begin{equation} \tag{GC} \label{eq:gc}
        \mathrm{GC}(\bar z) = \frac {\sum_{i, j \in [n]} |L_i \bar z_i - L_j \bar z_j|} {2 n \sum_{j \in [n]} (L_j \bar z_j)}.
    \end{equation}

    It evaluates to 0 when the stimuli are equally distributed and is $1 - 1/n$ when one node gets all the bailout amount. A useful optimization constraint is making the Gini coefficient at most $g \in [0, 1]$, or equivalently $\sum_{i, j \in [n]} |L_i \bar z_i - L_j \bar z_j| \le 2n g \sum_{j \in [n]} L_j \bar z_j$. We say that an allocation for which the Gini Coefficient is at most $g$ is a $g$-unfair allocation. 
    
    An advantage of this metric is that it is one widely used metric for calculating income inequality\footnote{For example see: \url{https://data.worldbank.org/indicator/SI.POV.GINI}.} and is widely used in policymaking. However, a possible disadvantage of this metric is that it does not take into account each individual node's debts, i.e. it treats all nodes on an equalized basis. This issue is mitigated by the \eqref{eq:sgc} metric which is presented below that takes into account the financial ties between the nodes.  
    
    \noindent \textbf{Property Gini Coefficient.} The collection of real-world data from SafeGraph and the US Census we present in \cref{sec:experiments} consists of attributes characterize nodes. One of the key attributes in these datasets is the minority status of the owner of a business, if such business participates at the network as a node, or the demographic characteristics of a group of people, for instance the fraction of people belonging to a minority group within a Census Block Group under which we want to impose fairness constraints. (That is, to measure the relative assistance between different groups, in an approximate way). 
    
    This type of data motivates the following metric: We introduce the (assymetric) Property Gini Coefficient \eqref{eq:pgc} in which nodes may have a \emph{property} of interest (such as the demographic group in the SafeGraph or Census data) along which we want to apply an equity analysis.  We model this by a \emph{property vector} $q \in [0,1]^n$, where each element $q_j$ corresponds to the probability that node $j \in [n]$ has this property, as follows: We let $n_q = \sum_{j \in [n]} q_j$ and $n_{\neg q} = n - n_q$ be the total weights of the (soft) bipartition. For every node $j \in [n]$ is inequality subject to being in the minority group is given as $\tfrac {1} {2n_{\neg q}} \sum_{i \in [n]} (1 - q_i) | L_i \bar z_i - L_j \bar z_j |$. The sum over all $j$ with weight $q_j$ give the numerator of the \eqref{eq:pgc}. The denominator of the \eqref{eq:pgc} is $\sum_{j \in [n]} q_j L_j \bar z_j$: 

    \begin{equation*} \tag{PGC} \label{eq:pgc}
        \mathrm{PGC}(\bar z; q) = \frac {\sum_{j, i \in [n]} q_j (1 - q_i) |L_i \bar z_i - L_j \bar z_j| } {2 \left ( n - n_q \right ) \cdot \sum_{j \in [n]} q_j L_j \bar z_j}.
    \end{equation*}

    Note that taking $q = \tfrac 1 2 \cdot \one$ reduces \eqref{eq:pgc} to the conventional GC. Moreover, for $L = \ell \cdot \one$ and $q^T \bar z = 0$ we observe that \eqref{eq:pgc} becomes unbounded since the denominator goes to zero. One case where this happens, and further justifies the correctness of the criterion is when $q$ and $\bar z$ are a 0/1 vector and where the entries of $q$ are 1 the entries of $\bar z$ are 0, and vice-versa, where the entries of $q$ are 0 the entries of $\bar z$ are 1, which corresponds to giving all the bailouts to the majority group. We say that an allocation which achieves a \eqref{eq:pgc} at most $g$ subject to a property $q$ is $(g, q)$-unfair. Note that the \eqref{eq:pgc} constraint can be combined with the \eqref{eq:gc} constraint as follows: given $g_{\mathrm{between}}, g_{\mathrm{within}} \ge 0$ we seek to find an allocation that respects both \emph{between-fairness}, i.e. $\mathrm{PGC}(\bar z; q) \le g_{\mathrm{between}}$, and \emph{within-fairness}, i.e. $\mathrm{GC}(\bar z \odot q) \le g_{\mathrm{within}}$ and $\mathrm{GC}(\bar z \odot (\one - q)) \le g_{\mathrm{within}}$. 
    
    An advantage of this fairness measure is that it asserts that all (fuzzy) groups within the network are treated \emph{almost equally}, whereas the measure suffers from the same problem that \eqref{eq:gc} has, i.e. there is no dependency on the actual financial ties.
     
    \noindent \textbf{Spatial Gini Coefficient.}  To make the GC take into account network effects, we define its spatial analogue, the \eqref{eq:sgc}, to be 

    \begin{equation} \tag{SGC} \label{eq:sgc}
        \mathrm{SGC}(\bar z; A) = \frac {\sum_{(j, i) \in E} a_{ji} |L_j \bar z_j - L_i \bar z_i |} {2 \sum_{j \in [n]} \beta_j L_j \bar z_j}
    \end{equation}

    A simplified version of \eqref{eq:sgc} appears in work regarding school funds allocation~\cite{mota2021fair} where the graph is assumed to have unit weights. In our case, the role of the unweighted graph plays the relative liability matrix $A$. Since $A$ is substochastic the total weight of each row is $\beta_i < 1$ and the contribution of edge $(i, j)$ is $a_{ij}$. Normalizing by the sum $\sum_{j \in [n]} \beta_j L_j \bar z_j$ allows for comparing different population groups and different bailout magnitudes. When the bailouts are distributed equally \eqref{eq:sgc} is 0. If $A = A^T$ and one node gets all the bailouts, then the \eqref{eq:sgc} is bounded by 1. We say that an allocation which achieves an \eqref{eq:sgc} of at most $g$ is $(g, A)$-unfair. We note here that unlike \eqref{eq:gc}, the \eqref{eq:sgc} metric takes into account each node's debt, that is a node $j$ with a significant (compared to its neighbors) liability to node $i$, i.e. it has $a_{ji} \approx \beta_j$, then this deviation will get a higher weight in the calculation of the coefficient compared to $j$'s deviation from the rest of its neighbors.

\subsection{Optimization Formulation} 

We formulate the following relaxations to the optimization problems involving the aforementioned fairness metrics for a target fairness metric upper bound $g \ge 0$. First, we consider the following optimization problem that extends~\eqref{eq:generalized_bailouts} by adding the GC-dependent constraints (resp.~\eqref{eq:generalized_bailouts_relaxation} on the $\til p, \til z$ variables):

\begin{equation} \label{eq:generalized_bailouts_gini} \tag{GC-Problem}
\begin{split}
    \max \quad & \evx {f(\bar p)} \\
    \text{s.t.} \quad & (\bar p, \bar z) \in \text{\eqref{eq:generalized_bailouts} constraints} \\
    & \sum_{i, j \in [n] \times [n]} |L_i \bar z_i - L_j \bar z_j | \le 2 n g L^T \bar z, \qquad \qquad \qquad \qquad \qquad  \\
\end{split}
\end{equation}

\begin{equation} \label{eq:generalized_bailouts_gini_relaxation} \tag{GC-Relaxation}
\begin{split}
    \max \quad & \evx {f(\til p)} \\
    \text{s.t.} \quad & (\til p, \til z) \in \text{\eqref{eq:generalized_bailouts_relaxation} constraints} \\
    & \sum_{i, j \in [n] \times [n]}  \til \varpi_{ij} \le 2 n g L^T \til z \\
    & \til \varpi_{ij} \ge 0, \; - \til \varpi_{ij} \le L_i \til z_i - L_j \til z_j \le \til \varpi_{ij}  \qquad (i, j) \in [n] \times [n]. 
\end{split}
\end{equation}

The optimization problem relaxations can be formulated mutatis mutandis for \cref{eq:pgc,eq:sgc} for a fairness bound $g \ge 0$. The optimal fractional solutions to these problems consider fractional allocations whereas their discrete counterparts consider discrete allocations. Using the same rounding scheme as presented in \cref{sec:approx_lp}, we can deduce approximation guarantees for the~\eqref{eq:generalized_bailouts_gini}. We present such a result for \eqref{eq:gc}.  
As in the previous section, we refer to \cref{sec:ommited} for all proofs in this section.

\begin{theorem} \label{theorem:gini_approximation}
    Let $\til z^*(g)$ be the optimal solutions to~\eqref{eq:generalized_bailouts_gini}, for $f$ being the~\eqref{eq:linobj} objective, and $g > 0$ being a fairness constraint. Then, if all bailouts are equal, i.e. $L = \ell \cdot \one$ with $k = \tfrac \Lambda \ell$, and for $g > 1 - \tfrac k n$, the GC problem admits an $\left ( \tfrac {(1 - \beta_{\max})(1 - k / n)} {g \zeta} - o(1) \right )$-approximation algorithm.
    
\end{theorem}

One can extend the runtime analysis of this algorithm (\cref{theorem:runtime}) with the incorporation of the finite differences inequality~\cite{doob1940regularity} to devise high probability bounds for the deviation constraint. We believe that further runtime analysis lies beyond the main points of this paper. 

\subsection {Price of Fairness} 

A natural question we might ask is, \emph{``What is the maximum effect of these fairness constraints on the welfare objective function?''}. We define the \emph{Price of Fairness} (PoF)~\cite{bertsimas2011price} to be 

\begin{equation*}
    \mathrm{PoF} = \frac {\text{Optimal Value when sans Fairness Constraint}} {\text{Optimum value when the solution is $g$-unfair (resp. $(g, q)$-unfair / $(g, A)$-unfair)}}.
\end{equation*}

It is evident that the since the value in the numerator refers to an optimization problem with a larger feasible region than the one in the denominator that always $\mathrm{PoF} \ge 1$. At glance, a natural question arises: Do there exist instances for which PoF is \emph{unbounded/bounded} for \emph{discrete/fractional} allocations? The answer to this question is affirmative for the discrete allocations case:

\begin{theorem}[Unbounded Discrete PoF Instances] \label{thm:unbounded_pof_discrete}
    There exist finite instances where the  PoF is unbounded for \textbf{discrete allocations}, for all the constraints defined by~\eqref{eq:gc}, \eqref{eq:pgc}, and \eqref{eq:sgc}, and any linear objective given by a vector of coefficients $v > \zero$. 
\end{theorem}

A natural follow-up question that arises is the following: \emph{Does there exist a bound on the PoF regarding fractional allocations?} An answer to this question provide the following Theorems

\begin{theorem}[Fractional PoF Boundedness] \label{theorem:fractional_pof_bounded} 

For every increasing objective $f$ with $f(\zero) = 0$, $\| f \|_\infty < \infty$, and for every $g \ge 0$ the PoF for \textbf{fractional} bailouts is bounded (for all fairness metrics).
    
\end{theorem}

In the sequel, the next question is the derivation of a PoF bound for the fractional case. The following result gives such a bound for the~\eqref{eq:gc} case and the \eqref{eq:sgc} case when the relative liability matrix is symmetric and the network $G$ is connected and $A = A^T$. The proof of this Theorem depends on two steps. Firstly, given that all the optimal fractional allocations consume a certain budget $\mu \in (0, \Lambda]$ there could be two cases: either the budget is distributed equally or at least one node gets a budget different than $\tfrac {\mu} {nL_j}$. In the former case, to get the desired PoF bound we create a feasible solution by \emph{perturbing} the values of the optimal solution allocations so they fall into the second case (i.e. create a suboptimal solution) establishing, this way, an upper bound to the PoF. We state the Theorem below:

\begin{theorem}[Fractional PoF Bounds] \label{theorem:fractional_pof_bounds}
    Given a \textbf{finite} instance $\mathcal I$ of the EN model with $g > 0$ the following upper bounds are true for any linear objective given by a coefficient vector $v > \zero$: There exists a finite $C(\mathcal I) > 0$ such that:
    
    \begin{compactitem}
        \item For the~\eqref{eq:gc} constraint
        
        \begin{equation*}
            \mathrm{PoF}_{g} \le C ( \mathcal I ) \cdot \frac {2 \zeta g (\| c \|_1 + \Lambda) \sqrt n} {(1 - \beta_{\max})}.
        \end{equation*}

        \item For the~\eqref{eq:sgc} constraint, if $A = A^T$ and $G$ is connected, we have that

    \begin{equation*}
        \mathrm{PoF}_{(g, A)} \le C ( \mathcal I ) \cdot \frac {\sqrt 2 \zeta g (\| c \|_1 + \Lambda) \beta_{\max}} {(1 - \beta_{\max}) \phi(A^{(2)})}.
    \end{equation*}

    Where $\phi (A^{(2)})$ is the conductance of the graph with adjacency matrix $A^{(2)}$.

    \end{compactitem}

\end{theorem}

The above Theorem relates the PoF for a $(g, A)$-unfair allocation when $A$ is symmetric with two quantities of a topological nature: the former one is the conductance (of $A^{(2)}$), a quantity widely used in spectral graph theory in order to justify the mixing time of random walks, and the latter one is $\beta_{\max}$, i.e. the maximum row sum of $A$. Note that the conductance of $A^{(2)}$ and $\beta_{\max}$ are not independent of one another. Another useful bound that relates the conductance of $A$ with \eqref{eq:sgc} is given by the following Theorem, which stems from an alternative expression of conductance. \cref{sec:conductance} contains an algebraic proof of the argument, which is of independent interest. We state the Theorem: 

\begin{theorem} \label{theorem:conductance}
    Let $A$ be a symmetric and connected network and let $\til z$ be any allocation for which $\mathrm{SGC}(\til z; A) \neq 0$. Then $$\frac {\phi(A)} {2} \le \mathrm{SGC}(\til z; A).$$
\end{theorem}

\cref{theorem:conductance} states that for $A = A^T$ (connected) \eqref{eq:sgc} is at least an $1/2$ factor away from $\phi(A)$. Moreover, note that in general for \cref{theorem:fractional_pof_bounds,theorem:conductance} we have that $\phi(A) \neq \phi(A^{(2)})$. 

In addition to these general bounds, it is also interesting to ask about the behaviour of the PoF for simple topologies, the conductance of which we can control in a standardized manner. 
With this in mind, we examine following example created from two equal-sized cliques with randomly sampled edges between:

\begin{figure}
    \centering
   \includegraphics[width=0.4\textwidth, ext=.png]{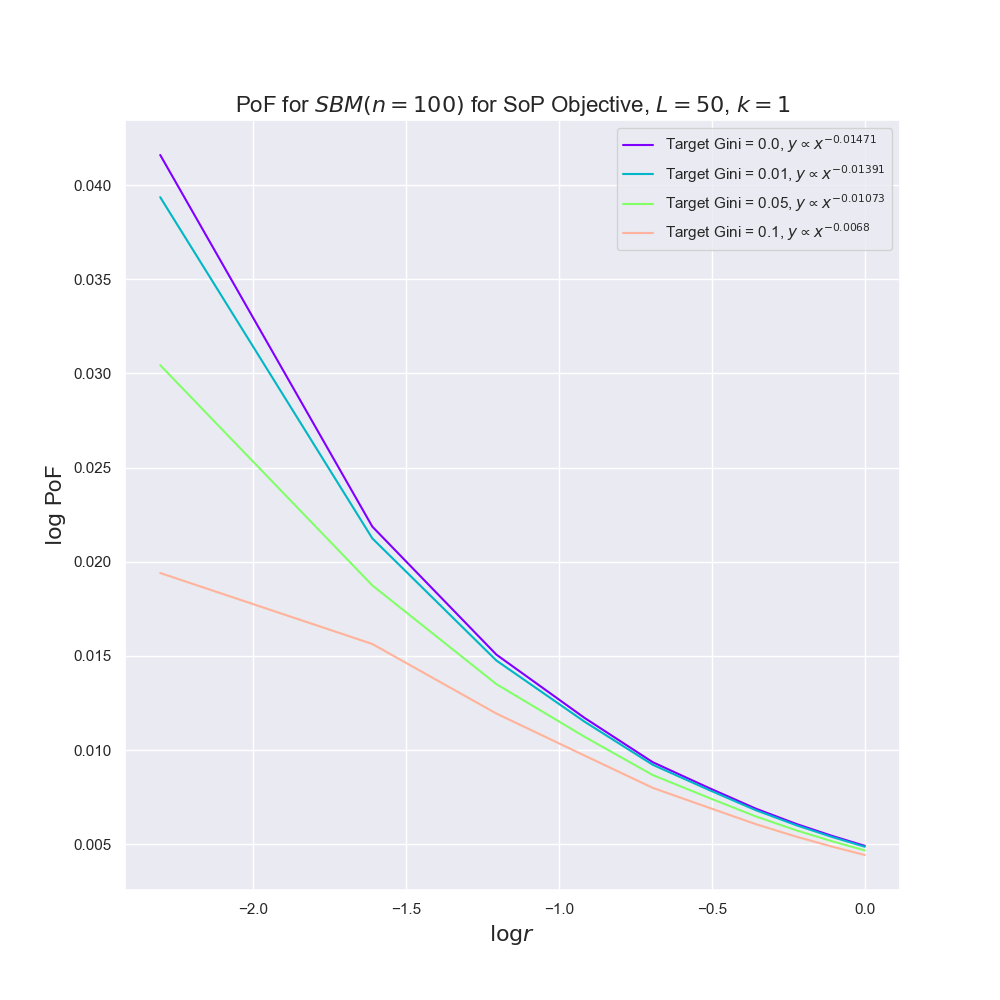}
    \caption{Randomized Construction of \cref{example:sbm} with log-log fits.}
    \label{fig:sbm}
\end{figure}

\begin{example} \label{example:sbm}
     A graph on $n$ nodes is built from 2 cliques of size $n / 2$ between which edges are generated randomly and independently with probability $r \in (0, 1]$. Each edge $(j, i)$ is endowed with a liability $p_{ji} = 1$, and each node $j$ has $c_j = n, b_j = 1$, $L_j = n / 2$, and $X_j = \one \{ j \le n / 2 \}$. A log-log plot relating the $\log$-PoF with $\log r$ is presented in \cref{fig:sbm}. We observe that as the nodes become more densely connected the PoF decreases. Moreover, note that for two target values $g_1 \le g_2$ for the \eqref{eq:sgc} constraint, the PoF curve relating to $g_1$ is always above the PoF curve relating to $g_2$. This is expected, as the latter problem has a larger feasible region and thus solutions of higher welfare can be achieved.
\end{example}

\section{Experiments} \label{sec:experiments}

\subsection{Data} 

In this section, we evaluate our methods on public datasets from two kinds of sources:
\emph{high-level granularity} data, among nodes corresponding to financial institutions in a country, financial institutions between countries, or financial interactions between different financial sectors of the same country; 
and \emph{lower granularity} data, among nodes corresponding to anonymized  groups of people defined by the US Census.
In the latter case, the relevant public datasets are constrained through anonymization or aggregation due to the privacy considerations of the individuals.

We now present the datasets we analyze in this study. A more extensive data analysis of the datasets can be found in \cref{sec:data_analysis}.

\begin{compactitem}
    \item \emph{German Banks.} Datasets from 22 German banks from the work of~\cite{chen2016financial} where the internal and external assets and liabilities are reported. 
    \item \emph{EBA.} Data from 76 banks that participated in the European Banking Authority (EBA) 2011 stress test, provided in~\cite{glasserman2015likely}. The paper reports only the column sums $a$ ($a_j$ are the total internal assets of $j$) and the row sums $l$ ($l_j$ are the total internal liabilities of $j$), assuming that $a = l$. The network structure is unknown and are inferred using the \texttt{systemicrisk} package, introduced in~\cite{gandy2017bayesian} which incorporates a Gibbs sampler to fit data where the row and column sums of the liabilities are provided. Prior to fitting and sampling, we perturb the liabilities vector $l$ to $\hat l$ as follows: We draw $n$ i.i.d. Gaussians $\epsilon_1, \dots, \epsilon_n$ with mean 0 and std 100 and we set each $\hat l_j = \left \lceil (l_j + \epsilon_j) \tfrac {\one^T l} {\one^T (l + \epsilon)} \right \rceil $ for $j \in [n - 1]$ and we finally set $\hat l_n = \one^T l - \sum_{j \in [n - 1]} \hat l_j$. To fit the model with \texttt{systemicrisk} we used the complete network on $n  = 76$ nodes.

    \item \emph{Venmo.} Data from the Venmo social payments application\footnote{The data has been public prior to this study, and available at: \url{https://github.com/sa7mon/venmo-data}. As of 2019, the API could be accessed with a GET request at \url{https://venmo.com/api/v5/public}.}. The data consists of 7M nodes and 7M transactions between users of Venmo. Each transaction is encoded by its sender $u$ and its receiver $v$, represented by a directed edge $(u, v)$ and a timestamp. The payment amounts are not available in the original data, for data privacy reasons. Finally, the data consists of hundred of thousands \emph{weakly-connected components}, many of them with very small size (less than 10 nodes). We keep the Venmo network of $n \approx 7 \mathrm{M}$ nodes and $m \approx 7 \mathrm{M}$ edges by dropping the timestamp column (i.e. assuming a static setting). The network has a power-law degree distribution (see \cref{sec:data_analysis} for more). We calculate all the weakly-connected-components (WCCs) of the network. We observed that most of the WCCs are very sparsely connected and small for which reason we decided to keep the component $G_i$ that has $n_i \ge 50$ nodes and has the maximum density $m_i / n_i^2$ that is equal to 4.44\%. Since the values of the entities were missing we generated values based on models of the existing literature and the individual analysis that we had performed on the data sources where data was available (see the distribution fits and regression plots in \cref{sec:data_analysis}). Using this knowledge, we generated i.i.d. internal liabilities\footnote{The exponential internal liability model is also followed in the work of~\cite{gandy2017bayesian}.} from $\mathrm{Exp}(1)$, (independent) external assets $c_j = \left (d^{\mathrm{in}}_j + d^{\mathrm{out}}_j \right ) \cdot \mathrm{Exp}(1)$, and external liabilities given by $b_j = 0.9 \cdot c_j$\footnote{A linear relation between $c_j$ and $b_j$ is reasonable as evidence presented in \cref{sec:data_analysis} underline.}. 
    
    \item \emph{SafeGraph.} Data generated based on mobility data from the SafeGraph platform during April 2020. The nodes in the financial network represent: \emph{(i)}~Points of Interest nodes (POI nodes) that represent various businesses categorized by their \emph{NAICS codes}\footnote{\url{https://www.naics.com/search}} to categories (i.e. grocery stores, banks, gas stations etc.) and the Census Block Group\footnote{A CBG is a unit used by the US Census. It is the smallest geographical unit for which the bureau publishes sample data, i.e. data which is only collected from a fraction of all households and contains 600-3K people.} (CBG) they are located at; \emph{(ii)}~CBG nodes that represent a set of households in a certain location. The dataset is constructed by access to an initial pair of geographical coordinates (i.e. latitude and longitude) and a number $k_{\mathrm{kNN}}$ of neighboring CBGs. The POI nodes are determined to be the businesses that are located in the $k_{\mathrm{kNN}}$-nearest neighboring CBGs based on the Haversine distance metric. Each POI provides data about the CBGs of its unique visitors\footnote{A unique visitor is a unique mobile device, i.e. each device is only counted once. We assume that each device represents a distinct person.} and the dwell times. For the source of its visitors we estimate the number of people that come from each CBG. From the \emph{dwell times} of devices that are available we determine the percentage of people that work on the POIs and the ones who visit the POIs. For the former category we create a financial liability edge from the POI to the CBG node to indicate the payment of a liability (i.e. a wage) and for the latter category we create a liability edge from the CBG node to the POI representing some form of expense (e.g. groceries). The weights are multiplied accordingly to represent the set of people that interact with each POI. The aforementioned process creates a bipartite network. Each CBG node is associated with multiple data from the US Census, as well as every POI node is associated with data from the US Economic Census. For each CBG node, we estimate the average size of households per CBG, the average income level and the percentage of people that belong to a minority group. 
    
    We use the above data to estimate the external assets and liabilities of the CBG nodes. For the bailouts of CBG nodes we calculate a bailout devised by the CARES act that considers as income the average income of the CBG and as size of household the average size of household multiplied by an estimate for the number of people in that CBG who interact with the POI nodes. Similarly, for the POI nodes we use data from the US Economic Census and NAICS to determine average wages, income and expenses. For the bailouts of the POI nodes we use loan data regarding loans that were given during April 2020 as part of the SBA Paycheck Protection Program (PPP) provided by SafeGraph, adjusted to the number of workers being present in the network and the span of one month. Moreover, the loan data included demographic characteristics about the businesses in question so we were able to determine (or estimate in the case of missing data) the minority status of a business, i.e. the probability of a certain business being a business with a minority owner. A complete description of the data generation process is presented in \cref{sec:data_analysis}.  

\end{compactitem}

\noindent \textbf{Shocks.} For all of our experiments we assume that the shocks $X_j$ for each $j \in [n]$ are independent and each shock $X_i$ is sampled from the uniform distribution with support $[0, c_i]$. We ran experiments with the shocks being scaled $\mathrm{Beta}(1/2, 1/2)$ distributions and the results were similar, and thus were omitted. 

\noindent \textbf{Heuristic Methods.} We use the following heuristic methods to allocate stimulus, where for each step we augment each set (based on the criteria below) maximally with respect to the budget constraint.

\begin{compactitem}
    \item \emph{Wealth Policy.} We sort the nodes in increasing  order of initial wealth $w_i$,  nodes with the lowest wealths each time. Note that we performed the same experiments with decreasing order of wealths and the results were similar and thus omitted.
    \item \emph{Out-degree Policy.}  We take the nodes with the highest outdegrees. 
    \item \emph{PageRank~\cite{page1999pagerank}.} We take the $k$-top nodes in decreasing value of their PageRank values. The calculation of PageRank takes into account the directionality of the graph. 
    \item \emph{Eigenvector Centrality~\cite{freeman1977set}.} We take the nodes in decreasing value of their eigenvector centrality (i.e. values of the principal eigenvector of the corresponding random walk transition matrix). The computation of the eigenvector centrality measure ignores directionality.
    \item \emph{Random Permutation.} Baseline criterion that considers a random permutation of the nodes.
    
\end{compactitem}

All the above policies are well known benchmarks and have been used on similar tasks such as Influence Maximization~\cite{kempe2003maximizing}. For all experiments we also report the upper bound of $\opt_f$ which is the corresponding \emph{relaxation optimum $\optr_f$} that is an upper bound to the true optimum, which in our case is computationally intractable, that is if we used brute-force in the case where the stimuli were equal, we would need to search over $O(n^k)$ sets to find the optimal solution for each $k$. 

\subsection{Experiments} 

\noindent \textbf{Discrete Allocations.} Firstly, for various values of the stimuli vector $L$, either fixed or varying, we report the corresponding objective values averaged over multiple draws of shocks from the corresponding shock distribution, where we report both the empirical mean and standard deviation (std). We parametrize the available budget with a pair of parameters. The former parameter $\ell$ parametrizes the budget increase rate, and the latter parameter $k$ parametrizes the multiplicity of resources. We make assumptions about the bailouts that fall into two main categories: \emph{(i)} fixed bailouts: where $L = \ell \cdot \one$ and $\Lambda(k) = \ell \cdot k$ and the number of bailouts $k$ varies along the x-axis of the plots. This is equivalent with bailing out at most $k$ nodes on the network where every node gets a stimulus value of $\ell$.  \emph{(ii)} variable bailouts: in the SafeGraph experiments we determine the bailouts as discussed in the start of \cref{sec:experiments} and \cref{sec:data_analysis}. We assume that for each step $k$ the budget increases by some amount $\ell$ so, again, the available budget is $\Lambda (k) = \ell \cdot k$.  

\begin{figure}[t]
    \centering
    \subfigure[SoP Objective for $L = 10^6 \cdot \one$]{\includegraphics[width=0.45\textwidth]{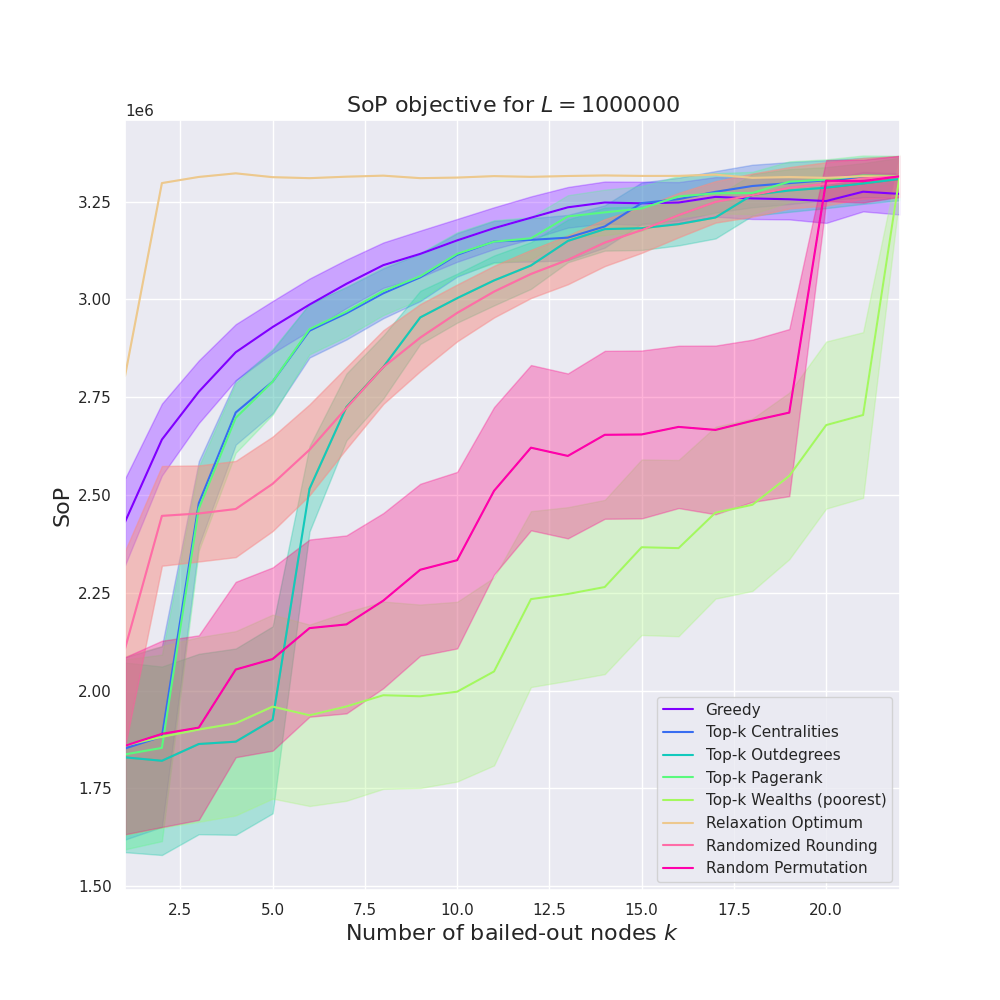}}
    \subfigure[SoIP Objective for $L = 10^6 \cdot \one$]{\includegraphics[width=0.45\textwidth]{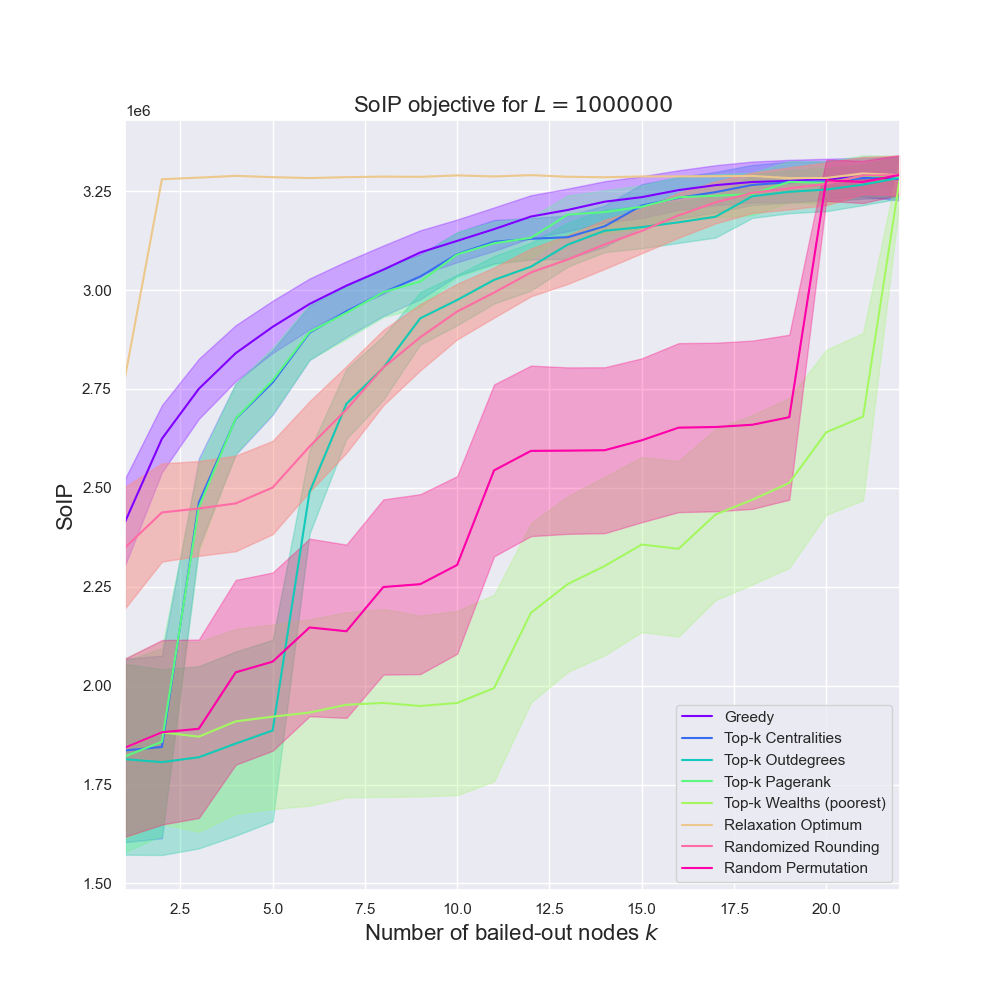}}
    \caption{German Banks Dataset results with uniform shocks. The simulations have been run for 1000 iterations. The error areas represent 1 std.}
    \label{fig:german_banks_results}
\end{figure}

\begin{figure}[t]
    \centering
    \subfigure[SoP Objective for $L = 10 \cdot \one$]{\includegraphics[width=0.45\textwidth]{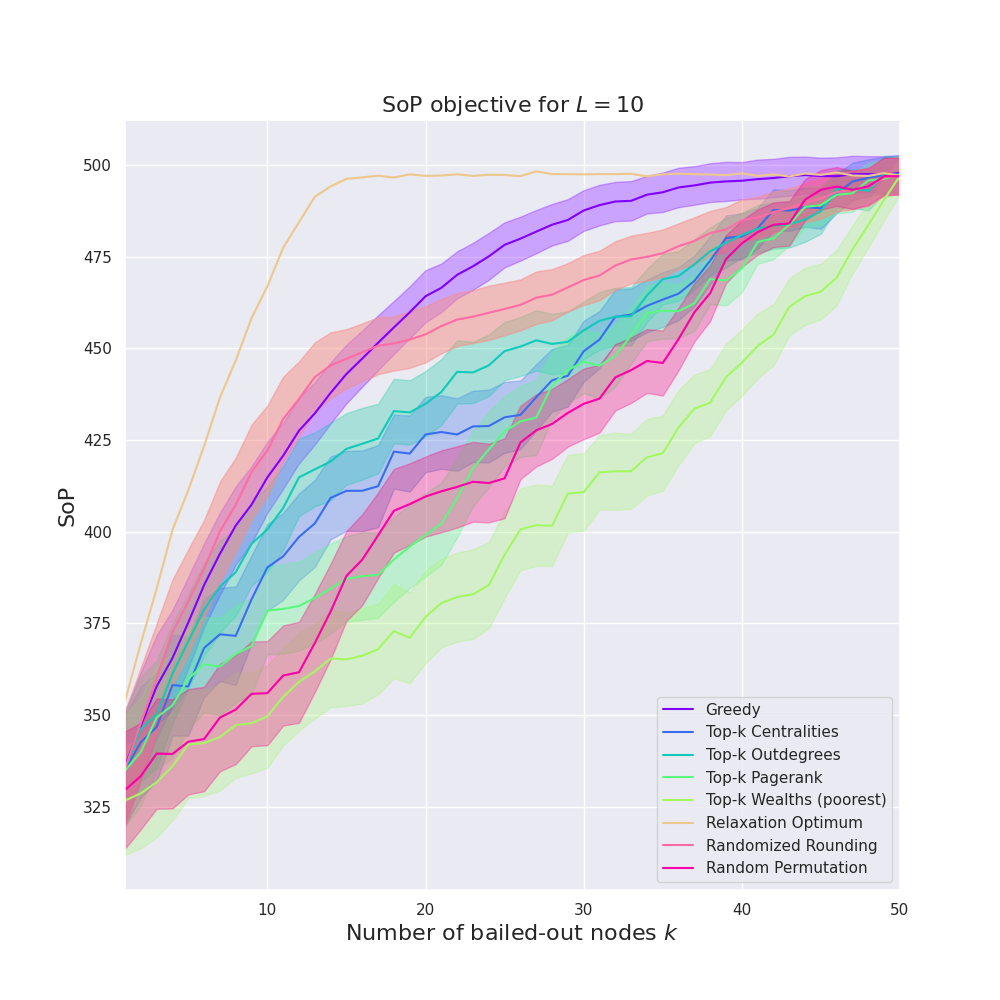}}
    \subfigure[SoIP Objective for $L = 10 \cdot \one$]{\includegraphics[width=0.45\textwidth]{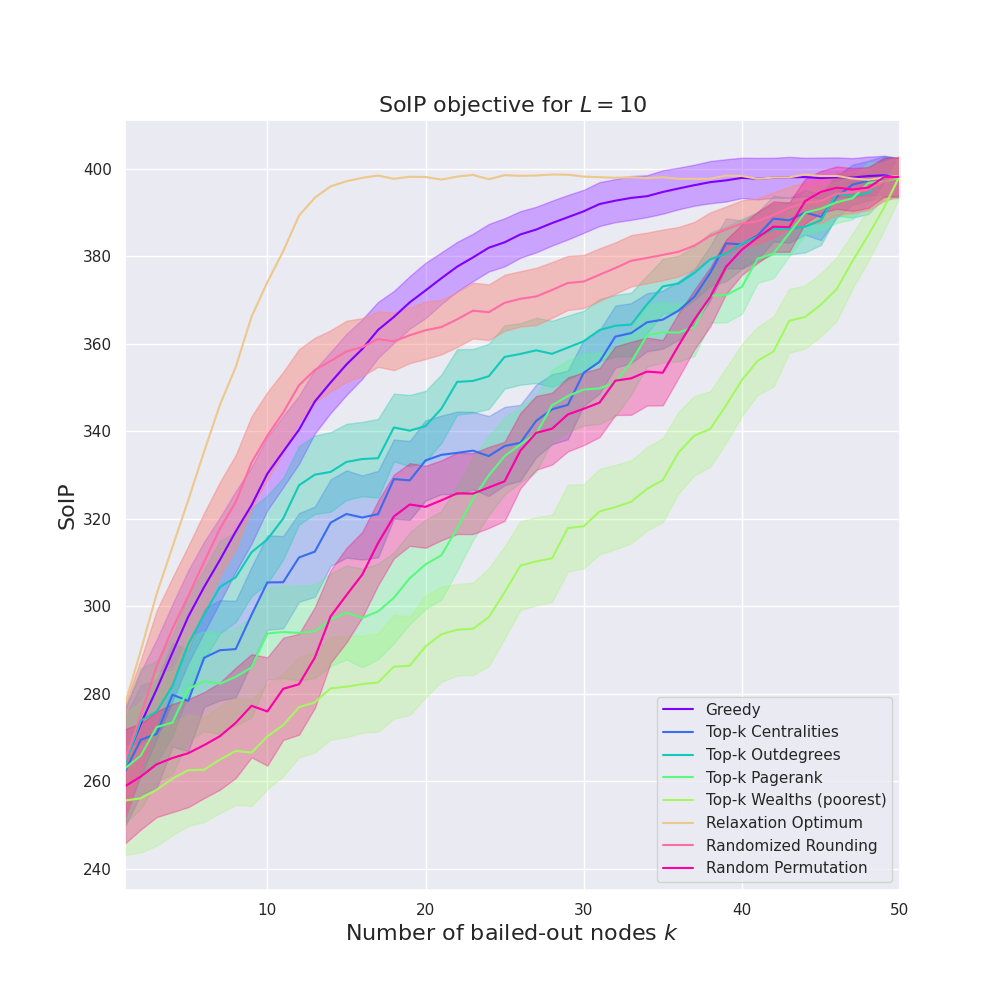}}
    \caption{Venmo Dataset results with uniform shocks. The Venmo network is taken to be the densest weakly connected component with at least 50 nodes from the Venmo network. The internal liabilities have been generated from $\mathrm{Exp}(1)$ for every $(i, j) \in E$. The external assets $c_i$ are given as $c_i = (d^{\mathrm{in}}_i + d^{\mathrm{out}}_i) \cdot \mathrm{Exp}(1)$, and the external liabilities are given as $b_i = 0.9 \cdot c_i$. The simulations have been run for 2000 iterations. The error areas represent 1 std.}
    \label{fig:venmo_results}
\end{figure}

\begin{figure}[t]
    \centering
    \subfigure[SoP Objective for $L = 10^6 \cdot \one$]{\includegraphics[width=0.45\textwidth]{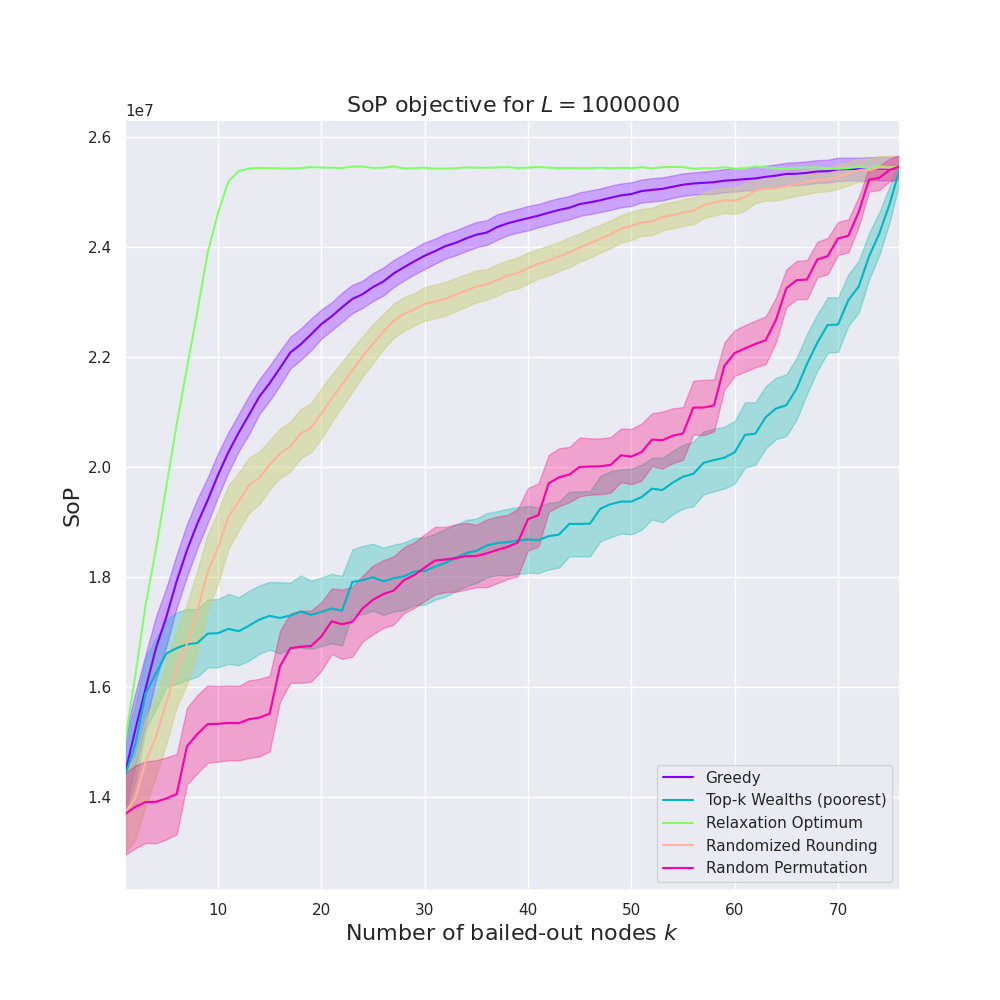}}
    \subfigure[SoT Objective for $L = 10^6 \cdot \one$]{\includegraphics[width=0.45\textwidth]{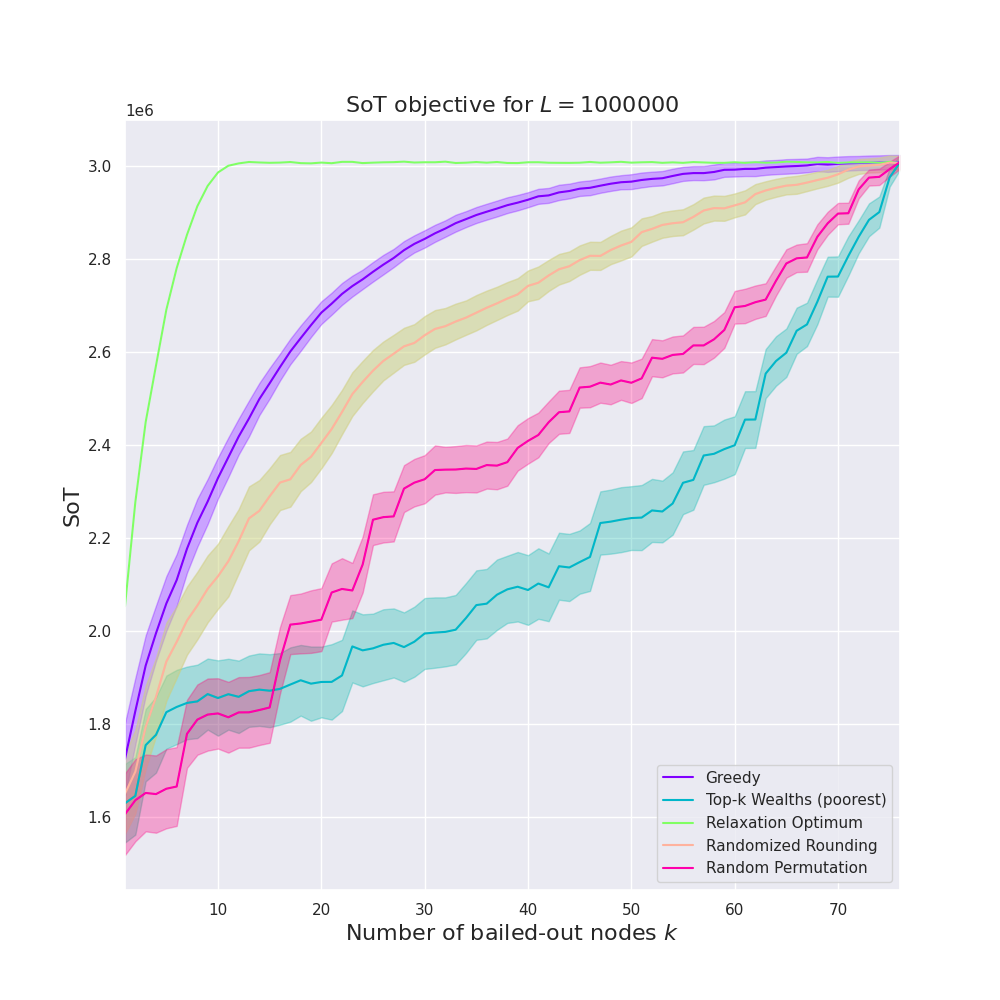}}
    
    \caption{EBA Dataset results. The simulations have been run for 1000 iterations. The error areas represent 1 std.}
    \label{fig:eba_results}
\end{figure}

\begin{figure}[t]
    \centering
    \subfigure[SoP Objective for $L$ estimated from real-world data with a budget given by $\Lambda(k) = 10^4 \cdot k$]{\includegraphics[width=0.45\textwidth]{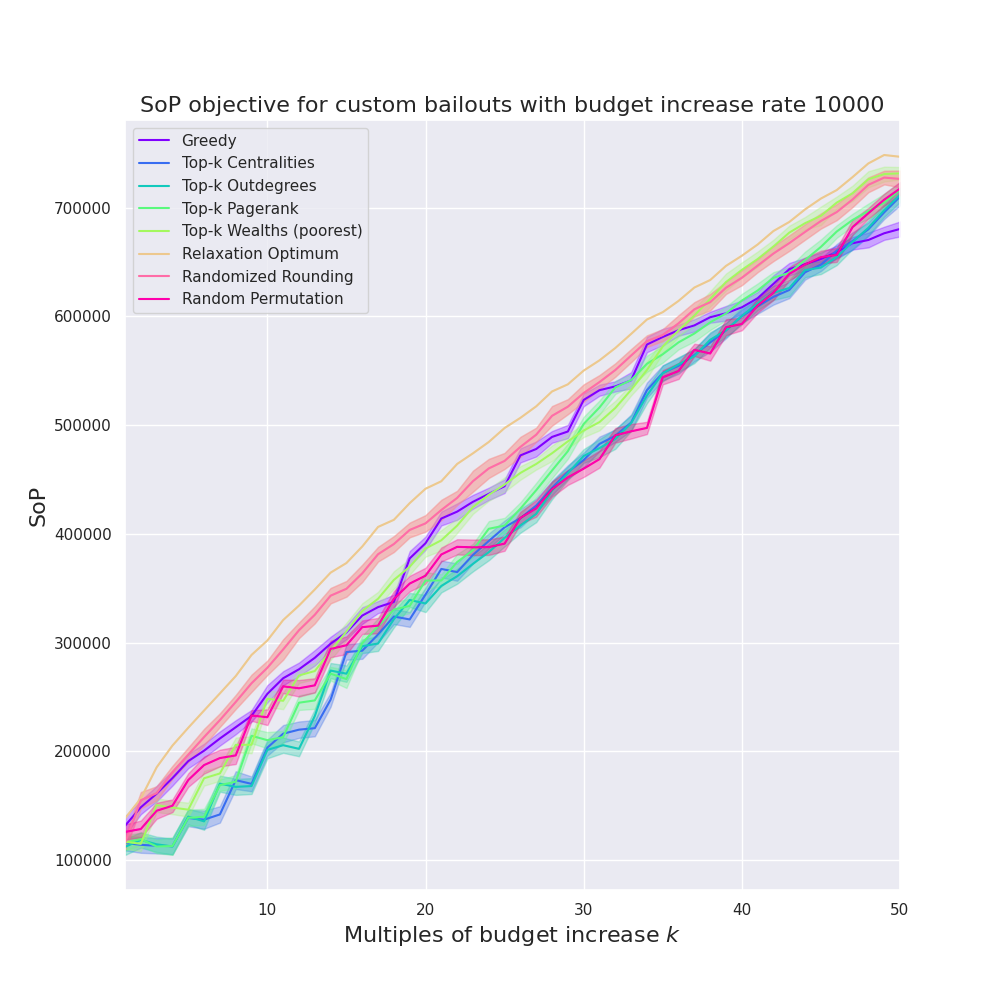}}
    \subfigure[SoT Objective for $L$ estimated from real-world data with a budget given by $\Lambda(k) = 10^4 \cdot k$]{\includegraphics[width=0.45\textwidth]{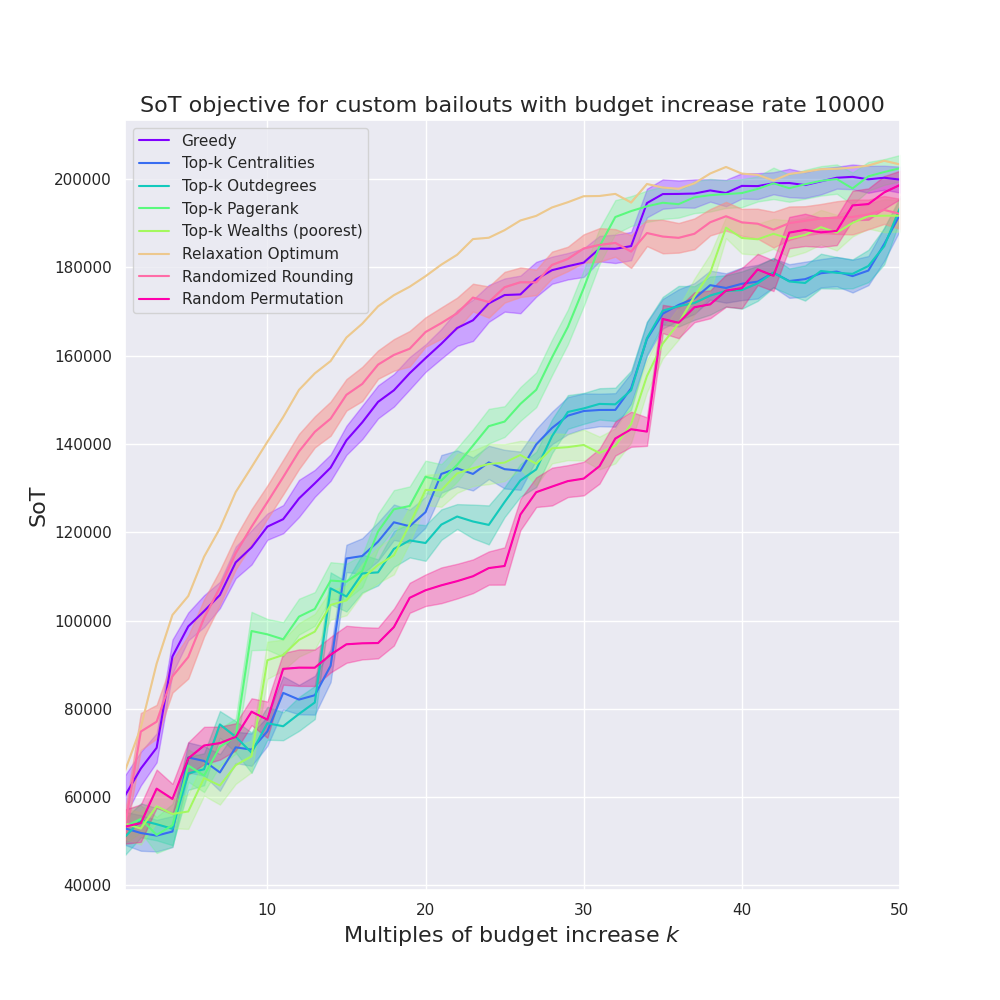}}
    \caption{Network generated from Safegraph data whose generation is described in \cref{sec:data_analysis} for businesses spanning the 3 nearest neighbor CBGs around the geographical location given by \texttt{(42.43969750363193, -76.49506530228598)} in a monthly basis. The shocks are uniform. The simulations have been run 50 times and the error areas represent 1 std.}
    \label{fig:safegraph_results}
\end{figure}

The results of these experiments for the various datasets are reported in~\cref{fig:german_banks_results,fig:venmo_results,fig:eba_results,fig:safegraph_results}. In brief, the greedy algorithm outperforms all other algorithms in all settings. Then, the rounding algorithm comes second, outperforming the other heuristics. Finally, note that the Wealth Policy performs very poorly due the fact that it is independent of the contagion process. This last observation is consistent with the theoretical observations that we made in \cref{sec:thresholds}). 

\begin{figure}[t]
    \centering
    \subfigure[Relaxation Optima and rounded values.]{\includegraphics[width=0.45\textwidth]{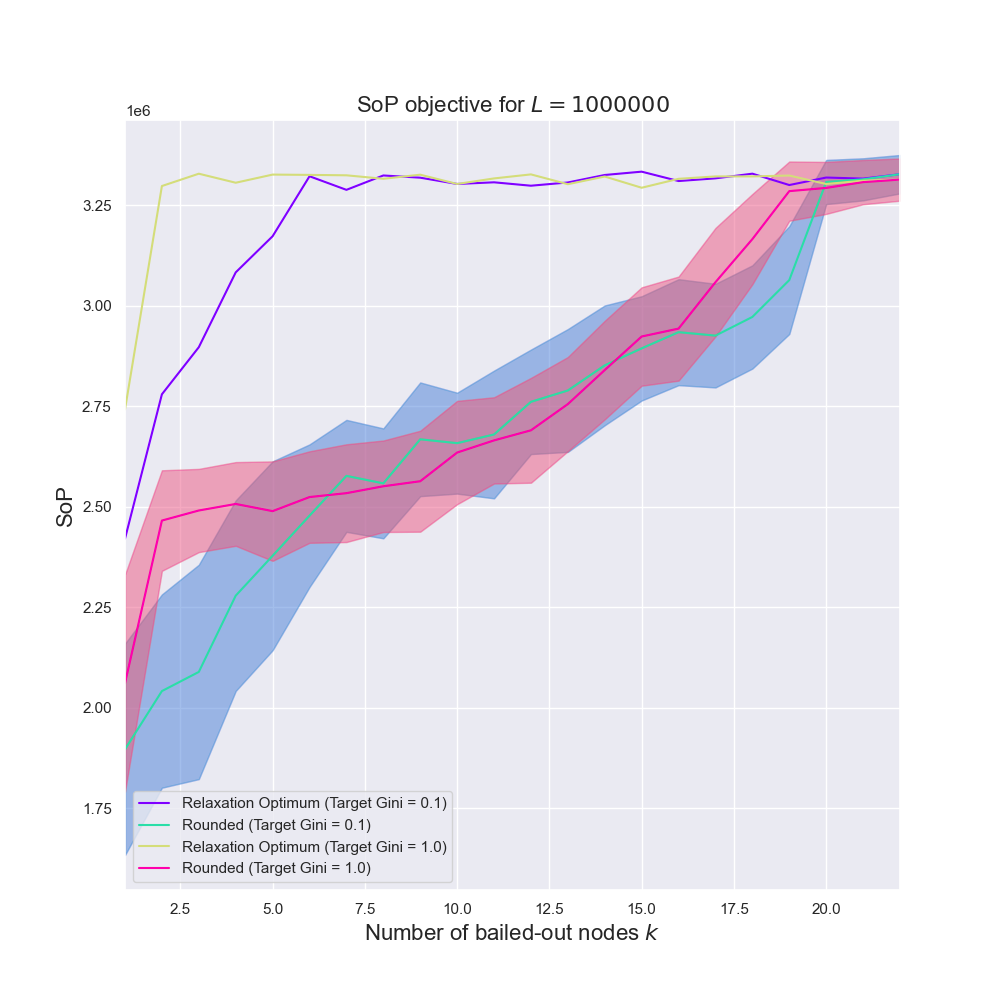}}
    \subfigure[Gini Coefficient calculated on the average values of the fractional solutions. ]{\includegraphics[width=0.45\textwidth]{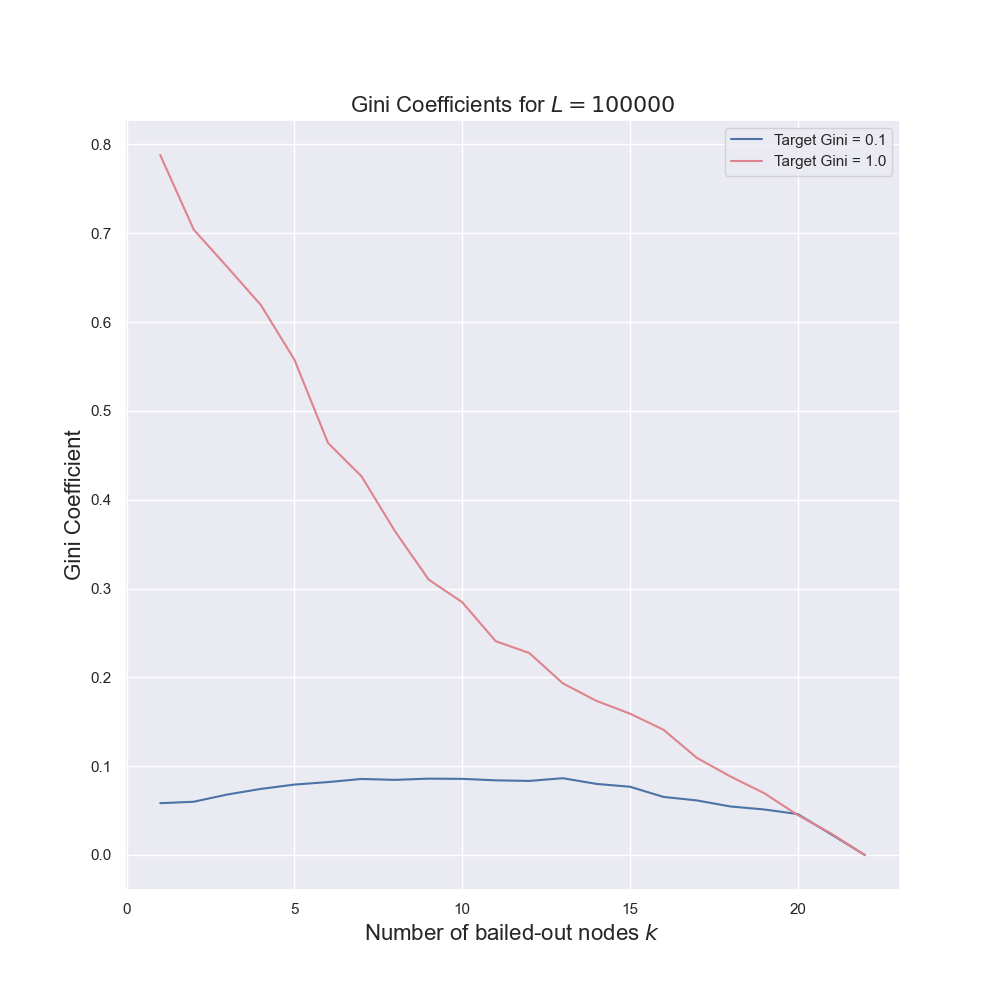}}
    \caption{Optimization subject to Gini Coefficient constraints on the German Banks dataset with $L = 10^6 \cdot \one$. Comparison of the unconstrainted problem (\texttt{Target Gini $\le$ 1}) and a fairness constrainted problem (\texttt{Target Gini $\le$ 0.1}). The left plot represents the values of the relaxation optimum for both values of the constraint, and the corresponding rounded solutions based on the uniform rounding scheme of \cref{sec:approx_lp}, and the right plot represents the actual value of the Gini coefficient calculated upon the optimization with values equal to $\bar z_{\mathrm{avg}, i} = \sum_{t = 1}^m \bar z_i^{*(t)} / m$ for $m = 1000$ simulations. Error areas represent 1 std.}
    \label{fig:gini_german_banks}
\end{figure}

\begin{figure}[t]
    \centering
     \subfigure[Relaxation Optima and rounded values for SafeGraph data]{\includegraphics[width=0.45\textwidth]{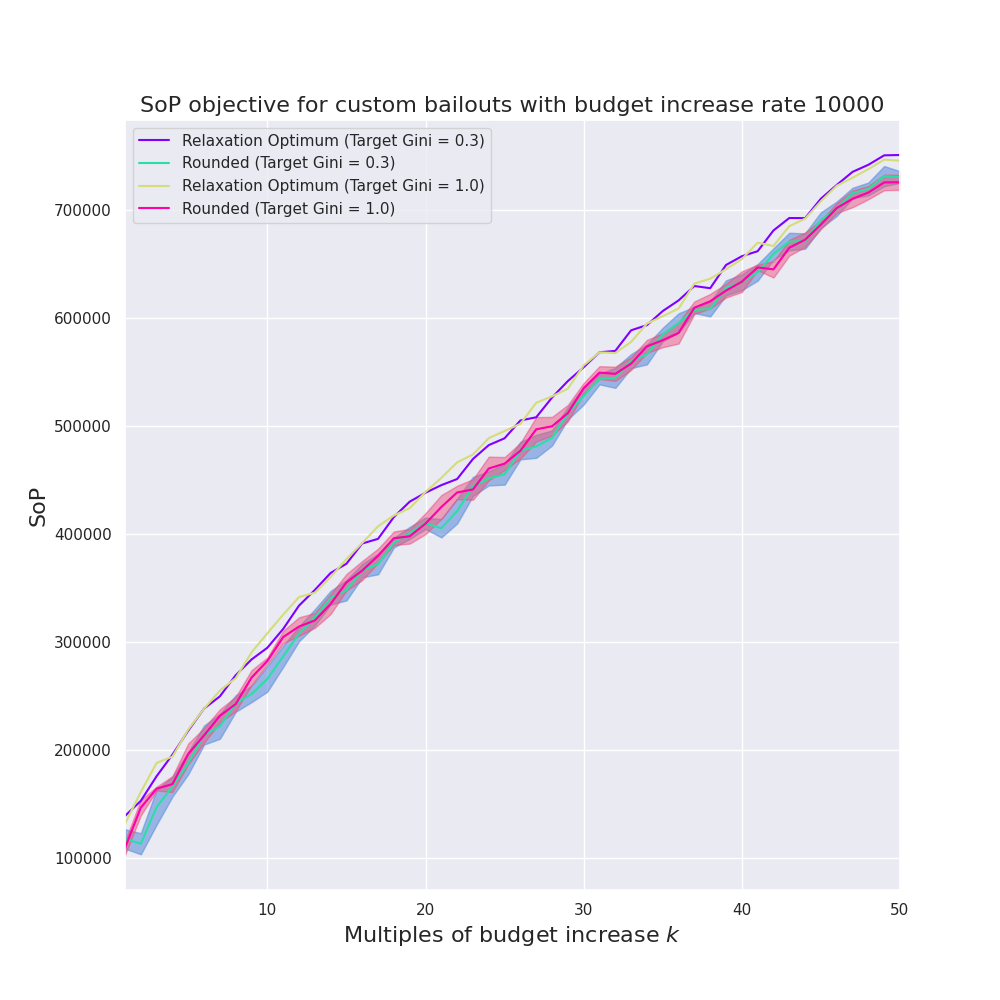}}
    \subfigure[Relaxation Optima and rounded values for German Banks data]{\includegraphics[width=0.45\textwidth]{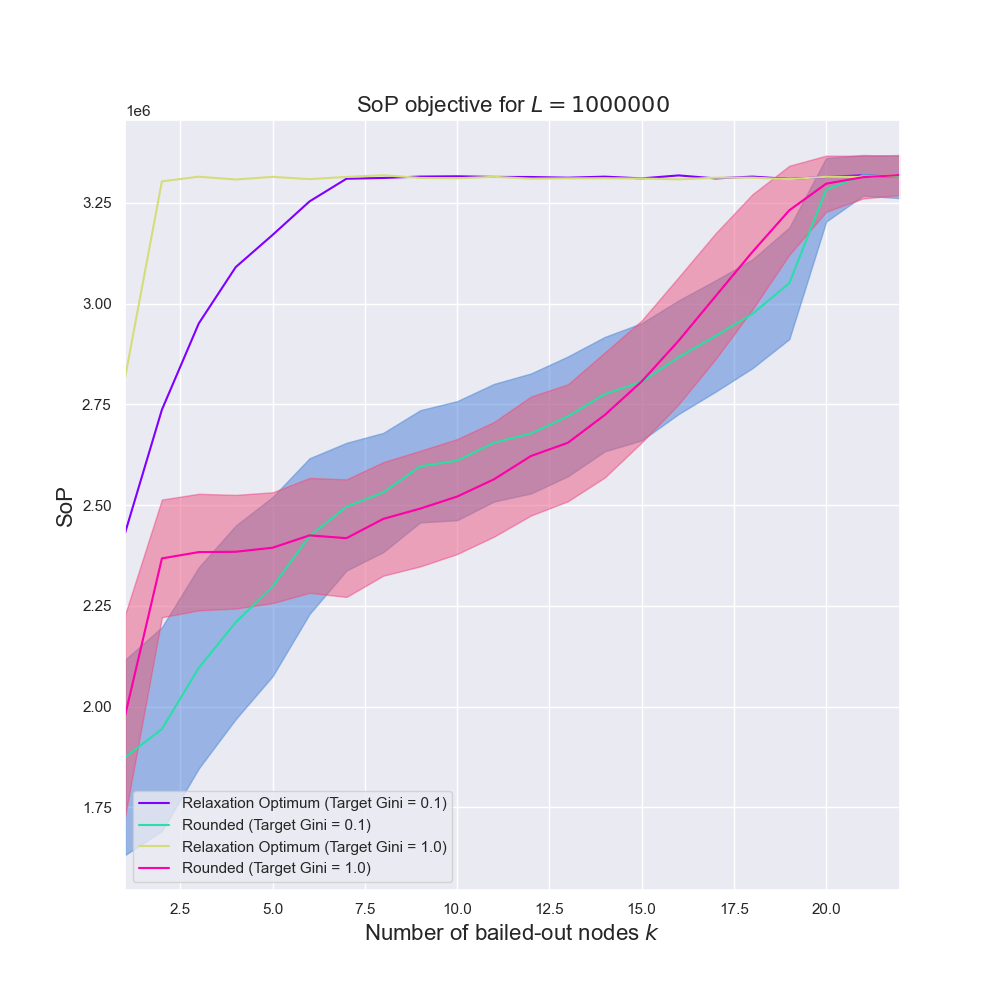}} 

    \caption{(i) Optimization subject to the Property Gini Coefficient constraints between minorities/non-minorities  data (with property data from the US Census and SafeGraph) with custom bailouts and a budget increase rate of $10^4$. Error areas represent 1 std after 50 simulations with constraint target values \texttt{0.3} and \texttt{1.0}. (ii) Optimization subject to the Property Gini Coeffifient on the German Banks data with property data drawn independently from $\mathrm{Beta}(2, 5)$ with constraint target values \texttt{0.1} and \texttt{1.0}. Error areas represent 1 std after 1000 simulations.}
    \label{fig:property_gini}
\end{figure}

\noindent \textbf{Fairness.} Secondly, we perform experiments where we constrain the fairness of the model as follows: 

\begin{compactitem}
    \item For the German Banks dataset we run the following problems \emph{constraining} the \eqref{eq:gc}. First, we run the unconstrained optimization problem (i.e. allowing Gini to be at most 1), and, second, we restrict the \eqref{eq:gc} to be at most 0.1. We report the relaxation optimum, the rounded solutions to the problem and the realized \eqref{eq:gc} after the optimization on~\cref{fig:gini_german_banks}. 
    \item For the Venmo dara we run the following problems \emph{constraining} the \eqref{eq:sgc}. First, we run the unconstrained optimization problem (we allow the \eqref{eq:sgc} to be at most 1 which in our experiments suffices to produce the optimal solution in the unconstrained version), and, second, we restrict the \eqref{eq:gc} to be at most 0.1. We report the relaxation optimum, the rounded solutions to the problem and the realized \eqref{eq:gc} after the optimization on~\cref{fig:gini_german_banks}. 
    
    \item For the SafeGraph and the German Banks data we use the fuzzy version of the \eqref{eq:gc} where the weights represent the probability that a node is a minority node. In other words, we want to impose constraints between \emph{minority} and \emph{non-minority} groups. For the former dataset, the weights $q$ are the minority scores for each CBG and business. For the latter dataset, the values of $q$ are sampled i.i.d. from $\mathrm{Beta}(2, 5)$. We report the relaxation optimum and the rounded solutions as well as the total income allocated to the minority group. We use a budget increase rate of $10^4$ for SafeGraph, similar to the one reported in~\cref{fig:safegraph_results}, and a bailout $L = 10^6 \cdot \one$ for German Banks. The results are reported on~\cref{fig:property_gini}.
    \item In \cref{fig:pof_target} we plot the relation between the upper bound on the \eqref{eq:sgc} coefficient and the PoF for the Venmo and German Banks Data.
    
\end{compactitem}

\noindent \textbf{Implementation and Environment.} The source code is developed in the Python language using NumPy for numerics, NetworkX for network analysis, and Google's or-tools for optimization. For solving linear programs we use the GLOP Linear Solver, and support multi-threading. The experiments were run on a server with 144 cores and 1.5TB of RAM. 

\noindent \textbf{Source Code.} \url{https://github.com/papachristoumarios/financial-contagion}


\begin{figure}[t]
    \centering
     \subfigure[Relaxation Optima and rounded values]{\includegraphics[width=0.45\textwidth]{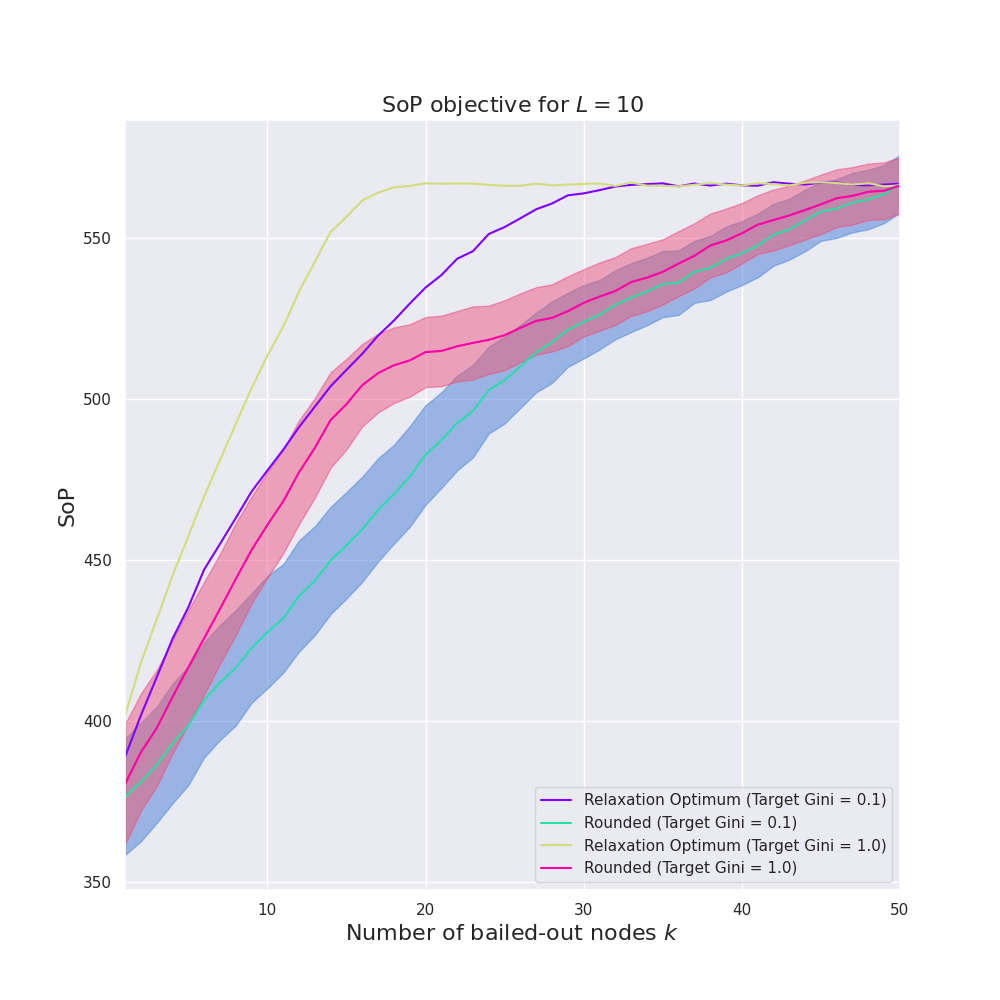}}
     \subfigure[Spatial Disparity for $k = 13$]{\includegraphics[width=0.45\textwidth]{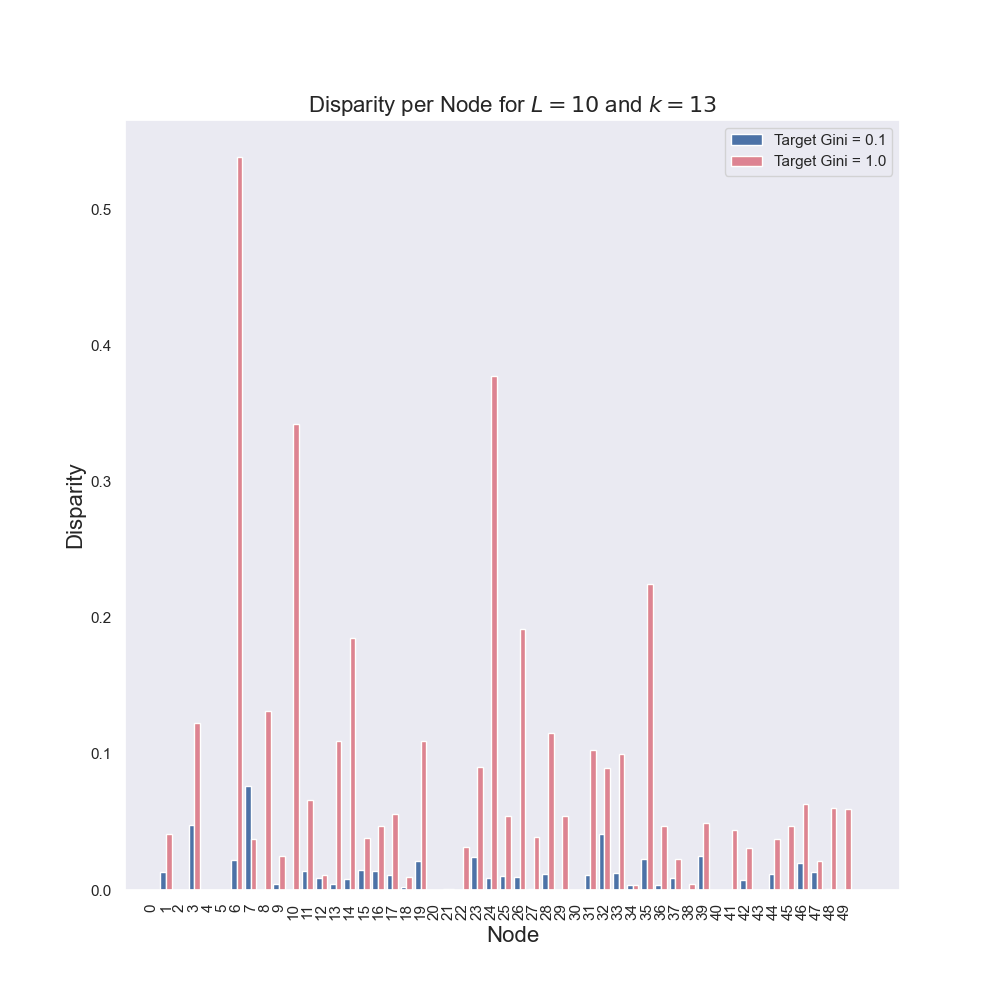}}

    \caption{Optimization subject to the \eqref{eq:sgc} constraints on the Venmo data with $L = 10 \cdot \one$. Comparison of the unconstrainted problem (\texttt{Target Gini $\le$ 1}) and a fairness constrainted problem (\texttt{Target Gini $\le$ 0.1}). The left plot represents the values of the relaxation optimum for both values of the constraint, and the corresponding rounded solutions based on the uniform rounding scheme of \cref{sec:approx_lp}, and the right plot represents the spatial inequality per node for $k = 13$. Error areas represent 1 std after 2000 simulations.}
    \label{fig:venmo_network_gini}
\end{figure}

\section{Discusion} \label{sec:discussion}

\subsection{Experimental Results}

\subsubsection{Discrete Allocations}

\cref{fig:german_banks_results} shows the performance of the various algorithms on the German Banks dataset. In both~\eqref{eq:sop} and~\eqref{eq:soip} the greedy algorithm outperforms all the other algorithms algorithms, and then come the centrality and PageRank-based heuristics, as well as the LP with randomized rounding. Note that this network is strongly connected and all edges have both directions so the eigenvector centrality and the PageRank rankings coincide. In the worst case, the greedy algorithm outperforms the LP-based one by approximately 15\%, whereas the PageRank and centrality-based heuristics are outperformed by greedy by approximately 58\% in the worst case. Finally, we note that in both objectives the random permutation, wealths and outdegree heuristic perform badly. The significantly decreased performance of the wealth heuristic is justified by the fact that nodes which are important for the bailout process and have priority (i.e. lower wealth) are not well-connected which is in accord with theory, where an instance can be constructed such that the approximation ratio of the wealth heuristic goes to zero. Similarly, one can argue about the other heuristics, i.e. in the case of the out-degree heuristic nodes with low equity may be well connected to the other nodes and thus giving them priority does not contribute substantially to the overall objective. That suggest that the uneven form of such curves represent the fact that these simple heuristics are not good candidates for this optimization problem. The results regarding~\eqref{eq:sot} are similar, so we omit them.

\cref{fig:venmo_results} shows the results on the simulations run on the Venmo network. We observe that both the greedy and the  randomized rounding algorithm outperform the other algorithms by a margin of approximately 13\%. The difference between the greedy and randomized rounding algorithms is approximately 3.2\%. Again we note that the other heuristics do not perform as well as the greedy or randomized rounding algorithms for the reasons outlined in the previous paragraph. More precisely, the PageRank and centrality-based heuristics perform substantially bad perhaps due to the connectivity characteristics of the network (i.e. the network being too sparse).  

\cref{fig:eba_results} displays results about the EBA dataset with inferred internal liabilities based on a complete network backbone. In this type of network, the outdegree, the PageRank, and centrality heuristics  are equivalent to a random permutation, since any tie-breaking mechanism would yield a different permutation. Again, the greedy algorithm outperforms the wealth heuristic by about 30\% in the worst case whereas the randomized rounding algorithm outperforms the wealth heuristic by about 23\% in the worst case. 

\cref{fig:safegraph_results} displays the results on the SafeGraph data for both the~\eqref{eq:sop} and~\eqref{eq:sot} objectives where the results are qualitatively similar in both plots, although in slightly different scales. In this simulation, the randomized rounding algorithm outperforms all benchmarks and is also very close to the greedy algorithm with a worst case diffeerence of about 7.1\% and being as far as approximately 40\% from the other heuristics in the worst case in the~\eqref{eq:sot} plot. In the~\eqref{eq:sop} plot the differences are  about 18\% in the worst case, however the greedy algorithm quickly approaches the randomized rounding algorithm. 

\subsubsection{Fairness}

Regarding the Gini-constrained problems,~\cref{fig:gini_german_banks} has a predictable behaviour. To be more specific, as the number of bailouts $k$ is small, the instance for which \texttt{Target Gini $\le$ 0.1} has a lower relaxation optimum as well as rounded value for $k \le 6$ and later approaches the unconstrained optimum (i.e. where \texttt{Target Gini $\le$ 1.0}). Moreover the realized Gini Coefficient calculated by averaging the relaxation solutions is dropping in the unconstrained problem and meets with the curve of the constrained problem at $k = 20$. In the constrained case, the Gini coefficient rises to its upper bound 0.1 until $k = 6$ and then starts to drop, that coincides with the objective value plots. The explaination for this phenomenon is rather simple: at first, when resources are scarce, selecting certain nodes on the network subject to these resources creates inequality which is mitigated by the constraint at the expense of the quality of the solution creating a wors PoF of about $1.2 \ge 1$. When $k$ is large enough, i.e. $k \ge 6$, the available resources allow the constrained version to create a solution close to the unconstrained version (by ``rebalancing'' some $\bar z_i^*$ values) which is reflected on both objective values, and eventuall the PoF reaches 1 when the two solutions eventually meet. Also in the unconstrained version, again due to the fact that the number of resources increases the Gini coefficient drops at almost a linear rate. In~\cref{fig:venmo_network_gini}, where we constrain the \eqref{eq:sgc} with the same values as~\cref{fig:gini_german_banks} we observe a worst-case PoF of about 1.09 at $k = 13$. Indeed, if we plot the spatial inequality for $k = 13$ we observe that the mitigation of disparity through the constraint is considerable in all of the nodes which yields an about 8\% lower~\eqref{eq:sop} value. In~\cref{fig:property_gini} the PoF for the SafeGraph data is approximately 1, meaning that all the resources have been equitably allocated between minority/non-minority groups  subject to the respective constraint(s). For the German Banks data, the PoF is approximately 1.16 in the worst case ($k = 2$) and approaches 1 at $k = 6$. Lastly, in \cref{fig:pof_target} we observe that the PoF drops quickly to 1 in most cases and a decent trade-off between fairness and optimality can be achieved when $g = 0.4$. 

\subsection{Societal Implications}

Our work suggests a set of societal implications. The first consideration is how the features of individuals related to (i) the economic shocks and (ii)  the financial interactions, are gathered, a problem which is also highlighted in~\cite{abebe2020subsidy}. While many attributes may be accessible (or at least can be accurately inferred) by public datasets (such as the US Census, the US Economic Census, the Poverty Tracker Dataset~\cite{povertytracker} and so on), many features correlated to income shocks and financial interactions may touch on sensitive and private attributes of the individuals, which suggest that work needs to be done in the lens of these models where there are missing or noisy data. 

Moreover, carefully choosing the objective and the fairness constraint (if any) suggests a separate issue informing our work. While we mostly study linear objectives of the clearing variable $\bar p$, as well as maximizing the number of solvent nodes, there are more objectives that one can study in this context. One of them is a \emph{maximin} (egalitarian) objective  similar to the one of~\cite{abebe2020subsidy} which, in our case, corresponds to maximizing the worst-possible ratio of the clearing vector ratio over total liabilities which can be thought of ``fractional default'', namely $ \max_{\bar p \in \text{\eqref{eq:generalized_bailouts} }} \min_{j \in [n]} \bar p_j / p_j $. While we do not study this objective in the present paper, this objective can be thought of having built-in fairness and possibly methods like the ones presented in~\cite{tsang2019group} can be deployed.

\begin{figure}
    \centering
    \subfigure[Venmo Data, $L = 10 \cdot \one$]{\includegraphics[width=0.45\textwidth]{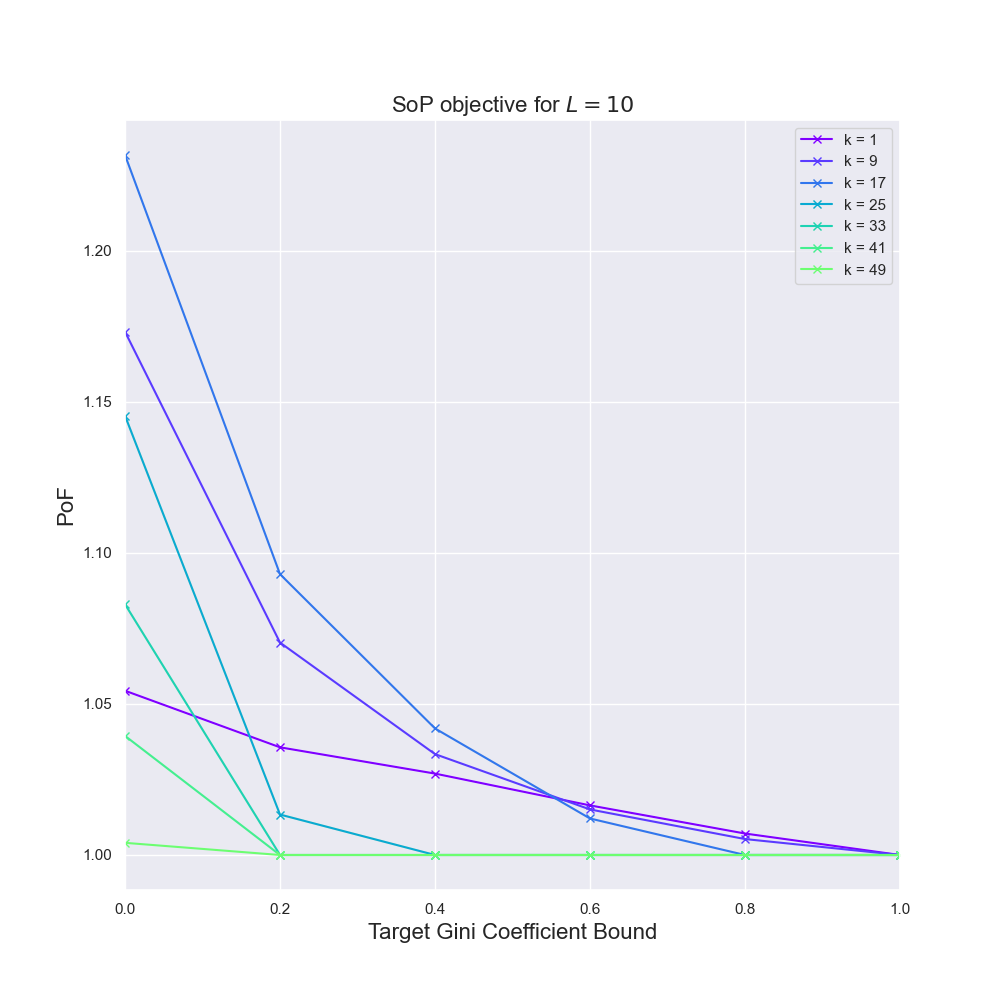}}
    \subfigure[German Banks Data, $L = 10^5 \cdot \one$]{\includegraphics[width=0.45\textwidth]{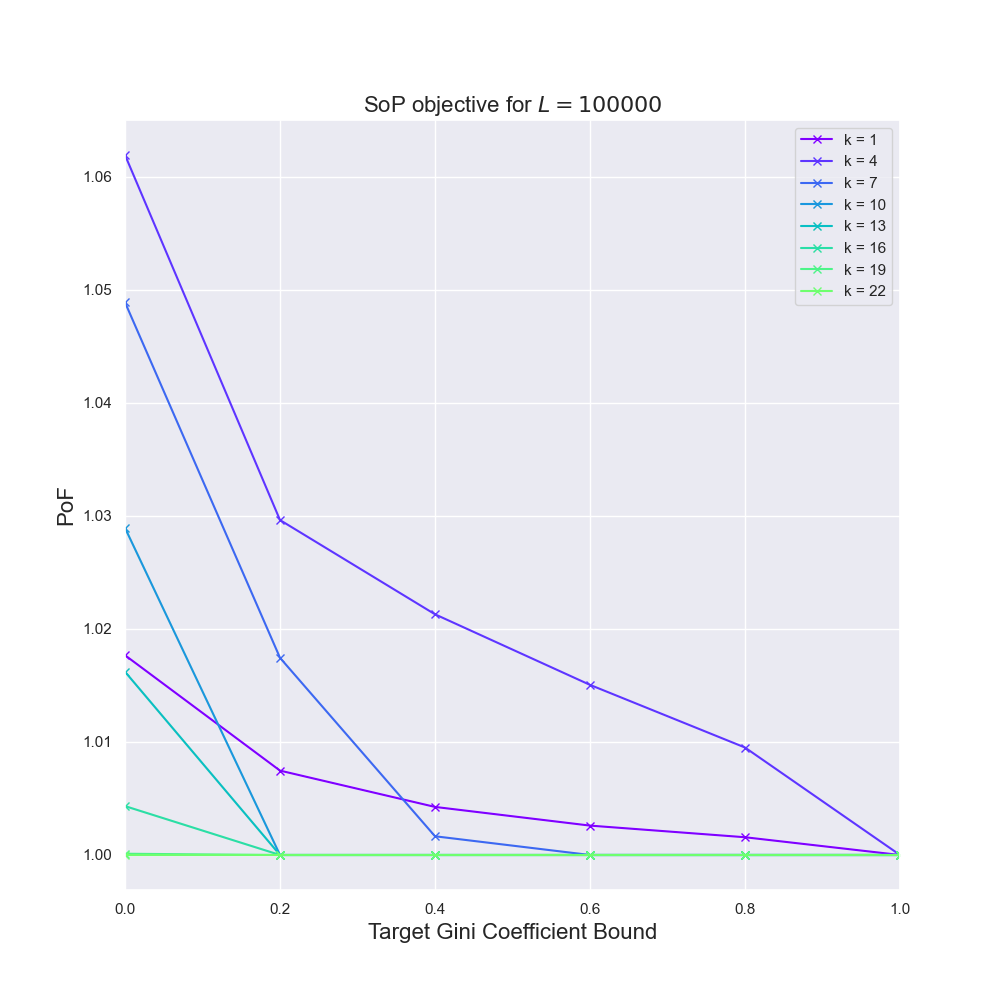}}
    \caption{Relation between fractional PoF and the upper bound $g$ on the \eqref{eq:sgc} for varying resources.}
    \label{fig:pof_target}
\end{figure}

From the perspective of policymakers, there are interesting phenomena that emerge in our work and can be subject to more extensive discussion. Perhaps the most ubiquitous one is that wealth-thresholding policies, i.e. allocating stimulus to people below (or, theoretically, even above) a certain income, does not seem to perform well when contagion happens on a network. This result \emph{juxtaposes} conventional rules and underlines that in order for the cumulative welfare to be maximized bailouts may not be redirected towards both the poorest or, vice-versa, the wealthiest nodes. An intuitive explanation for the bad performance of this criterion in presence of an underlying network that propagates contagion, is that  such nodes may not have many connections to the network, so the effect of bailing them out is less than bailing out more central nodes. Ending, we observe that the PoF showcases a quick phase transition to being 1 (\cref{fig:pof_target}) suggesting that a \emph{decent trade-off} between being fair (within our framework) and optimal can be achieved, even with moderate resources.

\subsection{Further Related Work} \label{sec:related_work}

\noindent \textbf{Financial Networks and the Eisenberg-Noe Model.} The EN model introduced in~\cite{eisenberg2001systemic} models a financial network where each node has assets and liabilities both with respect to the internal network, namely the other nodes of the network, and the external sector, namely the node may, for instance, owe taxes to the government, or has loans to external creditors, as well as may get social security benefits from the government. According to the EN model, when a node \emph{defaults}, namely is not able to pay out its creditors (internal and external), it rescales its obligations proportionally\footnote{This property is also known as absolute priority.} and pays the rescaled responsibilities in full\footnote{This property is also known as ``liability over equity''.}. The EN model is used to calculate these payments, which are usually called \emph{clearing payments}\footnote{The general problem has multiple equilibria, however, our paper, for simplicity, examines the cases where the equilibrium payments are unique.}, by computing a solution to a fixed-point problem, or, equivalently, an optimal solution to a mathematical program with a strictly increasing objective.  

In the presence of shocks~\cite{glasserman2015likely}, the nodes similarly become default due to external shocks that disrupt their assets. The works of~\cite{banerjee2018dynamic, feinstein2019dynamic} investigate dynamic variants of the EN model, where the liabilities and assets evolve as functions of time. Our work investigates the problem from a static viewpoint, and the hardness results obviously hold for the dynamical version. Moreover, \cite{jackson2020credit, ahn2019optimal, egressy2021bailouts}, investigate \emph{optimal bailouts} in the cases of defaulting. More specifically,~\cite{jackson2020credit}, investigates scenaria for the existence of multiple equilibria and provides necessary and sufficient conditions for solvency under any equilibrium. Besides this, their paper investigates optimal bailouts that brings the system into solvency and provides computational intractability results for the minimum bailout problem by a reduction from the partition problem. The subsequent work of \cite{egressy2021bailouts} considers the structure of optimal discrete bailouts under the RV model (a model with bankruptcy costs) of \cite{rogers2013failure} and proves hardness results for objectives that are similar to ours using different approaches where there are non-zero bankruptcy costs. Their paper does not consider  stochastic shocks, but rather a fixed state of the network and a fixed non-zero percentage regarding default costs. Additionally, their work does not include approximation algorithms, fairness-constrained optimization, and empirical work. They prove that for \emph{non-zero} bankruptcy costs maximizing the number of solvent nodes cannot be approximated within a factor of $n^{1 / (\log \log n)^C}$ for some universal constant $C > 0$ that agrees with our improved inapproximability result for the EN model.


An extension of the model includes credit default swaps, that is triads of entities enter contracts with one another whereas the default of a third entity in the network forces a bilateral transaction between the other two. Computing a clearing vector in such models was shown to be PPAD-Complete by~\cite{schuldenzucker2017finding}.    \cite{papp2020sequential} investigates the problem of \emph{sequential defaulting} in networks with debt contracts and credit default swaps and shows hardness results regarding identifying the number of default banks on the best-possible and worst-possible orders at which banks announce their defaults. 

The work of~\cite{ahn2019optimal} considers optimal intervention methods under budget constraints under the extended model of~\cite{glasserman2015likely} and formulate optimization problems that minimize the systemic losses as well as minimizing the number of defaulting institutions and apply their methods on publicly available data on the Korean financial system. Their work considers \emph{fractional intervention} policies which differentiates it from ours. More specifically, we consider \emph{discrete interventions} (i.e. a node gets the full bailout or does not) in which case the model of~\cite{ahn2019optimal} can be seen as the fractional relaxation of the optimization objectives we study. Secondly, we experiment both with high-granularity financial institutions (banks) as well as view the problem from a societal lens, namely modeling the entities of the system as ``societal nodes'' (businesses, households, individuals). 

Finally, the work of~\cite{feinstein2019obligations} generalizes the classical EN model to a multi-layered one, where the interconnected entities of the financial network have trades in multiple assets, develops a financial contagion 
model with fire sales that allows institutions to both buy and sell assets subject to some utility, and studies its equilibria. While the exact contributions of~\cite{feinstein2019obligations} are tangential to ours generalizing our model in their framework it provides an excellent pathway for extending the current work. 

\noindent \textbf{Income Shocks and Systemic Risk.} The theory of \emph{income shocks} has a long history financial mathematics. One of the papers that is most closely connected to our work is the work of~\cite{glasserman2015likely} where financial contagion in the presence of shocks is investigated, and results are provided regarding the probability of contagion of a subset of nodes due to a shock on a specific node. The setting of this paper is extensively discussed at~\cref{sec:eisenberg_noe}. 

The work of~\cite{abebe2020subsidy} studies subsidy allocations at the presence of income shocks where there is only temporal but no spatial information about the nodes. The analysis in their paper is based on \emph{ruin processes} and considers two main objectives, one is minimizing the expected number of nodes that are economically ruined (min-sum objective), and the other objective considers minimizing the worst ruin probability of an node in the network (min-max objective). As they mention in their paper, a crucial point that highlights income shocks in a societal level is the households have different abilities to withstand income shocks. The phenomenon disproportionately affects low-income families and can bring them into long-lasting poverty~\cite{desmond2016evicted, atake2018health, shapiro2004hidden}. Our paper can be seen as a part and extension of this work direction which uses optimization methods for decision-making, and a possible combination of the model of both~\cite{abebe2020subsidy} and ours presents an interesting research pathway. 

The contagion effect in a financial network can trigger a \emph{financial crisis} through contagion \cite{allen2000financial, freixas2000systemic}, with a vivid example being the 2007-2009 financial crisis which followed after the collapse of Lehman Brothers because of the decline of the housing market~\cite{acharya2009restoring}, that initiated a cascade of failures of financial institutions around the world. This domino of effects prompted government bailouts around the world, with the lending interests booming, which turned many nations, such as Greece, unable to service their mounting debts.



\noindent \textbf{Optimal Stimulus Allocation During COVID-19.} There are multiple recent works regarding stimulus allocation and the recent COVID-19 economic crisis response. Most notable plans for economic relief are the ones of the United States of America \cite{act2020cares}, and the European Union \cite{europeanrelief} which contain stimulus packages to be distributed to individuals by the governments. 

The work of \cite{carroll2020modeling} models the response to the US CARES act on consumption by extending existing models of consumption to incorporate features of the COVID-19 crisis, where spending has to be limited, and predicts, that under a short lockdown (as of April 2020) the stimulus payments will be sufficient to make a recovery. Their work does not incorporate devising a policy to allocate stimulus checks and investigating issues of fairness, which is the main part of our paper.

Data-driven approaches have been followed by the works of~\cite{baker2020income}\footnote{\url{https://github.com/econ-ark/Pandemic}} and~\cite{chetty2020did} where the authors used large datasets in order to study how the stimulus payments were allocated and found out that poorer consumers and consumers who had lower liquidity spent a significantly higher amount of their stimulus checks quickly. In contrast, our work studies the assignment problem and its fairness implication, and not the consumption response to the pandemic. 

Finally, a work tangential, but fundamentally different from ours is the work of~\cite{nygaard2020optimal, nygaard2021optimal} where the authors use a \emph{life-cycle} model and build an efficient algorithm to find the optimal policy that maximizes aggregate consumption. The main difference with~\cite{nygaard2020optimal} is that their work provides a policy based on a completely different model and does not account for the \emph{network effects} of the economy and shocks, and consequently do not know how they affect the bailouts. 

\noindent \textbf{Influence Maximization.} While not directly related to financial networks, the allocation problem we study in this paper has very close ties to \emph{Influence Maximization (IM)}. The work of~\cite{kempe2003maximizing} introduced the influence maximization problem as follows: given a network in which each edge can transmit information (e.g. disease, marketing information etc.) with probability $p$ independently of the other edges, the IM problem asks whether there exists a set $S$ with $|S| = k$ such as the number of influenced nodes is maximized. Based on the submodularity properties of the influence function, the authors devise an $(1 - 1/e - o(1))$-approximation algorithm for approximating the optimal influence set. This work has been greatly extended by a series of works that optimize its algorithm (e.g. see~\cite{borgs2014maximizing, goyal2011celf++} and the references therein), and adapt it to different contexts (see e.g. \cite{leskovec2007cost, leskovec2007dynamics, chen2020time}). Another recent interesting work related to ours is the concept of \emph{Fair Influence Maximization}~\cite{tsang2019group, rahmattalabi2020fair} which discusses the maximization of influence subject of fairness constraints.


\section{Conclusions and Future Work} \label{sec:conclusion}

In this work we propose a model of discrete subsidy allocation based on the Eisenberg-Noe model which assumes a collective of interconnected individuals which posses features such as external assets and liabilities, internal assets and liabilities, demographic features, and experience shocks on their assets leading a part of them to default. Our work addresses the problem of averting as much damage as possible from the shocks subject to a total budget constraint by relying in optimizing a welfare objective. The subsidy allocations are discrete and can be determined from each node's features (such as e.g. the number of children, the income level, the poverty status etc.). Policymakers are faced with similar planning problems when attempting to optimize some welfare measure on a population. Although exact network interactions are difficult to be inferred on a realistic world, our model and the consequent data-driven experiments based on individual level and business and households-level granularity offer a simple proxy of studying such problems in real-world scenarios, and provide useful insights with respect to the effectiveness of certain strategies. For instance, thresholding by an ex-ante wealth criterion (i.e. income level), which is a de facto pathway in policymaking,  yields worse results in our model than the greedy or the LP-based policies for the linear objectives in question. Furthermore, we show that maximizing the (absolute) number of solvent nodes cannot be approximated within a poly-time computable factor. Closing, we show how fairness can be incorporated on the model and study the price of fairness both theoretically and empirically. 

There are numerous ways to extend the horizons of the present work. First of all, there are two very interesting objectives that can be studied: the former is a maximin objective similar to the one presented in~\cite{abebe2020subsidy} subject to the bailout scheme we study in this paper; the latter one is the systemic loss objective as presented in~\cite{glasserman2015likely}. For such objectives, it could be an important contribution to develop approximation algorithms, or study the approximation algorithms presented in this work on the corresponding objectives. Besides this, it is interesting to   examine whether there exist algorithms with better approximation ratios, or prove the absence thereof, than the ones developed in this paper, and examine whether more efficient variants of the current algorithms can be implemented (for instance, using the ideas of~\cite{leskovec2007cost} and the subsequent works could improve the runtime of the greedy algorithm). Finally, there is a need to analyze additional real-world datasets at the societal level. The study,  modelling, and inference of income shocks and financial interactions from real-world datasets suggest  a cardinal direction for follow-up work. Note, however, that  obtaining such data may be a difficult task due to data protection legislation, pointing toward possible future synergy between the academic community, policymakers, and companies. 

\bibliographystyle{alpha}
\bibliography{references}

\newpage

\appendix

\section*{Acknowledgements}

The authors would like to thank Emma Pierson, and Serina Chang regarding discussions involving the SafeGraph data, and the anonymous referees for their constructive feedback.

\section{Omitted Proofs} \label{sec:ommited}

\noindent \textbf{Proof of ~\cref{lemma:comparison}.} The polytopes $\mathcal P_1(A, c_1, b)$ and $\mathcal P_2(A, c_2, b)$ satisfy the relation $\mathcal P_2 \subseteq \mathcal P_1$. Using the fact that the fixed point operator is increasing~\cite{eisenberg2001systemic} we get that $\bar p_1 \ge p \wedge \left ( A^T \bar p_2 + c_1 \right ) \ge \bar p_2$. Thus $\bar p_1 \ge \bar p_2$. Moreover for every vector $v \ge \zero$, trivially $v^T \bar p_1 \ge v^T \bar p_2$. To find a direction $\delta \ge \zero$ such that $\bar p_1 \ge \bar p_2 + \delta$, let $R_1, D_1$ be the sets of solvent and default nodes for assets $c_1$ and $R_2, D_2$ be the sets of solvent and default nodes for assets $c_2$. Then we know from above that $D_1 \subseteq D_2$ and consequently $R_2 \subseteq R_1$. So $R_1 \cap R_2 = R_2$, and $D_1 \cap D_2 = D_1$. Thus we calculate $\delta$ to have components

\begin{equation*}
    \delta_j = \begin{cases}
        0 & j \in R_2 \\
        \sum_{i \sim j} a_{ij} (\bar p_{1i} - \bar p_{2i} ) + c_{1j} - c_{2j} & j \in D_1 \\
        \left ( p_{1j} -\sum_{i \sim j} a_{ij} \bar p_{2i} - c_{2j} \right )^+ & j \in R_1 \cap D_2
    \end{cases}.
\end{equation*}

We also get that a feasible solution is given by $\tilde \delta = \one_{D_1} \odot (c_1 - c_2) \ge \zero$ such that $\tilde \delta \le \delta$. \qed

\noindent \textbf{Proof of~\cref{theorem:hardness}.} We first start by designing a polynomial-time reduction from a 3-Set-Cover instance to an instance of the decision version of the~\eqref{eq:as} objective.
    
    \emph{Reduction.} We fix a constant $\alpha \in (0, 3)$. We create a bipartite network $H$ with a node set partitioned into $V$ and $U$ where $V = \{ v_1, \dots, v_m \} $ is the node set that represents all the ``set nodes'' and $U = \{ u_1, \dots, u_n \}$ is the node set that represents all the ``item nodes'' of the corresponding 3-Set-Cover instance. For each set node $v_j$ and each item $u_i$ we add an edge $(v_j, u_i)$ if and only if $u_i \in S_j$. We set the parameters of the EN model as follows. First, for each ``set node'', there is an external influx of $c_j = 3$, and an external outflux of $b_j = \alpha$. For each edge $(v_j, u_i)$ created there is an obligation of $p_{ji} = 1 - \alpha / 3$ from the set to the corresponding item. Thus $\beta_{\max} = 1 - \alpha / 3 < 1$. Each item node has no external influx, i.e. $c_i = 0$, and has external liabilities of $b_i = 1 - \alpha / 3$. The shock distribution is taken to be the point-mass distribution that has $X_j = 3$ if $X_j$ is a ``set node'' and is zero otherwise. We also take $L_j =  3$ for all $j \in [n]$, $\Lambda = 3k$ and $r = k + n$. After the shock, the influx of cash is disrupted and all nodes default to having $\bar p_i = 0$, namely no-one can meet his/her obligations. The reduction creates a network with $n + m$ nodes and $n + 4m$ edges and runs in polynomial-time.  
    
    \emph{Forward Direction.} Suppose that $\mathcal C$ is a set cover with $|\mathcal C| = k$. Giving aid $L$ to set nodes $v_i$ such that $i \in \mathcal C$ we have that every ``set node'' in $\mathcal C$ becomes solvent, since it can pay the obligations of $p_j = 3$ to the ``item nodes'', and the corresponding items that it covers become solvent themselves. Since $\mathcal C$ is a set cover, every item becomes solvent, and all the set nodes $v_i$ for $i \notin \mathcal C$ remain default, hence we have a total of $k + n$ solvent nodes. 
    
    \emph{Reverse Direction.} Suppose that there are at least $r = k + n$ solvent nodes, therefore there are at most $d = m - k$ default nodes. Let $J \subseteq U \cup V$ be the set with $|J| = k$ elements such that every element $j \in J$ gets an aid of $L$. We will show that $J \subseteq V$. Suppose that $J$ has $\ell$ elements on $U$ and $k - \ell$ elements on $V$. Then, the number of solvent nodes is at most $\ell + k - \ell + k - \ell = 2k - \ell$, and the number of default nodes is therefore at least $(n - k + \ell) + (m - k)$. In order for this to hold we must have that $n - k + \ell \le 0$, which implies that $\ell = 0$. Therefore $J \subseteq V$. Therefore, every item node is solvent and thus $J$ must be a set cover for the original problem. The budget constraint is also satisfied.
    
\emph{\eqref{eq:linobj} objective.} The reduction construction for the~\eqref{eq:linobj} problem is very similar: instead of letting $r = k + n$ we let $r = 3k + n$. Moreover, we let $v = \one$.

\qed    

\noindent \textbf{Proof of \cref{lemma:approximation_greedy_induction}.} \condref{assumption:small_bailout} helps us establish a \emph{lower bound} on the marginal gain. We state the following Corollary which is obtained by applying ~\cref{lemma:comparison} with asset vectors $c_1 = c - x + L \odot\one_{S_t \cup \{ u_t \}}$ and $c_2 = c - x + L \odot\one_{S_t} $.

\begin{corollary} \label{corollary:lower_bound}
    Under \condref{assumption:small_bailout} and a fixed shock $X = x$, the marginal gain at each iteration $t \le k$ where node $u_t$ is chosen is at least $v_{\min} L_{u_t}$. 
\end{corollary}

Next, we prove the following upper bound regarding bailing out two sets $S, T$ with $S \subseteq T$.

\begin{lemma}[Marginal Gain Upper Bound Lemma] \label{lemma:gain_upper_bound}
    Let $f(\bar p) = v^T \bar p$ for some $v > \zero$ and let $\emptyset \subseteq S \subseteq T \subseteq [n]$ be two sets, and fix a shock $X = x$. Then if $\bar p_T$ is the clearing vector after bailing out $T$ and $\bar p_S$ is the clearing vector after bailing out $S$, then $v^T \bar p_T - v^T \bar p_S \le \tfrac {v_{\max} \sum_{j \in T \setminus S} L_j} {1 - \beta_{\max}}$.
\end{lemma}

\noindent \emph{Proof of ~\cref{lemma:gain_upper_bound}.}
    
    \begin{align*}
            \one^T \bar p_T - \one^T \bar p_S & = \| \bar p_T - \bar p_S \|_1 \explain{\cref{lemma:comparison}} \\
            & = \| p \wedge \left ( A^T \bar p_T + c - x + L \odot\one_T \right ) - p \wedge \left ( A^T \bar p_S + c - x + L \odot\one_S \right ) \|_1 \explain{Fixed Point Op.} \\
            & \le \| A^T \bar p_T + c - x + L \odot\one_T - A^T \bar p_S - c + x - L \odot\one_S \|_1 \explain{Non-expansion of $p \wedge ( \cdot )$} \\
            & \le \| A^T \|_1 \| \bar p_T - \bar p_S \|_1 + L^T \one_{T \setminus S}  \explain{Norm Consistency, $\one_T - \one_S = \one_{T \setminus S}$ for $S \subseteq T$} \\
            & \le \beta_{\max} \| \bar p_T - \bar p_S \|_1 + L^T \one_{T \setminus S} \explain{$\| A^T \|_1 = \beta_{\max} \in (0, 1)$}. 
    \end{align*}
    
    Rearranging we get $\one^T \bar p_T - \one^T \bar p_S \le \tfrac {\sum_{j \in T \setminus S} L_j} {1 - \beta_{\max}}$. Therefore for a linear objective $v^T \bar p$ with $v > \zero$ since the EN model would yield the same clearing vector $\bar p$ \cite{eisenberg2001systemic} we have
    
    \begin{equation*}
        v^T (\bar p_T - \bar p_S) \le v_{\max} \one^T (\bar p_T - \bar p_S) \le \frac {v_{\max} \sum_{j \in T \setminus S} L_j} {1 - \beta_{\max}}.
    \end{equation*}
    
\qed

We use the above lemma to state the following Corollary which results as a combination of ~\cref{corollary:lower_bound} and ~\cref{lemma:gain_upper_bound}. 

\begin{lemma} \label{lemma:gain_upper_bound_greedy}
    Let $S$ be a set and $S^*$ be the optimal bailout set set under a fixed shock $X = x$. Then, under \condref{assumption:small_bailout} we have that $v^T \bar p_{S^*} - v^T \bar p_S \le \tfrac {\zeta k} {1 - \beta_{\max}} \max_{u \notin S, u \text { feasible}} \left \{ v^T \bar p_{S \cup \{ u \}} - v^T \bar p_S \right \} $.
\end{lemma}


\noindent \emph{Proof of ~\cref{lemma:gain_upper_bound_greedy}.} Let $S$ be a set and let $S^*$ be the optimal bailout set. Let $S^* \setminus S = \{ j_1, \dots, j_q \}$ for $q \le k$. We have 

\begin{align*}
    v^T \bar p_{S^*} & \le v^T \bar p_{S \cup S^*} \explain{Monotonicity, $S^* \subseteq S \cup S^*$} \\
    & = v^T \bar p_S + \sum_{l = 1}^q v^T \left ( \bar p_{S \cup \{j_1, \dots, j_{l} \}} - \bar p_{S \cup \{ j_1, \dots, j_{l -1} \}} \right ) \explain{Telescoping Sum} \\
    & \le v^T \bar p_S + \frac {v_{\max}} {1 - \beta_{\max}} \sum_{l = 1}^q L_{j_l} \explain{\cref{lemma:gain_upper_bound}} \\
    & \le v^T \bar p_S + \frac {v_{\max} k} {1 - \beta_{\max}} \max_{l \in [q]} L_{j_l} \explain{Bound by Maximum} \\
    & \le v^T \bar p_S + \frac {v_{\min} \zeta k} {1 - \beta_{\max}} \max_{l \in [q]} L_{j_l} \explain{$\zeta = v_{\max} / v_{\min}$} \\
    & \le v^T \bar p_S + \frac {\zeta k} {1 - \beta_{\max}} \max_{e \notin S} \left \{ v^T \bar p_{S \cup \{ e \}} - v^T \bar p_{S}  \right \} \explain{\cref{assumption:small_bailout}}. \\
    \iff & \max_{e \notin S, e \text{ feasible}} \left \{ v^T \bar p_{S \cup \{ e \}} - v^T \bar p_{S}  \right \} \ge  \frac {1 - \beta_{\max}} {k \zeta} \left ( v^T \bar p_{S^*} - v^T \bar p_S \right ). \explain{Rearrangement}
\end{align*}

\qed

\noindent \emph{Proof of Main Theorem (\cref{lemma:approximation_greedy_induction}).} Let $S_t$ be the solution at set $t$ and let $k$ be the total number of iterations of the algorithm. Let $\rho = (1 - \beta_{max}) / (k \zeta)$ Then, for a fixed shock $X = x$

\begin{align*}
    v^T \bar p_{S_k} & \ge \rho v^T \bar p_{S^*} + (1 - \rho) v^T \bar p_{S_{k - 1}} \explain{\cref{lemma:gain_upper_bound_greedy}} \\
    & \ge v^T \bar p_{\emptyset} + \rho v^T \bar p_{S^*} \sum_{0 \le l \le k} (1 - \rho)^l \explain{Induction} \\
    & \ge \frac {\rho (1 - \rho)^k} {\rho} \cdot  v^T \bar p_{S^*} \explain{$v^T \bar p_{\emptyset} \ge 0$ and Geometric Sum} \\
    & \ge \left (1 - e^{- k \rho} \right ) \cdot v^T \bar p_{S^*} \explain{$1 + x \le e^x$} \\
    & = \left ( 1 - e^{- (1 - \beta_{max}) / \zeta} \right ) \cdot v^T \bar p_{S^*} \explain{Definition of $\rho$}.
\end{align*}

Taking expectations over $X \sim \D$ in the above expression we get the desired result.

\qed

\noindent \textbf{Proof of \cref{theorem:greedy_runtime}.} \emph{Different Bailouts.} The proof follows similar philosophy as Lemma 3.6 of~\cite{borgs2014maximizing}. We construct the set of feasible solutions $\mathcal S = \{ S \in 2^{[n]} : S \text{ is a feasible allocation set} \}$. A set $S \in \mathcal S$ can have at most $\Lambda / L_{\min}$ elements. Applying the Chernoff bound for $m = O \left ( \tfrac {\log n \cdot \Lambda / L_{\min}} {\varepsilon^2} \right )$ each $S \in \mathcal S$ deviates at most an additive factor of $\varepsilon \opt_f$ from its expectation with probability at least $1 - 2^{-\Lambda / L_{\min}} n^{-1}$. The probability that there exists $S \in \mathcal S$ that violates this event is given by the union bound on $\mathcal S$, with $| \mathcal S | \le 2^{\Lambda / L_{\min}}$, and is at most $1 / n$. Conditioned on this event which happens with probability $1 - O(1/n)$ and letting $\overline {\opt_f}$ denote the sample average of the optimal solutions we have 

\begin{equation*}
    \sol_f \ge \overline {\sol_f} - \varepsilon \opt_f \ge (1 - \gamma_f) \overline {\opt_f} - \varepsilon \opt_f \ge \left [ (1 - \gamma_f) (1 - \varepsilon) - \varepsilon \right ] \opt_f \ge (1 - \gamma_f - 2 \varepsilon) \opt_f
\end{equation*}

Therefore, the average estimator achieves an $(1 - \gamma_f - 2 \varepsilon)$-approximation. The total number of iterations is therefore $O \left ( m \tfrac {\Lambda}  {L_{\min}} \right ) = \til O \left ( \tfrac {\Lambda^2} {\varepsilon^2 (L_{\min})^2} \right )$. Assuming that each iteration is solved by the fixed point iteration in time $\mathcal T_{\mathrm{op}} = O \left ( (n + |E|) \tfrac {\log (1 / \eta)}{\log (1 / \beta_{\max})} \right )$ within relative accuracy $0 < \eta < 1$ via using a sparse-matrix multiplication (or adjacency lists) the total runtime becomes 

\begin{equation*}
    O \left ( \frac {(n + |E|) \log n \Lambda^2 \log (1 / \eta)} {L_{\min}^2 \varepsilon^2 \log(1 / \beta_{\max})} \right ) = \til O \left ( \frac {(n + |E|) \Lambda^2} {L_{\min} \varepsilon^2 }\right ),
\end{equation*}

where the final solution is within an $(1 - \gamma_f - 2 \varepsilon)$ approximation factor from the optimal solution and modulo a (very small) relative approximation error of $\eta$.

\qed

\noindent \textbf{Proof of~\cref{theorem:apx_guarantees}.} To prove \cref{theorem:apx_guarantees} we first prove the following two technical helper Lemmas:

\begin{lemma}[Upper Bound Lemma] \label{lemma:bounding}
    Let $\til \xi = (\til p, \til z)^T$ be any feasible solution to the \eqref{eq:generalized_bailouts_relaxation} problem, and let $X = x$ be a point-mass shock. Then, for every $S \subseteq [n]$ the following inequality holds: $\sum_{j \in S} (1 - \beta_j) \til p_j \le  \sum_{j \in S} (c_j - x_j + L_j \til z_j)$. 
\end{lemma}

\noindent \emph{Proof of ~\cref{lemma:bounding}.} The proof of the Lemma is quite of technical nature, and involves summing the relative liabilities constraint $\bar p \le A^T \bar p + c - x + L \odot\bar z$ over the elements of the set~$S$.  More formally we consider nodes of the set $S$ and sum over $S$, i.e.

\begin{equation*}
    (\one - \beta)_S^T \bar p \overset {\text{\cite[(13)]{glasserman2015likely}}} {\le} \one_S^T (I - A^T) \bar p \le  \one_S^T (c - x) + (L \odot \bar z)^T  \one_S.
\end{equation*}
 
Thus $\sum_{j \in S} (1 - \beta_j) \bar p_j \le \sum_{j \in S} (c_j - x_j + L_j \bar z_j)$. 

\qed

\begin{lemma}[Lower Bound Lemma]
\label{lemma:lower_bound}
    Let $\til \xi^* = (\til p^*, \til z^*)^T$ be an optimal solution to the~\eqref{eq:generalized_bailouts_relaxation} problem under a point-mass shock $X = x$ and let $j$ be a default node. Then $\til p_j^* \ge c_j - x_j + L_j \til z_j^* $. Moreover if $Z_j \sim \mathrm{Be}(\til z_j^*)$ and $\bar p_j$ is the corresponding (feasible) clearing solution of~\eqref{eq:generalized_bailouts} after rounding, then, conditioned on the fact that $j$ is default we have that $\bar p_j \ge c_j - x_j + L_j Z_j$, and subsequently $\evz {\bar p_j} \ge c_j - x_j + L_j \til z_j^*$.  
\end{lemma}

\noindent \emph{Proof of ~\cref{lemma:lower_bound}.} Let $\til \xi^* = (\til p^*, \til z^*)$ be an optimal solution to the relaxation problem, and let $j$ be a default node under this solution. Then $j$ satisfies the following equation (due~to~absolute~priority), 
\begin{equation*}
    \til p_j^* = \sum_{i \sim j} a_{ij} \til p_i^* + c_j - x_j + L_j \til z_j^* \ge c_j - x_j + L_j \til z_j^*,
\end{equation*}

where we have used the fact that the optimal solution is feasible and that $\til p^* \ge \zero$, and $a_{ij} \ge 0$. The result for the rounded solution is inferred exactly the same way. 

\qed

\noindent \emph{Proof of Main Theorem (\cref{theorem:apx_guarantees}).} We now proceed with the main Theorem: Let $f(\bar p) = v^T \bar p$ for some $v > \zero$ (i.e. a strictly increasing linear function of $\bar p$). The rounded variables are sampled from $\mathrm{Be}(\til z^*)$. So, for a fixed shock $X = x$

\begin{equation*}
    \evz {\sol_f(x)} = \sum_{\bar z \in \{0, 1\}^n} \prz {Z = \bar z} \cdot \evz {\sol_f(x) | Z = \bar z}.
\end{equation*}

Conditioned on the event $\{ Z = \bar z \}$, we use the sets $D_{\sol(\bar z)}$ and $R_{\sol(\bar z)}$ to denote the sets of default and solvent nodes under the assignment $\bar z \in \{ 0, 1 \}^n$. We therefore break the above sum as 

\begin{equation*}
    \evz {\sol_f(x) | Z = \bar z} = \evz{\sum_{j \in D_{\sol(\bar z)}} v_j \bar p_j + \sum_{j \in R_{\sol(\bar z)}} v_j \bar p_ j \bigg | Z = \bar z} 
\end{equation*}

We treat the two components of the partition of $[n]$ as follows: 

\begin{compactenum}
    \item \emph{Solvent Nodes.} Every solvent node $j$ satisfies $\bar p_j = p_j \ge \til p_j^* \ge \tfrac {1 - \beta_{\max}} {\zeta} \til p_j^*$. Summing over the set of solvent nodes we get 
    
    \begin{equation*}
        \evz {\sum_{j \in R_{\sol(\bar z)}}  v_j \bar p_j \bigg | Z = \bar z} \ge \left ( \frac {1 - \beta_{\max}} {\zeta} \right ) \cdot \evz {\sum_{j \in R_{\sol(\bar z)}}  v_j \til p_j^* \bigg | Z = \bar z}.
    \end{equation*}
    
    \item \emph{Default Nodes.} We have
    
    \begin{align*} \label{eq:sop_apx_ratio}
         \evz {\sum_{j \in D_{\sol(\bar z)}} v_j \bar p_j \bigg | Z = \bar z } & \ge \evz {\sum_{j \in D_{\sol(\bar z)}} v_j(c_j - x_j + L_j Z_j) \bigg | Z = \bar z }  \explain{\cref{lemma:lower_bound}} \\
        & \ge \evz {\sum_{j \in D_{\sol(\bar z)}} v_{\min} (1 - \beta_j) \til p_j^* \bigg | Z = \bar z} \explain{\cref{lemma:bounding}} \\
         & \ge \evz {\sum_{j \in D_{\sol(\bar z)}} \frac {1 - \beta_{\max}} \zeta v_{\max} \til p_j^* \bigg | Z = \bar z} \explain{$\zeta = v_{\max} / v_{\min}$} \\
        & \ge \left (\frac {1 - \beta_{\max}} \zeta \right ) \cdot \evz {\sum_{j \in D_{\sol(\bar z)}}  v_{j} \til p_j^* \bigg | Z = \bar z} \explain{$v_{\max} \ge v_j$}.
    \end{align*}

\end{compactenum}

Thus, since the LP bound is an upper bound to the optimal we get  $\evz {\sol_f(x)} \ge \left ( \tfrac {1 - \beta_{\max}} {\zeta} \right ) \cdot \evz {\opt_f(x)}.$ \qed

\noindent \textbf{Proof of ~\cref{theorem:runtime}.}
    The proof of the argument is straightforward in general, and follows the classical analysis pathway for this kind of algorithms~\cite{williamson2011design}. We start by bounding the probability of failure of this step. Let us fix a sample $X = x^{(i)}$ where $i \in [m]$ and let $\mathcal F_f^{(i)}$ be the event that the $i$-th random assignment fails. Then we know that 

\begin{equation*}
\begin{split}
    \prz {\mathcal F_f^{(i)}} & = \prz { \left \{ \sol_f^{(i)} \le (1 - \gamma_f - \varepsilon^2) \opt_f^{(i)} \right \}  \cup \left \{  L^T Z^{(i)} \ge (1 + \delta_{\Lambda} ) \Lambda \right \}  }\\
    & \le \prz { \sol_f^{(i)} \le (1 - \gamma_f - \varepsilon^2) \opt_f^{(i)} } + \prz{ L^T Z^{(i)} \ge (1 + \delta_{\Lambda}) \Lambda },
\end{split}
\end{equation*}

by virtue of the union bound. We bound the first probability by (reversed) Markov's Inequality

\begin{equation*}
    \prz { \sol_f^{(i)} \le (1 - \gamma_f - \varepsilon^2) \opt_f^{(i)} } \le \frac 1 {1 + \varepsilon^2 / \gamma_f}  \le 1 - \frac {\varepsilon^2} {2 \gamma_f}  \le 1 - \frac {\varepsilon^2} 2.
\end{equation*}

We distinguish the following two cases for our analysis: the former case considers equal bailouts and the latter one considers unequal bailouts:

\begin{compactenum}

\item \emph{Equal bailouts.} For the second inequality, we know that all the bailouts have the same value $\ell$. Let $k = \Lambda / \ell$. Then we know from the Chernoff bound that $\prz {\one^T Z \ge (1 + \delta_k) k} \le e^{-\delta_k^2 k / 3}$ and thus letting $\delta_k = \sqrt {3 \log (4 / \varepsilon^2) / k}$ we know that the above probability is at most $\varepsilon^2 / 4$. Thus a deviation of $k + \til O(\sqrt k)$ (resp. $\Lambda + \til O(\sqrt \Lambda)$ for the total budget) is allowed with high probability for the case of \emph{equal bailouts}.

Combining the two events above we end up with $\prz { \mathcal F_f^{(i)} } \le 1 - \varepsilon^2 / 4$. Repeating the procedure $T$ times boosts the probability of failure to at most $(1 - \varepsilon^2 / 4)^T \le e^{-\varepsilon^2 T / 4}$. Failing in any of the $m$ shocks happens with probability at most $me^{-\varepsilon^2 T / 4}$. Finally, by the one-sided Chernoff bound we have that for the sample average  $\overline {\opt_f}$  of the corresponding optima

\begin{equation*}
    \prx {\overline {\opt_f} \ge (1 - \varepsilon) \opt_f } \le e^{-m \varepsilon^2 \opt_f^2 / \| f \|_\infty^2}.
\end{equation*}

We want to make the probability at most $e^{-\varepsilon^2 T / 4}$ so choosing $m = \frac {T \opt_f^2} {4 \| f \|_\infty^2}  $ yields the desired result. Note that $m \le T / 4$ since $\opt_f \le \| f \|_\infty$, so repeating the process for $T / 4$ steps yields the desired guarantee. The final probability of failure is $O(T e^{-\varepsilon^2 T / 4})$. Setting $T = 4 (\log n) / \varepsilon^2$ we get that the failure probability is $O(\log n / (\varepsilon^2 n))$. So the algorithm succeeds with probability $1 - O(\log n / (\varepsilon^2 n))$.  The total runtime  is 

\begin{equation*}
    O \left ( \frac {\log n} {\varepsilon^2} \left ( \mathcal T +  \frac {k \log n} {\varepsilon^2} \right ) \right ) = \til O \left ( \frac {\mathcal T} {\varepsilon^2} + \frac k { \varepsilon^4} \right ).  
\end{equation*}

Finally, with probability $1 - O(\log n / (\varepsilon^2 n))$ we have that for $\overline {\sol_f} = \frac 1 m \sum_{i \in [m]} \sol_f^{(i)}$

\begin{equation*}
        \overline {\sol_f} \ge (1 - \gamma_f - \varepsilon^2) \overline {\opt_f} \ge (1 - \gamma_f - \varepsilon^2) (1 - \varepsilon) \opt_f \ge (1 - \gamma_f - 2 \varepsilon) \opt_f.
\end{equation*}

\noindent \emph{Remark on the value of $\varepsilon$.} A suitable value for $\varepsilon$ can be $\varepsilon = \Theta(n^{-1/4})$ for which we get a runtime of $\til O(\mathcal T \sqrt n + kn)$, a success probability of $1 - O(\log n / \sqrt n)$ and an approximation guarantee of $1 - \gamma_f - \Theta (n^{-1/4})$. 

\item \emph{Adaptation of the Rounding Scheme of~\cite{srinivasan2001distributions} for general bailouts.} We can adapt the dependent rounding scheme of~\cite{srinivasan2001distributions} to perform randomized rounding such that (i) we obtain the same approximations as in the case of independent rounding; (ii) the constraints of the problem are satisfied with high probability. The problem that the independent rounding scheme has is that there exist choices of the entries of $L$ (e.g. $L_j = 2^{j-1}$) such that $L^T Z$ is anti-concentrated. In fact, when $L_j = 2^{j - 1}$ and $Z \sim \mathrm{Be} (1/2 \cdot \one)$ then $L^T Z \sim \mathrm{Uniform}(\{ 0, \dots, 2^n - 1 \})$. To mitigate this problem we will use the main result of ~\cite{srinivasan2001distributions} as follows: We first solve the LP relaxation and get the optimal fractional bailouts $\til z^*$ as in the independent rounding case.  Then we define a vector $U$ as a sample from the oracle of~\cite{srinivasan2001distributions} (described in Section 2, and Section 3.2 that modifies the original algorithm for a non-integral sum) 

\begin{equation*}
\small
    U \sim \D_{\mathrm{dep}} \left ( \underbrace {\frac {L_1} {\| L \|_{\infty}} (1 - \til z_1^*), \dots, \frac {L_n} {\| L \|_{\infty}} (1 - \til z_n^*), \left \lceil \frac {\| L \|_1 - \Lambda} {\| L \|_{\infty}} \right \rceil -  \frac {\| L \|_1 - \Lambda}  {\| L \|_{\infty}}}_{\text{probabilities}}; \underbrace {n + 1}_{\text{\# variables}} ; \underbrace{\left \lceil \frac {\| L \|_1 - \Lambda} {\| L \|_{\infty}} \right \rceil}_{\text{total weight}} \right ),
\end{equation*}

and let $Z = \one - U$. For (i) we have that $\ev {} {Z_j} = \Pr [ Z_j = 1] = 1 - \Pr [ U_j = 1] = 1 - \tfrac {L_j} {\| L \|_{\infty}} (1 - \til z_j^*) \ge \til z_j^*$ and thus we can achieve the desired approximation guarantees of~\cref{theorem:apx_guarantees}. For (ii) we know that $\sum_{j} L_j U_j$ are concentrated and thus we can easily show that $\sum_{j} L_j Z_j$ are concentrated as well, and thus with high probability the rounded bailouts are at most $\Lambda + \til O(\sqrt \Lambda)$. Finally, the runtime can be similarly deduced to be $\til O \left ( \tfrac {\mathcal T} {\varepsilon^2} + \tfrac {\Lambda} {L_{\min} \varepsilon^4} \right )$ for a success probability of $1 - O \left ( \tfrac {\log n} {n \varepsilon^2} \right )$. Thus, again, for $\varepsilon = \Theta (n^{-1/4})$ we get an approximation ratio of $1 - \gamma_f - \Theta (n^{-1/4})$ with probability $1 - O \left ( \tfrac {\log n} {\sqrt n} \right )$.  
\end{compactenum} \qed

\noindent \textbf{Proof of~\cref{thm:integrality_gap}.}
    We start with the complete network on $n$ nodes $K_n$, with $c_i = b_i = 1$ and $p_{ij} = 1$, and a point-mass shock of $X_i = \varepsilon$ for some $\varepsilon \in (0, 1)$. We also let $L = n \varepsilon / k \cdot \one$ and $v > \zero$. In the fractional optimum  we have that $\til z_i^* = k / n$ and the optimum value of $\| v \|_1 \cdot  n$ is achieved. When we choose to round to $k$ nodes, we do that uniformly (since $\til z_i^* =  k / n$) over all $k$-sets, that is the probability of each $k$-set being selected is $\tfrac 1 {\binom n k}$. Thus the value optimum is just the value of the optimum given that the values of any $k$-set are set to 1 (w.l.o.g we can choose the $k$-set $\{ 1, \dots, k \}$). The rounded nodes' bailouts would suffice to avert the shock, so they will be truncated to have a value of $\varepsilon$, from the problem constraints (debts must only be paid in at most their full value). Since all nodes are marginally default (that holds for the bailed-out nodes with the truncated values as well), we have from 
    ~\cref{lemma:bounding} (with equality since every node satisfies the constraint with tight equality) that $(1 / n) \sum_{j \in [n]} \bar p_j = n - n \varepsilon + k \varepsilon$ 
    which implies that $\opt_{f} = v^T \bar p \le \| v \|_\infty \left [ n (n - \varepsilon (n - k)) \right ]$. 
    We thus have $\sigma_f \ge \tfrac {\zeta^{-1}} {1 - \varepsilon (1 - k / n)}$.
     So for $k = o(n)$ (small) and $n$ large the gap tends to $\tfrac {\zeta^{-1}} {1 - \varepsilon}$ which can be arbitrarily bad for $\varepsilon \to 1$.  
\qed

\noindent \textbf{Proof of~\cref{theorem:as_inapproximability} (see \cref{fig:inapproximatbility}).} Let $a (| \mathcal I|)$ be a poly-time computable function on the input size $| \mathcal I |$, and $\alpha \in (0, 3)$. We construct the same proof as the hardness reduction of ~\cref{theorem:hardness} with the only change that instead of adding one copy of ``element nodes'' we add $a (| \mathcal I |)$ copies of ``element nodes'' resulting in a network with a polynomial number, i.e. $m + a(| \mathcal I | ) \cdot n$, of nodes. We connect, with direction from left to right,  the element nodes with liabilities $\tfrac {1 - \alpha / 3} n$ (i.e. a fully-connected network) and we set the external liabilities of the $a(| \mathcal I|)$-th level to be $1 - \alpha / 3$. We distinguish the following cases: If the answer to the 3-Set-Cover problem is YES then there are at least $k + a(| \mathcal I|) \cdot n$ solvent nodes. Else, if the answer to the 3-Set-Cover problem is NO then there are at most $k + n$ solvent nodes. That is, if we were able to distinguish between the~\eqref{eq:as} being between $k + n$ and $k + a(| \mathcal I|) \cdot n$ in polynomial time we will be able to solve the 3-Set-Cover in polynomial time, which is a contradiction assuming that P $\neq$ NP. The approximation gap is $\tfrac {k + n \cdot a(| \mathcal I |)} {k + n} \ge \tfrac {a(| \mathcal I |) \cdot n } {2n} = \tfrac {a(|\mathcal I|)} {2} = \Omega \left ( a (| \mathcal I |) \right )$. 

\qed

\noindent \textbf{Proof of~\cref{theorem:gini_approximation}.} 

\begin{compactitem}
\item \emph{GC Problem.} We omit the dependence on $g$ to lighten notation. We have that

\begin{align*}
    \evz {\sum_{j \in [n]} Z_j} & \ge \frac 1 {2ng} \evz { \sum_{i, j \in [n]} |Z_i - Z_j| } \explain {Gini Constraint} \\
    & = \frac 1 {2ng} \sum_{i, j \in [n]} \left ( \til z_i^* (1 - \til z_j^*) + \til z_j^* (1 - \til z_i^*) \right ) \explain{Definition of Expected Value} \\
    & \ge \frac 1 {ng} \left ( n \sum_{j \in [n]} \til z_j^*  - \sum_{i, j \in [n]} \til z_i^* \til z_j^* \right ) \explain{Reorder Summation} \\ 
    & \ge \frac 1 {ng} \left ( n \sum_{j \in [n]} \til z_j^*  - k \sum_{j \in [n]} \til z_j^* \right ) \explain{At most $k$ nodes are bailed out} \\
    & = \frac 1 g \left ( 1 - \frac k n \right ) \sum_{j \in [n]} \til z_j^*
\end{align*}

If $\tfrac 1 g \left ( 1 - \tfrac k n \right ) < 1$, or $g > \left ( 1 - \tfrac k n \right )$ then we follow the same analysis as in~\cref{theorem:apx_guarantees} and obtain an $\tfrac {(1 - \beta_{\max}) (1 - k  / n)} g$ approximation guarantee. We can use concentration inequalities to devise bounds about the runtime of the algorithm in a similar manner to~\cref{theorem:runtime}.







\end{compactitem}
\qed

\noindent \textbf{Proof of~\cref{thm:unbounded_pof_discrete}.} \noindent \emph{General Instance.} Suppose that we have the star network $S_n$ with node 1 being in the center and nodes 2 through $n - 1$ being the peripheral nodes. Let $b = \one$, $c = n \cdot \one_{\{ 1 \}}$, and $p_{1i} = 1$ for $i \neq 1$. The bailouts are $L = n \cdot \one$ and the total budget is $\Lambda = n$. 
The network is hit with a point-mass shock $X = c$. We will (w.l.o.g.) focus on the~\eqref{eq:sop} objective [It can be shown for any linear objective given by a vector of coefficients $v > \zero$ the same result holds]. 
The unbounded objective has value $2n - 1$ since bailing out the central node restores the network to its initial state.  Let $g = 0$. We have the following results: 

\begin{compactitem} 
    \item \emph{\eqref{eq:gc} Solution.} Subject to absolute equality (i.e. $g = 0$), it should hold that for the bailout indicator variables $\bar z_j$ that $\sum_{j \neq 1} | \bar z_1(g) - \bar z_j(g) | = 0$. Thus $\bar z_j(g) = 0$, therefore the objective value is $0$. That yields a PoF of $\tfrac {2n - 1} {0} = \infty$. 
    
    \item \emph{\eqref{eq:pgc} Solution.} Let $q = \one_{\{ 1 \}}$. Again, similarly to the previous case, we get $\bar z_j(g, q) = 0$ for all $j$. Thus the PoF is  $\infty$. 
    \item \emph{\eqref{eq:sgc} Solution.} We again get that $\sum_{(i, j) \in E} a_{ij} |\bar z_i(g, A) - \bar z_j(g, A)| = 0$ which yields $\bar z_j(g, A) = 0$ for all $j \in [n]$. Again, the PoF is $\infty$. 
    
\end{compactitem}
\qed

\noindent \textbf{Proof of~\cref{theorem:fractional_pof_bounded}.} We will prove the correctness of the Theorem for the~\eqref{eq:gc} metric since the other metrics have exactly the same proof. Fix some $g \ge 0$. The fractional PoF is equal to $f(\til p^*(1)) / f(\til p^*(g))$. Assume, for contradiction, that the PoF is unbounded. The following can happen:

\begin{compactenum}

    \item $f(\til p^*(g)) \neq 0$ and $f(\til p^*(1))$ is such that the limit of their ratio goes to infinity. This is a contradiction since $f$ can get values up to $\| f \|_\infty$. Thus, we arrive at a contradiction.
    
    \item $f(\til p^*(g)) = 0$. Since $f$ is strictly increasing with $f(\zero) = 0$ then $\til p^*(g) = \zero$. We will show that $\til p^*(g) = \zero$ if and only if $X = c$ w.p. 1 and $\til z^*(g) = 0$. We remind here that there are no isolated nodes (to the internal or the external sector) and hence $p > \zero$. The reverse direction is trivial since the fairness constraint is always satisfied (hence it does not affect the feasible region) and the fixed point operator is simply $\Phi(\til p^*(g)) = p \wedge (A^T \til p(g))$, so $\Phi(\zero) = p \wedge \zero = \zero$, i.e. $\til p^*(g) = \zero$. For the reverse direction, since $p > \zero$ the only way for $\til p^*(g)$ to be $\zero$ is (i) $g = \Lambda = 0$ which is impossible since by the definition of the problem $\Lambda > 0$, and (ii) the relative liability non-homogeneous part is zero, i.e. $(c - X) + L \odot\til z^*(g) = 0$ (the fairness constraint is trivially satisfied). Since $c - X \ge \zero$, $L > \zero$ and $\til z^* \ge 0$, the only way for the equation to hold is $X = c$ with probability 1 and $\til z^*(g) = \zero$, yielding a contradiction. Therefore, $f(\til p^*(g)) > 0$. 

\end{compactenum}
\qed

\noindent \textbf{Proof of \cref{theorem:fractional_pof_bounds}.} 

\begin{compactitem}

\item \emph{\eqref{eq:sgc} Solution.} Since $g > 0$ and $A$ is a connected network (i.e. no isolated nodes) a positive budget $\mu \in (0, \Lambda]$ is spent in total on bailouts. We let $$\B(\mu) = \{ (\til p'(g, A), \til z'(g, A))^T \text{ feasible $(g, A)$-unfair allocation} : L^T \til z' = \mu \}.$$

Subsequently, we define the following partition of $\B (\mu)$: Let $\B_1(\mu) = \B (\mu) \cap \{ \exists j \in [n]:  \til z_j(g, A) \neq \mu /  (n L_j) \}$, and $\B_2(\mu) = \B (\mu) \cap \{ \forall j \in [n] : \til z_j(g, A) = \mu /  (n L_j) \}$.
    We show that if $\B_2(\mu)$ is non-empty then $\B_1(\mu)$ must be non-empty. Let $\til z(g, A) \in \B_2 (\mu)$. Then since $g > 0$ for $0 < \delta_{ij} \le \min \{ \tfrac {\mu} {nL_i}, \tfrac {\mu} {nL_j}, \tfrac {g \beta_{\min} \mu} {\beta_i + \beta_j} \}$ we choose two arbitrary nodes $i$ and $j$ such that the former node's bailout is increased by $\delta_{ij}$ and the latter node's bailout is decreased by $\delta_{ij}$, that is absolute equality is achieved via the optimal allocation and we are allowed to increase up to $g$. The total budget constraint is not violated and the new bailout scheme belongs to $\B_1(\mu)$.  If the optimal bailout scheme belongs to $\B_2(\mu)$ perturb the bailouts of two nodes as we described above to get a sub-optimal assignment $(\til p'(g, A), \til z'(g, A))^T \in \B_1(\mu)$ since the optimal assignment is also $(0, A)$-unfair. The altered assignment is a lower bound on the original assignment per the objective. Below we bound the PoF for the~\eqref{eq:linobj} objective. We have: 

    \begin{align*}
        \frac 1 {\mathrm {PoF}_{(g, A)}} &= \frac {\sum_{j \in [n]} v_j \til p_j^*(g, A)} {\sum_{j \in [n]} v_j \til p_j^*(1, A)} \\
        & \ge \frac 1 \zeta \cdot \frac {\sum_{j \in [n]} \til p_j'(g, A)} { \sum_{j \in [n]} \til p_j^*(1, A)} \explain{Constrain to $\B_1(\mu)$} \\
        & = \frac 1 \zeta \cdot \frac {\sum_{j \in D(g, A)} \til p_j'(g, A) + \sum_{j \in R(g, A)} \til p_j'(g, A)} {\sum_{j \in D(1, A)} \til p_j^*(1, A) + \sum_{j \in R(1, A)} \til p_j^*(1, A)} \explain{Partition of $[n]$} \\
        & \ge \frac 1 \zeta \cdot \frac {\sum_{j \in D(g, A)} \til p_j'(g, A)} {\sum_{j \in D(1, A)} \til p_j^*(1, A) + \sum_{j \in R(1, A)} \til p_j^*(1, A)} \explain{Drop Summands} \\
        & \ge \frac {1 - \beta_{\max}} {\zeta} \cdot \frac {\sum_{j \in D(g, A)} (c_j - x_j + L_j \til z_j'(g, A))} {\sum_{j \in [n]} (c_j - x_j + L_j \til z_j^*(1 , A))} \explain{\cref{lemma:lower_bound,lemma:bounding}} \\
        & \ge \frac{1 - \beta_{\max}} {\zeta} \cdot \frac {\sum_{j \in D(g, A)} (c_j - x_j + \beta_j L_j \til z_j'(g, A))} {\sum_{j \in [n]} (c_j - x_j) + \Lambda} \explain{Budget Constraint, $\beta_j < 1$} \\
        & \ge \frac {1 - \beta_{\max}} {\zeta} \cdot \frac {\sum_{j \in D(g, A)} (c_j - x_j) + \frac 1 {2g} \sum_{(i, j) \in E} a_{ij} |L_i  \til z_i'(g, A) - L_j \til z_j'(g, A)|} {\sum_{j \in [n]} (c_j - x_j) + \Lambda} \explain{Gini Constraint} \\
        & \ge \frac {1 - \beta_{\max}} {2g \zeta} \cdot \frac {\sum_{(i, j) \in E} a_{ij} |L_i  \til z_i'(g, A) - L_j \til z_j'(g, A)|} {\| c \|_1 + \Lambda} \explain{$\zero \le c - x \le c$} \\
        & \ge \frac {1 - \beta_{\max}} {2g \zeta (\| c \|_1 + \Lambda)} \sqrt {\sum_{(i, j) \in E} a_{ij}^2 (L_i  \til z_i'(g, A) - L_j \til z_j'(g, A))^2} \explain{$\| x \|_1 \ge \| x \|_2$} \\
        & \ge \frac {1 - \beta_{\max}} {2g \zeta (\| c \|_1 + \Lambda)} \sqrt {\til \theta'(g, A)^T \L(A \odot A) \til \theta'(g, A)} \explain{$\til \theta'(g, A) = L \odot \til z'(g, A)$}. \\
    \end{align*}        
        
    We do a change of variables $\til y'(g, A) = \til \theta'(g, A) - \tfrac \mu n \one \neq \zero$ since $\theta'(g, A) \in \B_1(\mu)$. Since $\one$ is the eigenvector of the 0 eigenvalue we can center the aforementioned quadratic form to be equal to 
    
    $$\til y'(g, A)^T \L (A^{(2)}) \til y'(g, A) = \frac {\til y'(g, A)^T \L (A^{(2)}) \til y'(g, A)} {\| \til y'(g, A) \|_2^2} \cdot \| \til y'(g, A) \|_2^2.$$
    
    Moreover, 
    
    $$ \frac {\til y'(g, A)^T \L (A^{(2)}) \til y'(g, A)} {\| y'(g, A) \|_2^2} \cdot \| \til y'(g, A) \|_2^2 \ge \left ( \min_{\one^T y(g, A) = 0}  \frac { y(g, A)^T \L (A^{(2)}) y(g, A)} {\| y(g, A) \|_2^2} \right ) \cdot \| \til y'(g, A) \|_2^2.$$
    
    Using the Courant-Fischer Theorem~\cite{spielman2012spectral} the above expression evaluates to $\lambda_2(\L (A^{(2)})) \cdot \| \til y'(g, A) \|_2^2$, where 
    
    $$\| \til y'(g, A) \|_2^2 = \sum_{j \in [n]} \left ( L_j \til z_j' \right )^2 - \frac {\mu^2} n > 0.$$
    
    For $\lambda_2( \L (A^{(2)}) )$ we use the lower bound provided by Cheeger's inequality, i.e. 
    
    \begin{equation*}
        \lambda_2 ( \L (A^{(2)}) ) \ge \frac {\phi^2(A^{(2)})} {2 \max_{j \in [n]} \sum_{i \in [n]} a_{ji}^2} \ge \frac {\phi^2(A^{(2)})} {2 \max_{j \in [n]} \left ( \sum_{i \in [n]} a_{ji} \right )^2} \ge \frac {\phi^2(A^{(2)})} {2 \beta_{\max}^2},
    \end{equation*}
    
    where $\phi(A^{(2)}) = \min_{\emptyset \subset S \subset [n]} \frac {\sum_{j \in S, i \notin S} a_{ji}^2} {\min \left \{ \sum_{j \in S} \sum_{i : i \sim j} a_{ji}^2 , \sum_{j \in [n] \setminus S} \sum_{i : i \sim j} a_{ji}^2  \right \}}$ is the conductance of $A^{(2)}$. Thus we can bound the PoF by 
    
    \begin{equation*}
        \mathrm{PoF}_{(g, A)} \le \frac {\sqrt 2 g \zeta (\| c \|_1 + \Lambda) \beta_{\max}} {(1 - \beta_{\max}) \phi(A^{(2)}) \cdot \sqrt {\sum_{j \in [n]} (L_j \til z_j')^2 - \mu^2 / n}}.
    \end{equation*}
        
    Note that the denominator is always non-zero since $\beta_{\max} < 1$, $\phi(A^{(2)}) > 0$ and $\sum_{j \in [n]} (L_j \til z_j')^2 - \mu^2 / n > 0$ due to $\til z'(g, A) \in \B_1(\mu)$.
    
    \item \emph{\eqref{eq:gc} Solution.} The construction of the sets $\B(\mu), \B_1(\mu), \B_2(\mu)$ is exactly the same as in the previous case, as well as the derivation of the PoF bound. The only change that in place of $A^{(2)}$ we put the complete graph $K_n$ on $[n]$. Since $\lambda_2 (\L (K_n)) = n$ we do not need to employ Cheeger's inequality, and thus arrive at a bound of the form
        
    \begin{equation*}
        \mathrm{PoF}_{g} \le \frac {2 g \zeta (\| c \|_1 + \Lambda) \sqrt n} {(1 - \beta_{\max}) \sqrt {\sum_{j \in [n]} (L_j \til z_j')^2 - \mu^2 / n}}.
    \end{equation*}
   
    Note that there was a factor of $2n$ on the Gini Coefficient definition so the bound gets a $2 \sqrt n$ factor on the numerator. Again for a finite instance we have that $\mathrm{PoF}_g < \infty$. 
    
    
    
    
\end{compactitem}
\qed

\noindent \textbf{Proof of \cref{theorem:conductance}.} We have that 

\begin{align*}
    \mathrm{SGC}(\til z; A) = \frac {\sum_{i, j} a_{ij} |L_i \til z_i(g, A) - L_j \til z_j(g, A)|} {2 \sum_{j \in [n]} \beta_j L_j \til z_j(g, A)} = \frac {1} {2} \psi(L \odot z; A).
\end{align*}

where $\psi(x; A) = \frac {\sum_{i, j} a_{ij} | x_i - x_j |} {\sum_{j \in [n]} \beta_j |x_j|} $. In \cref{sec:conductance}, we prove that 

\begin{equation*}
    \phi(A) = \min_{x \neq \zero, \; \mathrm{median} (x) = 0} \psi(x; A).
\end{equation*}

Thus $\tfrac {\psi(L \odot z; A)} {2} \ge \tfrac {\phi(A)} {2}$. Therefore $\mathrm{SGC}(\til z; A) \ge \tfrac {\phi(A)} {2}$.

\section{Transformations} \label{sec:transform}

\subsection{Transforming Weakly-Increasing Objectives to allow clearing vectors} 

In the case that the~\eqref{eq:generalized_bailouts} problem consists of a function $f(\bar p)$ which is increasing (i.e. not necessarily strictly increasing), there is no guarantee that the solution that will be produced by the optimization procedure will correspond to a clearing vector (still if the spectral radius of $A$ is less than 1 the fixed point operator yields a unique solution). For that reason we consider, similarly to~\cite{ahn2019optimal} the transformed objective, parametrized by $\epsilon > 0$, 

\begin{equation*}
    \hat f(\bar p) = f(\bar p) + \frac {\epsilon (1 - \beta_{\max})} {2 \Lambda} \one^T \bar p.
\end{equation*}

It is obvious that for any $\bar p > 0$ we have that $\hat f(\bar p) > f(\bar p)$ since the added term contributes positively. Moreover $\hat f$ is strictly increasing and therefore an optimal solution $\hat p$ to $\hat f$ subject to the~\eqref{eq:generalized_bailouts} constraints is a clearing payment vector. Below we will show that if $\bar p^*$ is an optimal solution to $f$, and $\hat p^*$ is an optimal solution to $\hat f$ then $f(\hat p^*) - f(\bar p^*) \le \epsilon$. First, from optimality we have that 

\begin{equation*}
    \hat f(\hat p^*) - \hat f(\bar p^*) \ge 0.
\end{equation*}

Equivalently, 

\begin{equation*}
    f(\hat p^*) + \frac {\epsilon (1 - \beta_{\max})} {2 \Lambda} \one^T \hat p^* - f(\bar p^*) - \frac {\epsilon (1 - \beta_{\max})} {2 \Lambda} \one^T \bar p^* \ge 0.
\end{equation*}

Equivalently,

\begin{equation*}
    f(\hat p^*) - f(\bar p^*) \le \frac {\epsilon (1 - \beta_{\max})} {2 \Lambda} \| \hat p^* - \bar p^* \|_1 \le \frac {\epsilon (1 - \beta_{\max})} {2 \Lambda (1 - \beta_{\max})} \| L^T \hat z^* - L^T \bar z^* \|_1 \le \epsilon. 
\end{equation*}

Where the inequalities follow from rearrangements, the triangle inequality, \cref{lemma:gain_upper_bound_greedy}, and that the maximum distance between any two assignments can be at most $2 \Lambda$. Thus the clearing payment vector from solving the transformed objective yields an objective value which is $\epsilon$-close to the desired optimal value.

\subsection{A Property of Conductance} \label{sec:conductance}

To conclude the proof of \cref{theorem:conductance} we give a relation between the conductance and Gini-related Measures. For this subsection only, and for simplicity, we will assume a \emph{new notation} from the rest of the paper referring to an unweighted and undirected topology $G(V = [n], E)$ with $|V| = n$. The result we prove below is also of independent interest from the rest of the paper. The conductance of $G$, $\phi(G)$, is given as 

\begin{equation*}
    \phi(G) = \min_{\emptyset \subset S \subset V} \frac {e(S, \bar S)} {\min \{ |S|, | \bar S | \}} =  \min_{\emptyset \subset S \subset V, |S| \le n / 2} \frac {e(S, \bar S)} {|S|},
\end{equation*}

where $e(S, \bar S)$ is the number of edges going from $S$ to its complement  $\bar S = [n] \setminus S$. More specifically, we will show that 

\begin{align*} \tag{COND} \label{eq:cond}
    \phi(G) &  = \min_{x_1, \dots, x_n} & \frac {\sum_{(i, j) \in E} |x_i - x_j|} {\sum_{i \in [n]} |x_i|} \\
     & \text{subject to} & \mathrm{median}(x_1, \dots, x_n) = 0 \\
     & & \text{not all $x_i = 0$}.
\end{align*}

For that reason, we define the function 

\begin{equation*}
    \psi(x) = \frac {\sum_{(i, j) \in E} |x_i - x_j| } {\sum_{i \in [n]} |x_i|}.
\end{equation*}

Firstly, we will show that $\min_{x \in \eqref{eq:cond}} \psi(x) \le \phi(G)$. Let $S^*$ be the set that achieves $\phi(G)$. We let

\begin{equation*}
    x_i^* = \begin{cases}
    \frac {1} {\phi(G)} & i \in S^* \\
    0 & i \notin S^*
    \end{cases}.
\end{equation*}

Note that since $|S^*| \le n / 2$ the median of $x_i^*$ is 0. Moreover, not all $x_i^*$ are 0. We have that 

\begin{equation*}
    \psi(x^*) = \frac {\sum_{(i, j) \in E} | x_i^* - x_j^* | } {\sum_{i \in [n]} |x_i^*|} = \frac {\sum_{(i, j) \in E, i \in S^*, j \in \bar S^*} \frac {1} {\phi(G)}} {|S^*| \frac {1} {\phi(G)}} = \frac {e(S^*, \bar S^*)} {|S^*|} = \phi(G).
\end{equation*}

Thus we showed that $\min_{x \in \eqref{eq:cond}} \psi(x) \le \phi(G)$. We will now show that $\min_{x \in \eqref{eq:cond}} \psi(x) \ge \phi(G)$. In this case, we seek to find $x_1, \dots x_n$ such that $\psi(x)$ is minimized and the $x_i$'s have median 0. We define $y_i = \max \{ x_i, 0 \}$ and $z_i = \min \{ x_i, 0 \}$. It is elementary to show that $|x_i| = |y_i| + |z_i|$ and $|x_i - x_j| = |y_i - y_j| + |z_i - z_j|$. So 

\begin{equation*}
    \psi(x) = \frac {\sum_{(i, j) \in E} \left [  |y_i - y_j| + |z_i - z_j| \right ]} {\sum_{i \in [n]} \left [|y_i| + |z_i| \right ]}.
\end{equation*}

To show our claim, it suffices to show that $\psi(y) \ge \phi(G)$ and $\psi(z) \ge \phi(G)$. This is because for every $a, b, c, d, \alpha \ge 0$ with $\tfrac {a} {c} \ge \alpha$ and $\tfrac {b} {d} \ge \alpha$ then $\tfrac {a + b} {c + d} \ge \alpha$. By symmetry, we will constrain ourselves to proving $\psi(y) \ge \phi(G)$, whereas for $z$ the proof is identical. We let $y_1 = \dots = y_{n / 2} = 0$ and $y_i \ge 0$ for all $i \ge n / 2$. This guarantees that the median of $y_i$ is 0 and that $y_i \ge 0$ for all $i \in [n]$. For the numerator we have

\begin{figure}[h]
    \centering
    \begin{tikzpicture}
        \Vertex[size=0.8, x=-4, y=0]{1}
        \Text[x=-4, y=0, distance=0.5, position=below]{${y_1 = \dots = y_{n / 2} = 0}$}
        \Vertex[size=0.8, x=-2, y=0, label=${y_i}$]{2}
        \Text[x=-3,y=0]{$\dots$}
        \Vertex[size=0.8, x=-1, y=0, label=${y_{i + 1}}$]{3}
        \Vertex[size=0.8, x=-0, y=0, label=${y_{i + 2}}$]{4}
        \Text[x=1, y=0]{$\dots$}
        \Vertex[size=0.8, x=2, y=0, label=${y_j}$]{5}
        \Edge[color=red,bend=40](2)(3)
        \Edge[color=red,bend=40](2)(4)
        \Edge[color=red,bend=40](2)(5)
        \Edge[color=red,bend=40](2)(4)
        \Edge[color=red,bend=40](3)(4)
        \Edge[color=red,bend=40](3)(5)
        \Edge[color=red,bend=40](4)(5)
        \Vertex[Pseudo, x=0.75, y=1.5]{6}
        \Vertex[Pseudo, x=0.75, y=-1.5]{7}
        \Edge[color=gray,bend=30, style=dashed](6)(7)
        \Text[x=0.75,y=-1.5, position=right, distance=0.5,color=gray]{$C_\ell, \; \ell - 1 \ge n / 2$}

    \end{tikzpicture}

    \label{fig:conductance}
\end{figure}

\begin{align*}
    \sum_{(i, j) \in E, i < j} (y_j - y_i) & = \sum_{(i, j) \in E, i < j} \sum_{\ell = i + 1}^{j} \left ( y_{\ell} - y_{\ell - 1} \right ) \\ & =  \sum_{\ell} \left ( {\text {\# of copies of the $(\ell - 1, \ell)$ segment}} \right ) \cdot \left ( y_\ell - y_{\ell - 1} \right ) \\ & = \sum_{\ell} \left ( \text{\# of edges that cross over $(\ell-1, \ell)$} \right ) \cdot (y_{\ell} - y_{\ell - 1}) \\
    & = \sum_{\ell} \underbrace{\left ( \text {\# edges in the cut $C_\ell$ partitioning $[\ell - 1]$ from $[n] \setminus [\ell - 1]$ }\right )}_{| C_\ell |} \cdot ( y_{\ell} - y_{\ell - 1} )
    .
\end{align*}

A key observation is that $\tfrac {|C_\ell|} {n - \ell + 1} \ge \phi(G)$. Therefore the numerator satisfies

\begin{equation*}
    \sum_{\ell} |C_\ell| \cdot (y_{\ell} - y_{\ell - 1}) \ge \sum_{\ell} \phi(G) (n - \ell + 1) (y_{\ell} - y_{\ell - 1}) \ge \phi(G) \cdot \sum_{\ell} y_\ell = \phi(G) \cdot \sum_{\ell} |y_\ell|.
\end{equation*}

Thus $\psi(y) \ge \phi(G)$. We conclude that $\phi(G) = \min_{x \in \eqref{eq:cond}} \psi(x)$. \cref{theorem:conductance} is the weighted version (with weights $w_{ij} \ge 0$) of this identity where the proof is exactly the same, where we use the function $\psi(x; W) = \frac {\sum_{(i, j) \in E} w_{ij} |x_i - x_j|} {\sum_{i \in [n]} d_i |x_i|}$ and $d_i = \sum_j w_{ij}$. Finally, note that if we constrain $\psi$ in the domain in which the denominator is at most $\sigma'$, for some $\sigma' > 0$ then we modify an allocation $x$ with $\sum_{i \in [n]} |x_i| = \sigma > 0$ to an allocation $x' = \tfrac {\sigma'} {\sigma} x$ and the proof will remain valid since $\psi$ is scale-invariant, that is $\psi(x') = \psi \left ( \tfrac {\sigma'} {\sigma} x \right ) = \psi(x)$.






\section{Datasets Addendum} \label{sec:data_analysis}

\begin{figure}[t]
    \centering

    \subfigure[Internal liabilities distribution and Pareto fit]{\includegraphics[width=0.45\textwidth]{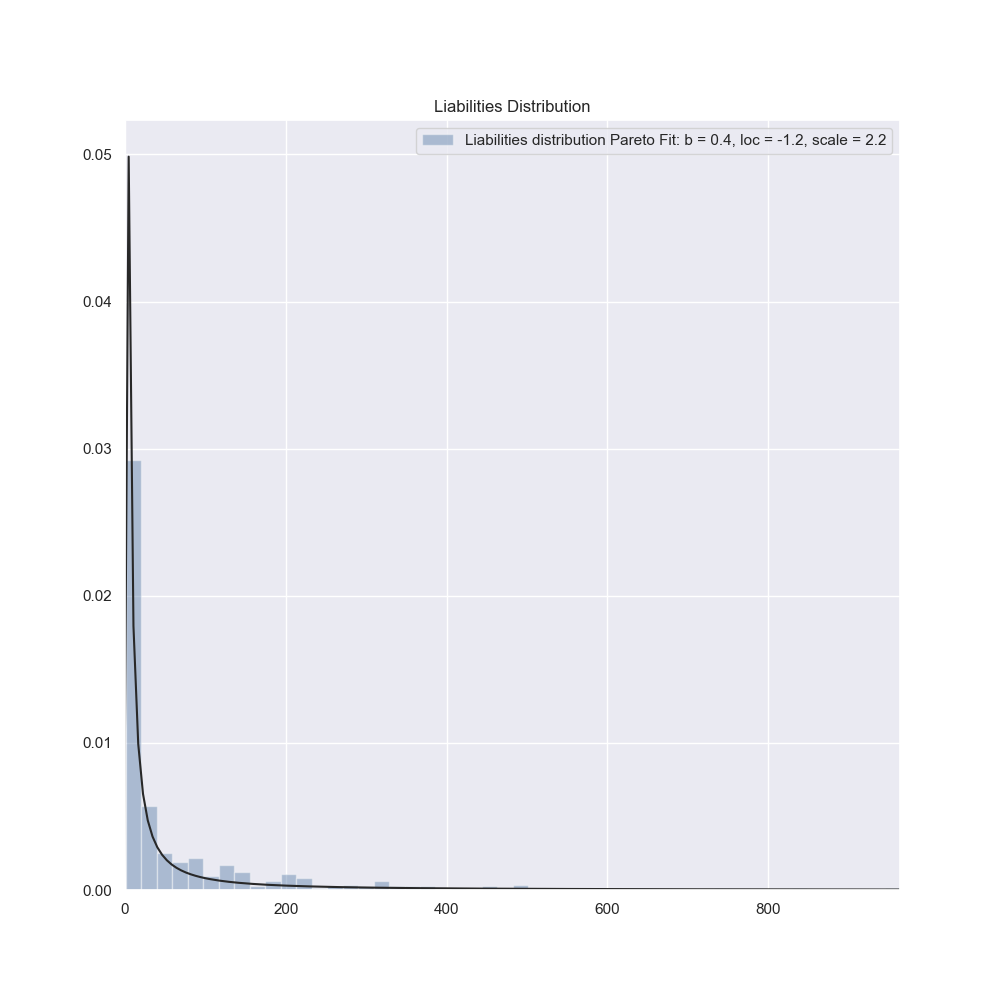}}
    \subfigure[External assets and external liabilities regression]{\includegraphics[width=0.45\textwidth]{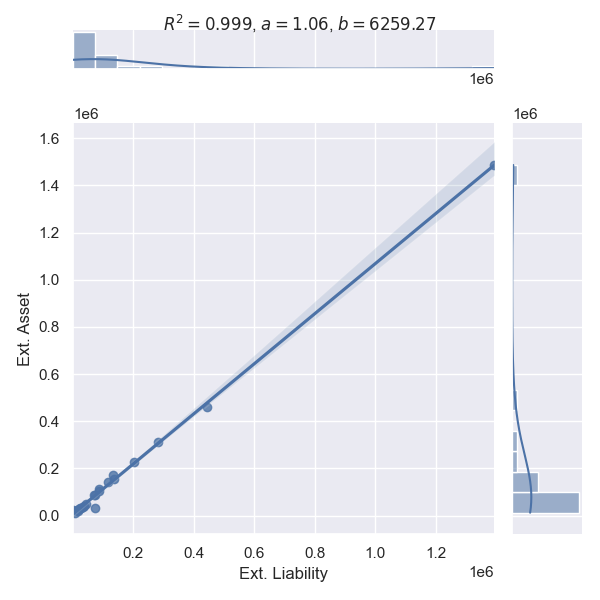}} 
    \subfigure[Financial network plot with wealths on log-scale]{\includegraphics[width=0.45\textwidth]{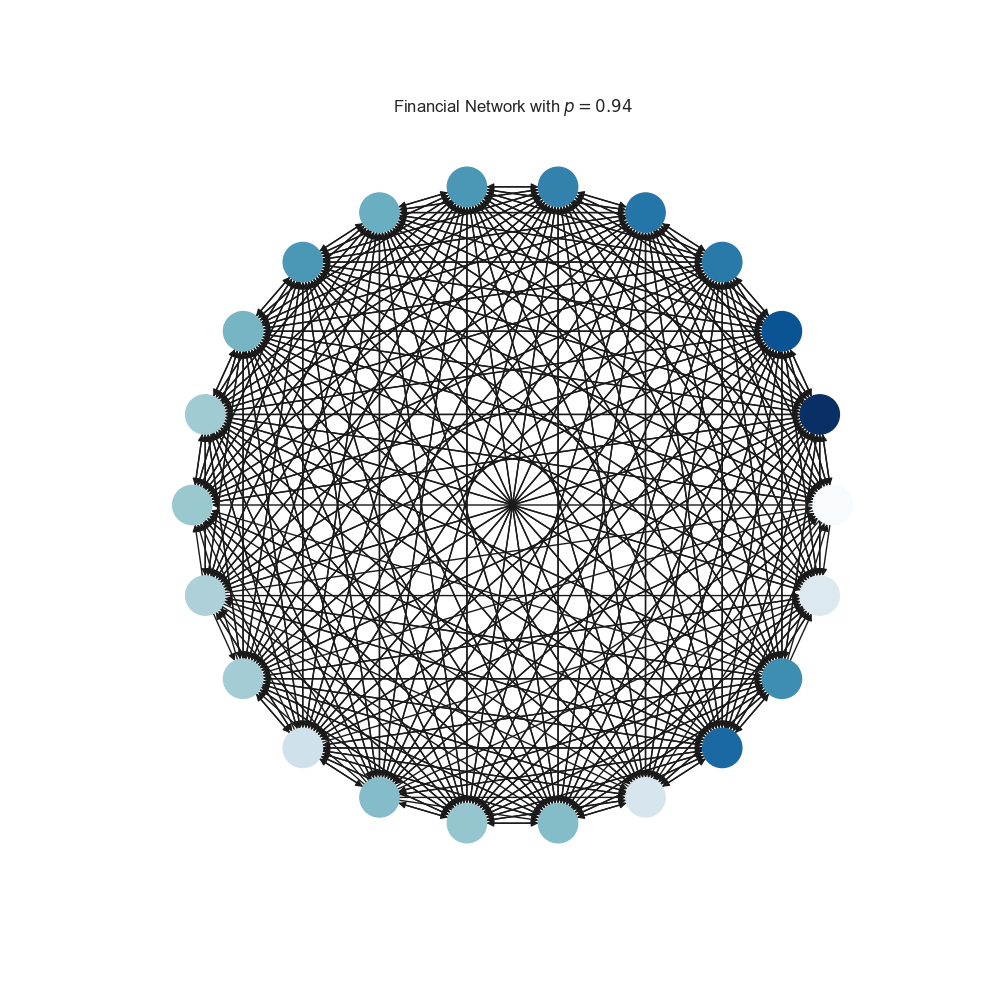}} 
    \subfigure[External assets and liabilities distributions and Pareto Fits]{\includegraphics[width=0.45\textwidth]{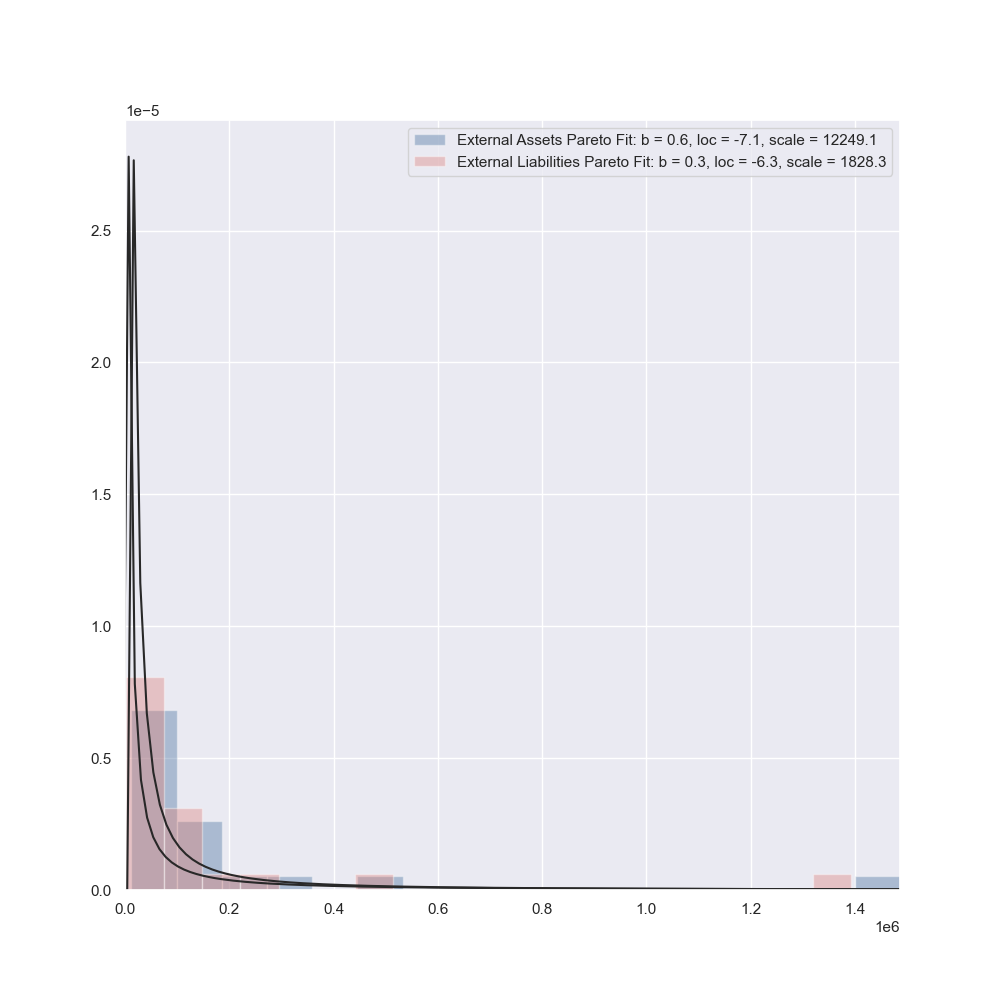}} 
    
    \caption{German Banks dataset statistics}
    \label{fig:german_banks_statistics}
\end{figure}

\noindent \textbf{German Banks Dataset~\cite{chen2016financial}.} The dataset contains $n = 22$ German Banks, connected with a liability matrix with $m = 435$ edges, with a mean outdegree $\bar d_{\mathrm{out}} = 19.8$. The network structure can be described by the ER network $G(n = 22, p = 0.94)$. Below we provide Pareto fits for the internal liabiliti-es distributions, a regression plot for external assets and external liabilities, a plot of the financial network with equities (wealths) displayed with colors on a log-scale, and a distribution plot between external assets and external liabilities with Pareto fits. Observe that the relation between external assets and external liabilities is a linear function with $R^2 = 0.999$. The visualization of the dataset appears in ~\cref{fig:german_banks_statistics}.

\begin{figure}[t]
    \centering
    \subfigure[Total Internal Assets distribution with Exponential fit]{\includegraphics[width=0.45\textwidth]{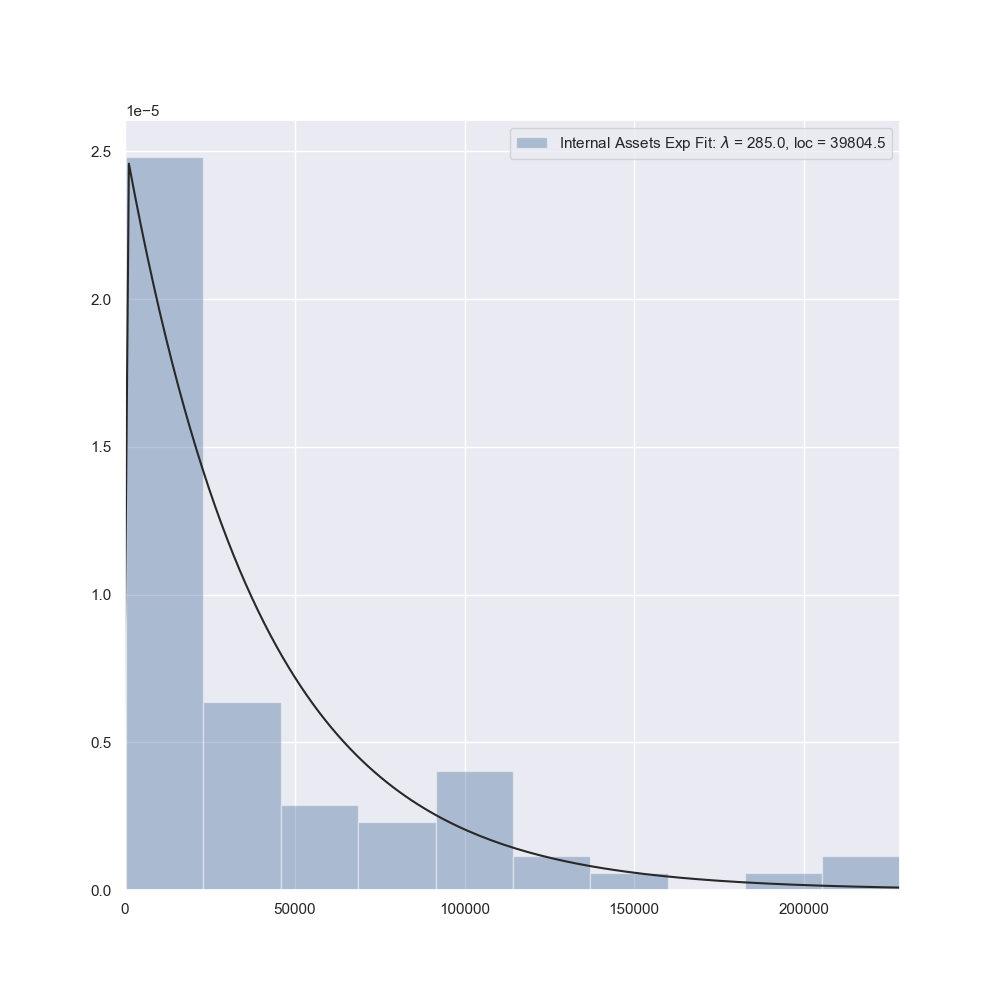}}
    \subfigure[External assets and external liabilities regression]{\includegraphics[width=0.45\textwidth]{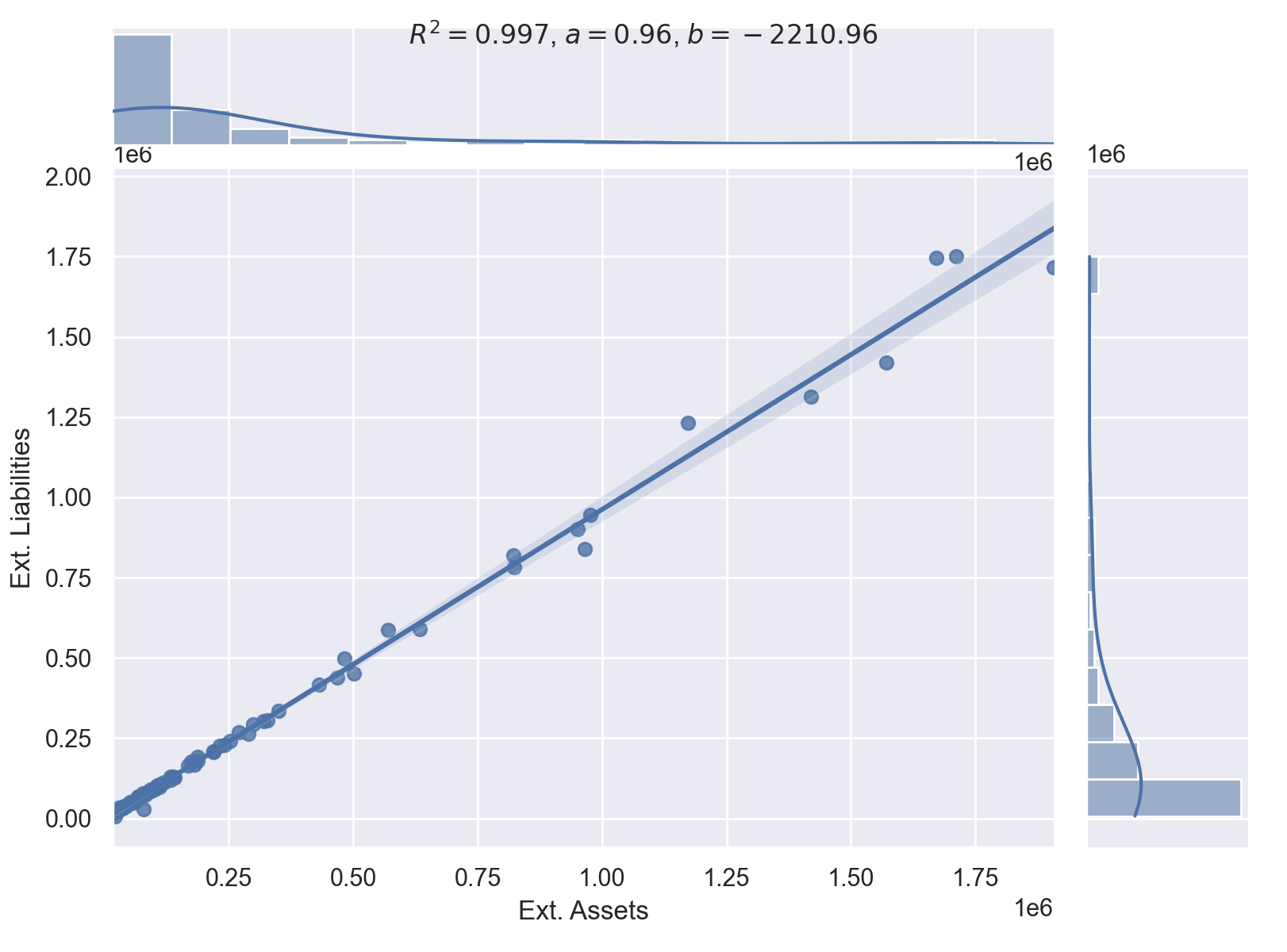}}
    \subfigure[External assets and liabilities distributions with Exponential fit]{\includegraphics[width=0.45\textwidth]{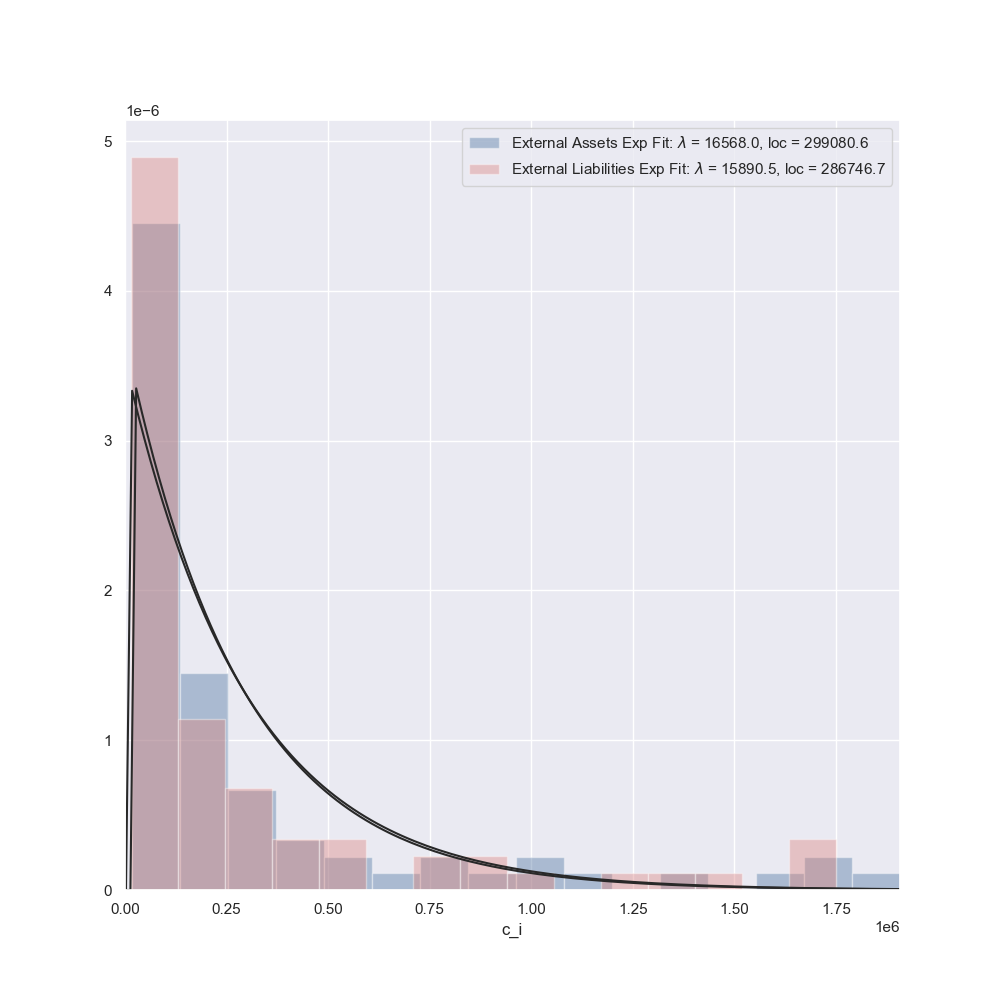}}
    \subfigure[Equity distribution and Exponential fit]{\includegraphics[width=0.45\textwidth]{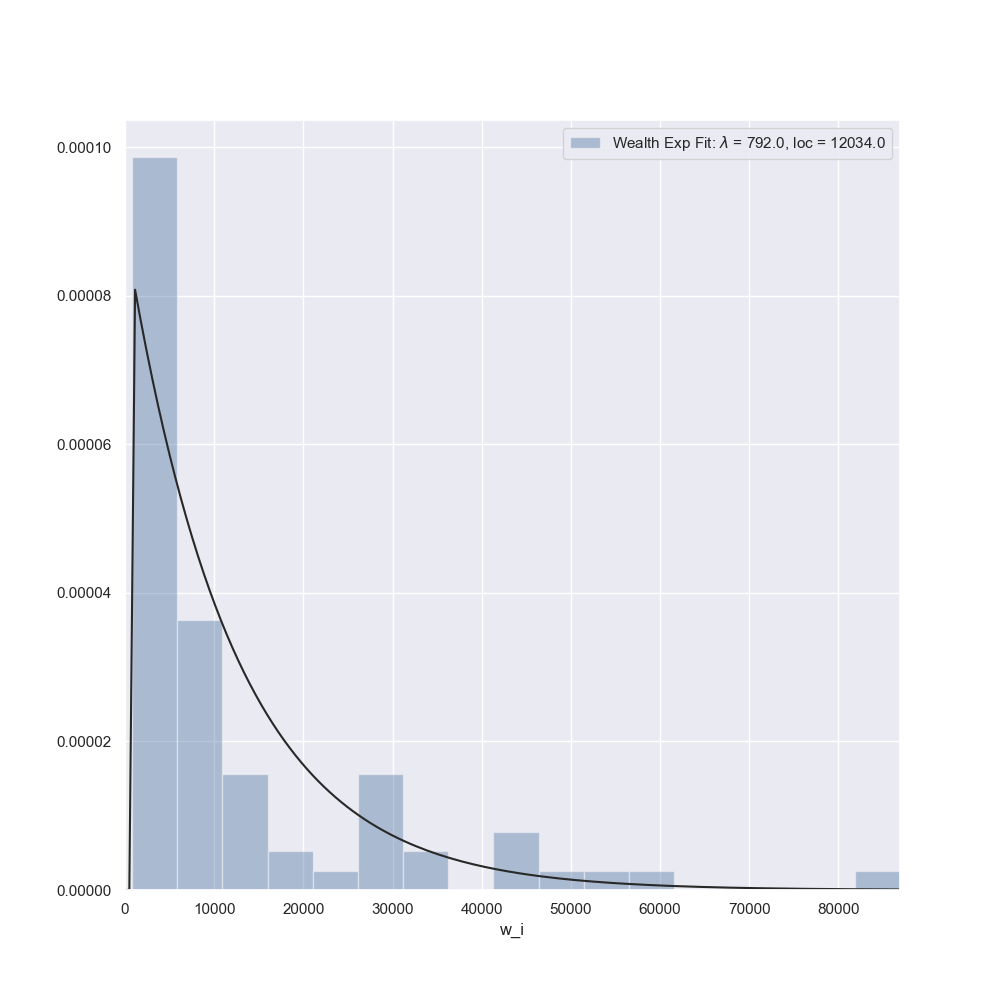}}
    \caption{EBA dataset statistics}
    \label{fig:eba_statistics}
\end{figure}

\noindent \textbf{EBA Dataset~\cite{glasserman2015likely}.} The EBA dataset contains information about the external assets and liabilities of each bank as well as the total internal assets of each bank, which by \cite{glasserman2015likely} are assumed to be equal to the total internal liabilities of each bank. In \cref{sec:experiments}, we have perturbed the total internal assets by adding some noise, and rebalancing them so that their total sum is invariant. The assets and liabilities can be approximated by exponential distributions. We report the distributions and the corresponding fits in ~\cref{fig:eba_statistics}. The external assets and liabilities are linearly correlated with $R = 0.997$.

\begin{figure}[t]
    \centering
    \subfigure[Indegree distribution]{\includegraphics[width=0.45\textwidth]{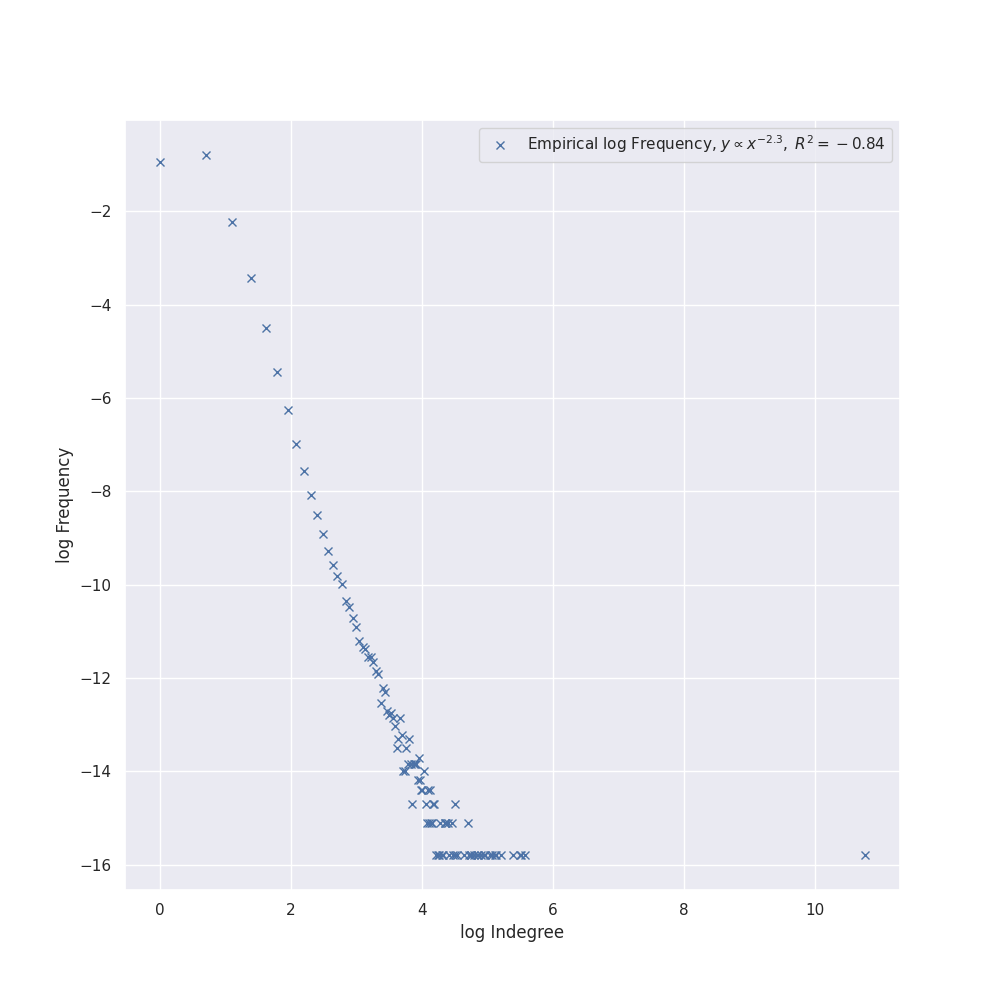}}
    \subfigure[Outdegree distribution]{\includegraphics[width=0.45\textwidth]{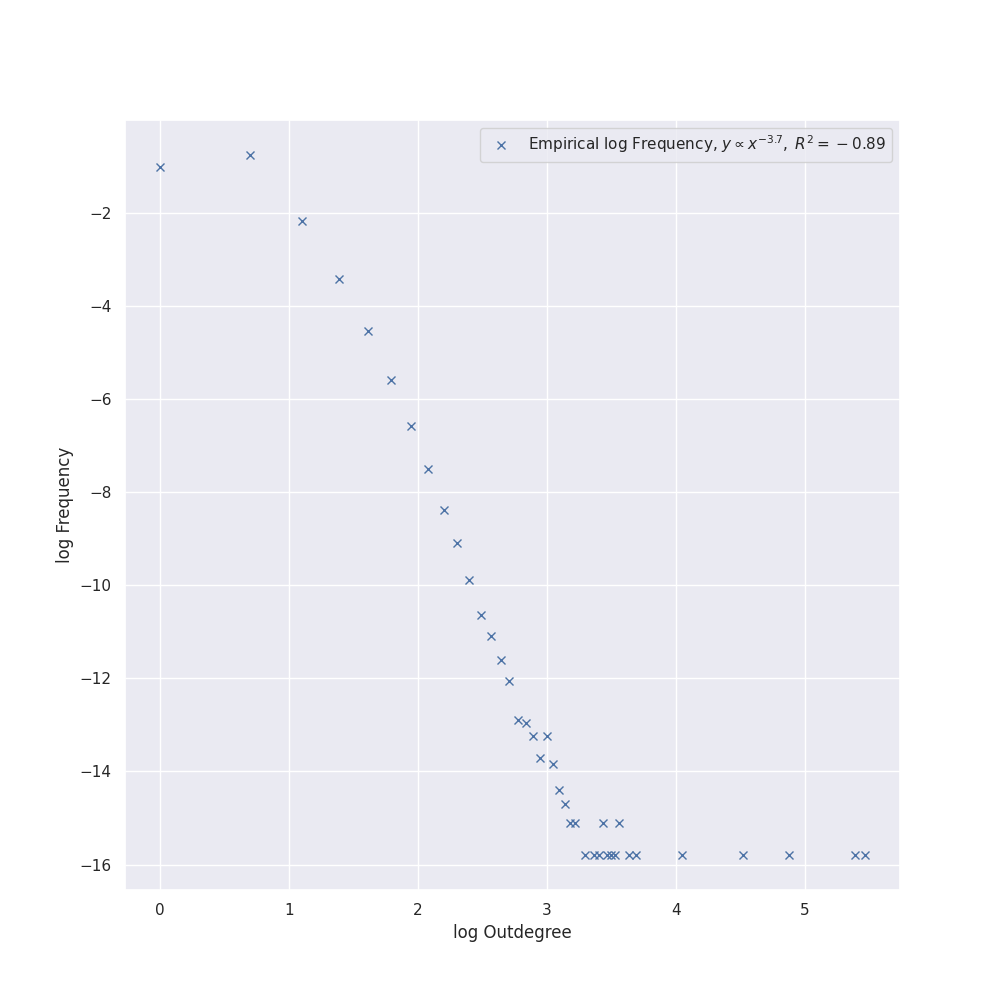}}

    \caption{Venmo dataset degree distributions with power-law fits.}
    \label{fig:venmo_degree_distributions}
\end{figure}

\noindent \textbf{Venmo Dataset.} The venmo dataset consists of $n = 7,178,381$ nodes and $m = 7,024,837$ edges (transactions) with an average (total) degree per user of 1.9572. The dataset consists of directed transactions from a sender node to a receiver node. There are no payment amounts data (for privacy reasons). ~\cref{fig:venmo_degree_distributions} displays the degree distributions of the dataset and the corresponding power-law fits. The indegree distribution law is $y \propto x^{-2.3}$ and the outdegree distribution law is $y \propto x^{-3.7}$ with corresponding Pearson coefficients $-0.86$ and $-0.89$ respectively. We have computed all the weakly-connected components of the Venmo transaction network and kept 19 weakly-connected components which contained more than $100$ nodes. We generated fake transactions using Pareto laws.

\noindent \textbf{SafeGraph Data.} We give the detailed steps we used and assumptions we relied on to build the SafeGraph data. We first describe the two different kinds of nodes, that form a bipartite graph where nodes of each kind lie in their corresponding set of the bipartition: 

\begin{compactitem}
    \item \emph{POI nodes.} We construct the POI nodes by setting a location on Earth and looking at the $k_{\mathrm{kNN}}$-nearest CBGs that contain POIs according to the Haversine distance. We choose the location \texttt{(42.43969750363193, -76.49506530228598)} and use $k_{\mathrm{kNN}} = 3$.   
    \item \emph{CBG nodes.} We use the Monthly Patterns data for the period between April and May 2020 to determine the CBGs that interact with the POIs in question. For each POI, we use the column \texttt{visitor\_home\_cbg} to list the CBGs and the total number of visitors from each CBG to each POI. These visitors correspond to the number of unique devices that interact with the POI. Moreover, we assume that each distinct device represents a different person. For each CBG node we use data from the US Census we also log demographic characteristics such as: \emph{(i)} race, \emph{(ii)} average income, \emph{(iii)} size of households, \emph{(iv)} unemployment rates. Using these data for each CBG we calculate: \emph{(i)} the probability of belonging to a racial minority group (i.e. non-White), \emph{(ii)} the average income, \emph{(iii)} the average size of a household on this CBG, \emph{(iv)} the probability of someone in the CBG being unemployed. 
    
\end{compactitem}

The above network consists of 152 nodes (for our choice of parameters), but can be different if e.g. the starting coordinates or $k_{\mathrm{kNN}}$ are changed. We create the edges and the internal liabilities $p_{ij}$ as follows: 

\begin{compactitem}
    \item We use the bucketed \emph{dwell times} to determine the percentage of people that are working in a business. More specifically, we consider a worker a device that has spent more than 240 min in the POI and we classify the person as a non-worker otherwise. Workers come from CBGs and are paid wages whereas non-workers come from CBGs and have expenses on the corresponding POI. For every CBG $j$ let $p^{\mathrm{worker}}_j$ be this percentage.  
    \item For every CBG $i$ that interacts with a POI $j$ with $n_{ij}$ people we add an edge $(j, i)$ referring to $n_{ij}^{\mathrm{workers}} = \lfloor n_{ij} \times p^{\mathrm{worker}}_j \rfloor$ workers, and an edge $(i, j)$ referring to $n_{ij}^{\mathrm{non-workers}} = n_{ij} - \lfloor n_{ij} \times p^{\mathrm{workers}}_j \rfloor $ non-worker nodes. We determine the weights of these edges (per unit) as follows: for the $(j, i)$ edge we take wage data by the US Economic Census and NAICS and create a liability by the business equal to the monthly wage of an employee which we multiply by $n_{ij}^{\mathrm{workers}}$. For the $(i, j)$ edge we use data from the Consumer Expenditure Survey conducted by the US Economic Census and add a liability regarding $n_{ij}^{\mathrm{non-workers}}$ equal to the average monthly expenditures due to the specific NAICS code of the business.  
\end{compactitem}

The number of workers for each POI $j$ is $n_j^{\mathrm{workers}} = \sum_{i \text{ is CBG}} n_{ij}^{\mathrm{workers}}$ (and the number of non-workers is defined respectively). Similarly we count the number of workers and number of non workers for each CBG $i$. We estimate the total number of households in the CBG that are related to interaction with the corresponding POIs\footnote{Of course the total number of people and households for each CBG is reported by the US Census the dataset would not be calibrated if we did not limit the transactions of each CBG with the corresponding POIs.}. We estimate the number of households as follows 

\begin{equation*}
    n_i^{\mathrm{households}} = \left \lceil \frac {\frac {n_i^{\mathrm{workers}}}  {p_i^{\mathrm{employed}}} + n_i^{\mathrm{non-workers}}} {\text{average size of household at CBG $i$}} \right \rceil .
\end{equation*}

The bailouts are determined as follows 

\begin{compactitem}
    \item For each CBG $i$ we use the average income of the CBG and the average size of the household to calculate the value of a bailout as it would be imposed by the CARES act (see \cite{act2020cares}). The bailout is multiplied by the estimated number of households calculated previously. 
    \item For each POI $j$ we have data from loans that were given on multiple businesses as parts of the COVID-19 relief plan in the US and are available via SafeGraph. The loan value is normalized to span a month and is also normalized by the true number of jobs reported in each business and then is multiplied by the number of workers at POI $j$ in the bipartite network. The loan data also report the race of its owner (owners can also choose not to answer) so we can determine if the business is a minority or a non-minority-owned business. To fill the missing data we use the percentage of minority businesses as a real-valued score for the corresponding business.  
\end{compactitem}

Finally, we determine the external assets and liabilities 

\begin{compactitem}
    \item For each POI $j$ we use the total assets earned annually from all establishments, normalized them by month,divide by the total number of workers for the specific NAICS code that the POI belongs, and multiply by the number of workers at the POI. Finally, from this amount we subtract the revenue due to nodes within the network (i.e. inbound edges) and take the positive part in case the result is negative. The external liabilities of the POI are determined in a similar way with the only change that the total weight of the outbound edges is subtracted in the end. To ensure that \asref{assumption:beta_max_lt_one} holds throughout the experiments (and, thus, we have a unique equilibrium) we assert a minimum of \$100 for the external liabilities. Such a number is smaller than the order of most quantities and thus does not significantly affect the results.
    
    \item For each CBG $i$ the process is similar where for external assets we use data regarding the average income as reported by the US Census, following a similar procedure as in the POI case, but now normalized with respect to the estimated number of households and the inbound connections from the POIs. Regarding the average expenses of a household, we the national US average of $\sim \! \$ 63,000$. 
\end{compactitem}

This completes the construction of the SafeGraph semi-artificial data experiment. It is important to note that all values are ``normalized'' with respect to entities participating in the network. This normalization process avoids scenaria like the following: there is a CBG where a small percentage of household heads are employed in the corresponding businesses so the weights of the network (internal liabilities) would have different scale than the external assets and liabilities resulting in diminishing network effects, namely a lot of liabilities are owned outside the network and hence the values of $\beta_i$ are very small and thus the effect of the relative internal liability matrix $A$ vanishes (i.e. $\beta_{\max} = \| A^T \|_1 \approx 0$) and thus the clearing payments become approximately $\bar p \approx b \wedge (c - x + L \odot\bar z)$; essentially eliminating network effects. In fact, without network effects, the problem is solvable in $O(n \log n)$ time: we calculate $\bar p_j' = b_j \wedge (c_j -x_j + L_j)$ in $O(n)$ time, we sort the values $v_j p_j'$ in decreasing order in $O(n \log n)$ time and then pick the first entries such that the budget constraint is respected. The correctness of this argument relies on the use of an elementary algebraic argument.  

\section{Data Ethics Statement} \label{sec:data_ethics}

All data in this paper comes from publicly available datasets derived and used in prior research.  The financial networks from banks can be found in the public domain and their corresponding publications. The Venmo network data are publicly available on GitHub, and do not contain transaction amounts. The mobility data were obtained from a public SafeGraph dataset in which privacy techniques have been applied by the data provider, and, thus, no human subjects are identifiable. Because all data consists of public, pre-existing datasets without identifiable individuals, the current work is exempt from IRB review.

\end{document}